\documentclass[12pt,a4paper,twoside]{book}

\usepackage{graphicx}
\usepackage{setspace}

\usepackage{amssymb}
\usepackage{amsmath}
\usepackage{comment}
\usepackage[retainorgcmds]{IEEEtrantools}
\usepackage{graphicx}
\usepackage{epstopdf}
\usepackage{float}
\usepackage{listings}
\usepackage{color}
\usepackage{xcolor}
\usepackage[titletoc]{appendix}
\usepackage{tensor}

\usepackage[section]{placeins}

\usepackage{afterpage}
\usepackage{mathtools}

\newdimen\CdotAxis
\newcommand*{\CdotAux}[3]{%
  {%
    \settoheight\CdotAxis{$#2\vcenter{}$}%
    \sbox0{%
      \raisebox\CdotAxis{%
        \scalebox{#1}{%
          \raisebox{-\CdotAxis}{%
            $\mathsurround=0pt #2#3$%
          }%
        }%
      }%
    }%
    \dp0=0pt %
    \sbox2{$#2\bullet$}%
    \ifdim\ht2<\ht0 %
      \ht0=\ht2 %
    \fi
    \sbox2{$\mathsurround=0pt #2#3$}%
    \hbox to \wd2{\hss\usebox{0}\hss}%
  }%
}

\def\be{\begin{equation}}
\def\ee{\end{equation}}
\def\ba{\begin{eqnarray}}
\def\ea{\end{eqnarray}}

\begin{document}

\frontmatter

\begin{titlepage}
\vspace*{\stretch{1}}
\begin{center}
\includegraphics{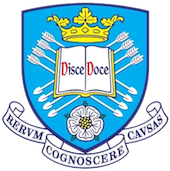}
\end{center}
\bigskip
\begin{center}\bfseries
{\LARGE Studies of inf\mbox{}lation and dark energy with coupled scalar f\mbox{}ields
}\\
\vspace{1.2cm}
{\LARGE Susan Vu}\\
\vspace{1cm}
{\Large Submitted for the degree of Doctor of Philosophy}\\
\bigskip
{\Large School of Mathematics and Statistics}\\
\bigskip
{\large September 2014}\\
\bigskip\bigskip\bigskip
{\large Supervisor: Prof. Carsten van de Bruck}
\end{center}
\vspace*{\stretch{3}}
\begin{center}
\textbf{\large University of Shef\mbox{}f\mbox{}ield}
\end{center}
\end{titlepage}

\newpage

\thispagestyle{empty}
\mbox{}


\newpage


\thispagestyle{empty}
\vspace*{\stretch{3}}
\begin{center}\scshape
Abstract
\end{center}
\vspace*{\stretch{1}}
\begin{quote}
Currently there is no def\mbox{}initive description for the accelerated expansion of the Universe at both early and late times; we know these two periods as the epochs of inf\mbox{}lation and dark energy.
Contained within this Thesis are two studies of inf\mbox{}lation and one in the context of dark energy.
The f\mbox{}irst study involves two noncanonical kinetic terms each in a two-f\mbox{}ield scenario, and their ef\mbox{}fects on the generation of isocurvature modes.
As a result, these terms af\mbox{}fect the isocurvature perturbations produced, and consequently the Cosmic Microwave Background.
In the following study, the impact of a sharp transition upon the ef\mbox{}fective Planck mass is considered in both a single-f\mbox{}ield and two-f\mbox{}ield model.
A feature in the primordial power spectrum arising from these transitions is found in single-f\mbox{}ield models, but not for two-f\mbox{}ield models.
The f\mbox{}inal model discussed is on the subject of dark energy.
A type of nonconformal coupling is examined namely the ``disformal'' coupling; in this scenario a scalar f\mbox{}ield is disformally coupled to matter species.
Two consistency checks are undertaken, the f\mbox{}irst to provide a f\mbox{}luid description and the second, a kinetic theory.
From this, observables are constructed and used to create constraints on the individual coupling strengths.
\end{quote}
\vspace*{\stretch{7}}


\newpage 
\newpage\null\thispagestyle{empty}\newpage
\thispagestyle{empty}
\vspace*{2cm}
\begin{center}
\emph{To my parents}
\end{center}

\doublespacing
\tableofcontents
\singlespacing

\newpage
\thispagestyle{empty}
\mbox{}

\newpage
\chapter{Preface}
This thesis is submitted in partial fulf\mbox{}ilment of the requirements for a degree of Doctor of Philosophy in Mathematics. 
The supervisor on the project has been Carsten van de Bruck. 
The thesis has been made for the greater part of the contents by the Author's original work.

\begin{itemize}
\item Chapter 1 contains introductory material including a brief summary of the discoveries made in Cosmology.
\item Chapter 2 is based on the published work in Physical Review D in February 2013 with Carsten van de Bruck \cite{vandeBruck:2012mr}.
All analytical and numerical calculations and production of f\mbox{}igures were done by the Author.
\item Chapter 3 is based on the published work in Physical Review D in November 2014 \cite{Ashoorioon:2014yua}, done in collaboration with Amjad Ashoorioon, Carsten van de Bruck and Peter Millington.
All analytical work, numerical calculations and f\mbox{}igures presented in this Chapter were conducted by the Author.
\item Chapter 4 is based on the published work in Physical Review Letters in October 2013 \cite{PhysRevLett.111.161302}, done in collaboration with Carsten van de Bruck and Jack Morrice.
All analytical work performed in Section~\ref{sec:chap4sec1} was performed by the Author.
Other contributions include numerical simulations and pieces of analytical work include those in Sections~\ref{sec:chap4sec2} and \ref{sec:chap4sec3}.
\item Chapter 5 is a summary of the three projects conducted. 
\end{itemize}

\newpage
\chapter{Acknowledgements}
First and foremost, I would like to thank Carsten for his supervision and guidance during my time in Shef\mbox{}f\mbox{}ield.
I cannot forget his tireless energy and enthusiasm during meetings, especially those about my Thesis. 
Above all, he gave me the conf\mbox{}idence to f\mbox{}inish my PhD.
%
I would also like to give special thanks to Caitlin for giving the advice I needed.
Thanks also go to Amjad, Peter and Jack for being wonderful collaborators.
\newline
\\
I would also like to thank the Science \& Technology Facilities Council (STFC) for the f\mbox{}inancial support over the course of the PhD and the Tottenham Grammar Foundation for their support through the Somerset Special Award.
\newline
\\
Finally, I could not have made it without the unconditional love and support from Tom and my family.

\cleardoublepage
\addcontentsline{toc}{chapter}{\listfigurename}
\listoffigures

\mainmatter

\doublespacing

\chapter{Introduction}
\noindent
Cosmological inf\mbox{}lation has been widely accepted as the theory giving rise to our Universe and over the decades, cosmologists have attempted to understand the mechanism behind this.
More recently, evidence of a second period of accelerated expansion in our Universe has been found.
However, the reasons behind this late-time acceleration are unknown.
It could be as a result of a new energy form, namely dark energy, or due to modif\mbox{}ications of gravity.
\newline
\\
This introductory chapter is presented as follows: f\mbox{}irst, a description of the three tests of General Relativity which is proceeded by a summary of the discoveries made in the f\mbox{}ield of cosmology. Following this, the problems in cosmology will be presented alongside its solution --- inf\mbox{}lation.
Lastly, a description of cosmological perturbation theory is given before ending with an outline of the Thesis.

\newpage
\section{General Relativity and the three tests}
The Theory of General Relativity by Einstein was a culmination of work attained during the period from 1905 to 1915.
Many papers detailing the technical nature were published during this time, but it was not fully presented until 1915 and 1916 \cite{Einstein:1915ca, Einstein:1916vd}.
However, tests were required to scrutinise his theory and three were proposed: the perihelion precession of the planet Mercury, the def\mbox{}lection of light by massive bodies and gravitational redshift of light near compact massive bodies.
\newline
\\
We will work with the mathematical framework provided by the Theory of General Relativity, the most successful theory describing gravity in curved spacetime.
In order to study models which are extensions of General Relativity, it must be ensured that these models satisfy the three historical tests.

\subsubsection{The f\mbox{}irst test --- Mercury's perihelion precession}
During the 19th Century, a French mathematician named Le Verrier undertook an extensive study into the nature of the planets.
Through the usage of Newtonian mechanics, Le Verrier had discovered that when all the motion of that surrounding planets had been accounted for, it was not enough to explain the perihelion precession of the planet Mercury.
The unaccounted precession was calculated to be approximately 38 arcseconds per century \cite{LeVerrier:1859}.
In his Letter of 1859, he gave possible reasons for this discrepancy which included an undiscovered planet and objects, such as asteroids, orbiting between the Sun and Mercury.
However it was not until November of 1915 that Mercury's missing perihelion precession would be solved by the Theory of Relativity.
In 1915, Einstein had calculated it to be 43 arcseconds per century, which was in agreement with the current-day observational values of $ 45'' \pm 5'' $ per century \cite{Einstein:1915bz}.

\subsubsection{The second test --- Light def\mbox{}lection by massive bodies}
Another test involved the def\mbox{}lection of light in the presence of a massive object.
This massive object induces the light def\mbox{}lection by causing a change in the curvature of space-time in its locality.
The bending of light near a celestial object, in this case the Sun, was f\mbox{}irst stated and shown by Soldner in 1804 through a calculation using Newtonian mechanics \cite{Soldner:1804}.
Through his calculation, Soldner found the amount of def\mbox{}lection of the light ray caused by the Sun is 0.84'', and this f\mbox{}igure was also calculated by Einstein in 1911 \cite{Einstein:1911vc} by using the equivalence principle.
The equivalence principle is the assumption that the gravitational force experienced by a local observer is the same as that of an accelerating reference frame.
However, Einstein corrected the calculation in 1916 by using his Theory of Relativity and from this, found that the amount of light def\mbox{}lection doubled to 1.75''.
This led to the solar eclipse experiments of May 29, 1919 performed by Eddington and Crommelin \cite{Dyson:1920} whose purpose was to conf\mbox{}irm if gravitation had an ef\mbox{}fect on light and if yes, whether Newtonian mechanics or Einstein's Theory of Relativity was the correct description of gravity.
It was noted that this eclipse would be particularly favourable due to the unexpected number of bright stars in the background, which gave many test stars to compare against.
The path of the eclipse was to pass over Northern Brazil and the Atlantic Ocean, eventually reaching S{\~a}o Tom{\'e} and Pr{\'i}ncipe, an island country of\mbox{}f the West Coast of Africa, before passing over the continent.
As a result, two expeditions were commissioned: one to be based in Sobral in Northern Brazil led by Crommelin, 
and the other on the island of Pr{\'i}ncipe to be led by Eddington.
Both experiments performed were found to be in agreement with the prediction made by Einstein's Theory of Relativity, with results from the experiment led by Crommelin found to be $ 1.98'' \pm 0.16'' $ and Eddington $ 1.61'' \pm 0.4'' $.

\subsubsection{The third test --- Gravitational redshift of light}
The gravitational redshift of light close to compact massive bodies was the last test proposed by Einstein.
It was conf\mbox{}irmed by Adams in 1925 \cite{Adams:1925im, 1925PNAS...11..382A} through the study of a binary star system in Sirius.
Previous to this, it was also Adams in 1915 who discovered that Sirius A had a faint companion, specif\mbox{}ically, a white dwarf - the f\mbox{}irst to be found, and was named Sirius B \cite{Adams:1915zz}.
It was Eddington in 1924 \cite{1924MNRAS..84..308E} that proposed that these white dwarfs are of very high density in comparison to other stellar objects; his statement was made whilst specif\mbox{}ically discussing the binary system of Sirius.
Both the ef\mbox{}fective temperature and absolute magnitude of Sirius was used to calculate the radius, and in combination with an assumption of a mass range, came to the conclusion that white dwarfs are extremely high in density.
Alongside this discussion, Eddington also gave an alternative explanation in which the stellar object has instead a very low ef\mbox{}fective temperature, and the ability to generate the same spectrum which an observer must be satisf\mbox{}ied with.
In order to determine one of these two ideas, the gravitational redshift of the spectrum generated by the white dwarf must be measured.
For Sirius B, a very high density stellar object, Eddington calculated the corresponding Doppler shift to be $ 20 \, \mathrm{km} \, \mathrm{s}^{-1} $.
In the 1925 paper by Adams, he had used a 100 inch ref\mbox{}lector telescope - an upgrade from the 60 inch telescope used to detect Sirius B, to make spectroscopic observations. 
These observations and taking into account the radial velocity of Sirius led to Adams calculating the displacement of the spectrum of Sirius B to be $ 21 \, \mathrm{km} \, \mathrm{s}^{-1} $ --- in agreement with Eddington.
\newline
\\
There are other tests of General Relativity aside from the three historical tests stated; one such example uses binary pulsars, with the f\mbox{}irst system discovered in 1974 by Hulse and Taylor \cite{Hulse:1974eb}.
Observations of binary star systems have shown that the pulsar's orbit contracts over time as a result of the emission of energy through gravitational waves.
\newline
\\
In 1917, Einstein's famous paper titled ``Kosmologische Betrachtungen zur allgemeinen Relativit\"{a}tstheorie'' \cite{1917SPAW.......142E} was published where he states the f\mbox{}ield equations and from this, obtains a model where the Universe is both closed and static, through the term which he introduced as the cosmological constant.
A positive cosmological constant was inserted into the f\mbox{}ield equations as a counter term to the ef\mbox{}fects of gravity --- in a Newtonian model of the Universe, all the structures within will collapse under gravity, and the term being positive ensured that the Universe was both closed and f\mbox{}inite.
Einstein and other physicists such as de Sitter \cite{1917KNAB...19.1217D} did not agree with the presence of the term.
However, the term has been conf\mbox{}irmed to be nonzero by supernovae observations performed in the late 1990s.
\newline
\\
We shall begin with a brief summary of the important discoveries made in cosmology.

\section{Discoveries in Cosmology}
\subsubsection{Expansion of space}
In Spring of 1917, an American astronomer named Slipher published a paper containing 4 years of study into the light emitted from surrounding galaxies \cite{Slipher:1917zz}.
In this study starting in 1912, Slipher had stated that the size of the telescope is irrelevant in obtaining the spectroscopic measurements of the galaxies but instead, it was the camera of the spectrograph that had the greatest inf\mbox{}luence, specif\mbox{}ically the lens speed. 
The spectrograph was attached to a 24 inch refractor telescope at the Lowell Observatory in Arizona, USA, and used to take between 40 and 50 spectrograms of 25 spiral galaxies.
By calculating the Doppler shift of the spectra lines, Slipher inferred the radial velocities of the galaxies and found that the average radial velocity was 570 $ \mathrm{km} \, \mathrm{s}^{-1} $.
In addition to this, most of the spiral galaxies were found to be moving away from the Solar System; Slipher had discovered that the light from the majority of the galaxies were shifted towards the infra-red section of the electromagnetic spectrum, known as redshift.
Hubble extended this study during the 1920s, resulting in his 1926 paper classifying the various types of galaxies in the Universe based on their shape, mass and luminosity \cite{Hubble:1926yw}.
Further study into the various galaxies' velocities and their distances led to the formation of the velocity-distance relation, presented in 1929 \cite{Hubble:1929ig}.
However, Hubble did not explicitly state the now-known relation $ v = H_0 d $ where $ v $ is the receding velocity, $ d $ is distance and $ H_0 $ is Hubble's constant, and it was in 1927, a Belgian Roman Catholic priest named Lema\^{i}tre who stated the relation between the recession velocity and distance \cite{Lemaitre:1927zz}, and added the statement that it is natural for the galaxies to have a velocity which is receding with an expanding Universe.
Furthermore in 1928, Robertson, an American physicist, also found a similar relation to Lema\^{i}tre \cite{Robertson:1928}; Robertson presented a review of the important f\mbox{}indings in relativistic cosmology in 1933 \cite{Robertson:1933zz}.
\newline
\\
Two extensive studies into both the static and expanding Universe were f\mbox{}irst presented in 1922 and 1924, respectively by a Russian mathematician named Friedmann \cite{Friedman:1922kd, 1924ZPhy...21..326F}.
With these two papers, he proposed a space-time which is allowed to grow and/or shrink depending on time, and proceeded to write down the equations to which we now know as the Friedmann equations.
The dif\mbox{}ference between the two papers is the sign associated with the curvature of the Universe; the f\mbox{}irst of the two papers, Friedmann considered a Universe with a positive constant curvature and hence, a static model with the Universe ending in a Big Crunch.
In the other paper, a negative constant curvature leading to an expanding model.
\newline
\\
In the very same paper of 1927, Lema\^{i}tre independently obtained the same results as Friedmann.
However, the work was written in French and published in an obscure journal in Brussels and hence, was relatively unknown to most of the scientif\mbox{}ic community; it was later published in English in 1931 \cite{Lemaitre:1931zza}.
The independent work performed by Lema\^{i}tre gave a greater appreciation for the two papers published by Friedmann.
Lema\^{i}tre was searching for a compromise between the solutions given by de Sitter \cite{1917MNRAS..78....3D} and Einstein; the de Sitter solution involved the assumption of no matter existing in the Universe and so the density equalling zero, whereas Einstein's solution contained matter and led to a relation between the matter density and the radius of the Universe.
In order to make a compromise between the two solutions, matter must exist in the Universe and adding to this, Lema\^{i}tre proposed that the radius of the Universe was allowed to vary.
The solution he obtained in the paper revealed that the Universe naturally expands without bounds. 
Towards the end of his work, he stated that there still a need to f\mbox{}ind the reason for this expansion, which Lema\^{i}tre would do in 1931.

\subsubsection{The hot Big Bang}
The idea of the Universe starting initially from a single point --- the Big Bang theory, was proposed by Lema\^{i}tre in his 1931 Letter \cite{Lemaitre:1931zzb} and the idea further developed in 1933 \cite{Lemaitre:1933, 1933ASSB...53...51L}.
\newline
\\
One of the consequences of a hot and dense Big Bang is a remnant of a cooled thermal background radiation.
The Cosmic Microwave Background (CMB) was discovered by Penzias and Wilson at Bell Laboratory in 1965, resulting in the award of the 1978 Nobel Prize for Physics \cite{Penzias:1965wn}.
A 20 foot horn ref\mbox{}lector antenna was used to make the measurements and from this, a temperature of the background radiation was recorded to be $ 3.5 \pm 1 \, \mathrm{K} $.
The result was independent of direction revealing the Universe to be isotropic.
The idea of a cooled thermal background was f\mbox{}irst postulated by Dicke and collaborators during the 1960s which also led to a 1965 paper on the CMB temperature  \cite{Dicke:1965zz}; Penzias and Wilson cited this paper as it provided a possible explanation for the existence of background radiation.
Dicke and collaborators measured the temperature using a radiometer with a receiving horn attached which has the ability to measure radiation at the wavelength of 3cm.
They found that with a wavelength of 3cm, the measured temperature of the cosmic blackbody radiation is $ 3.5 \pm 0.5 \, \mathrm{K} $.
In addition, the Letter was written with full acknowledgement of the work performed by Penzias and Wilson of Bell Telephone Laboratories.
Both their works, led to the acceptance of the Big Bang theory.
\newline
\\
\subsubsection{1980s --- Inf\mbox{}lation and new species (dark matter)}
There were three major problems of the Hot Big Bang model that surfaced by 1980; they were namely the f\mbox{}latness, horizon and magnetic monopole problems.
In order to solve the problems, the Universe must undergo a period of accelerated expansion. 
Guth in 1981 outlined the problems clearly and proposed the idea of cosmological inf\mbox{}lation \cite{Guth:1980zm} as a remedy for this; 
other authors who also independently worked on this concept are Albrecht and Steinhardt \cite{Albrecht:1982wi} and Linde \cite{Linde:1981mu, Linde:1983gd}.
We will come back later to the theory of inf\mbox{}lation.
\newline
\\
In the early 1930s Zwicky discovered that ordinary matter accounts for a very small proportion of the total matter in the Universe.
By measuring the velocity dispersion of galaxy clusters through the viral theorem \cite{Zwicky:1933gu}, he found that the visible matter was not enough to account for the fast movement of these galaxies.
He postulated that there must be some form of invisible matter contained within these clusters and this was later realised to be cold dark matter (CDM).
Several cosmologists in the 1980s including Peebles \cite{Peebles:1982ff}, Blumenthal \cite{Blumenthal:1984bp} and Davis \cite{Davis:1985rj}, presented their f\mbox{}indings which proposed that CDM was required for structure formation.
Up to now, the models that were considered only contained ordinary matter (baryons, radiation and massless neutrinos) and were unable to produce the observed primordial f\mbox{}luctuation amplitude.
The addition of the word ``cold'' was to af\mbox{}f\mbox{}irm that the dark matter was nonrelativistic during the lifetime of the Universe.
In the paper by Peebles in 1982 \cite{Peebles:1982ff}, CDM was included alongside ordinary matter and as a result, it caused the amplitude of the primordial f\mbox{}luctuations to change such that it could satisfy the current-day observations.

\subsubsection{Dawn of modern cosmological observations}
The Cosmic Background Explorer (COBE) in 1992 revealed that the background radiation of the Universe is uniform on large angular scales.
By the removal of various components such as the uniform background and dipole generated by the Earth's motion, primary temperature anisotropies were uncovered with an amplitude of $ \Delta T/ T \sim 10^{-5} $ \cite{Smoot:1992td, Wright:1992tf}.
The FIRAS instrument on COBE which investigated the blackbody nature of the CMB, produced the f\mbox{}irst precise measurement of the 
CMB temperature \cite{Mather:1993ij} recorded to be $ 2.726 \pm 0.010 \, \mathrm{K} $; the most recent measurement was taken in 2009 \cite{Fixsen:2009ug} with $ 2.72548 \pm 0.00057 \, \mathrm{K} $.
\newline
\\
In 1995, Goobar and Perlmutter, two members of the would-be Supernova Cosmology Project released a paper which discussed extending the apparent magnitude-redshift relation from $ z \sim 0.5 $ to $ z \sim 1 $ through the usage of Type Ia supernovae, and using this to measure the cosmological constant and the density parameter $ \Omega_0 $.
Other papers followed which described the techniques and searches for the Type Ia supernovae between the redshifts $ z = 0.35 - 0.65 $ \cite{1996NuPhS..51...20P, Perlmutter:1996ds}.
In 1998, observations of Type Ia supernovae showed that the cosmological constant is nonzero and hence, the Universe is currently experiencing a period of accelerated expansion. 
This was found by two groups, the Supernova Cosmology Project \cite{Perlmutter:1997zf, Perlmutter:1998np}, and Supernova Search Team \cite{Garnavich:1997nb, Riess:1998cb}, 
independently in 1998.
\newline
\\
In 2001, the Wilkinson Microwave Anisotropy Probe (WMAP) satellite was launched for a mission of over 12 years, to take precise measurements 
of the CMB \cite{Spergel:2003cb}; the f\mbox{}inal Nine-year WMAP data is presented in \cite{Hinshaw:2012aka}.
Recently, Planck, a satellite launched by the European Space Agency, released their results it found that the Universe is composed of 31.7\% matter (of the total Universe, 4.9\% is ordinary matter (baryonic) and 26.8\% is cold dark matter) and 68.3\% dark energy \cite{Ade:2013zuv}.
\newline
\\
Alongside this observational evidence, the theories developed in the early 1980s by Peebles, Blumenthal et al. \!and Davis, this led to the establishment 
of the $\Lambda \mathrm{CDM}$ model; the concordance model in cosmology. 
$\Lambda$ is known as the cosmological constant representing dark energy in the Universe. 
It has been accepted that dark energy is the cause for the accelerated late-time expansion of the Universe.
\newline
\\
Up to now, there is no understanding of the mechanism that initiated inf\mbox{}lation nor knowing the true nature of dark energy.
Due to this, cosmologists have turned to alternative theories in order to explain both these predicaments.

\section{Preliminaries}
\label{sec:chap1prelim}
We shall use the mathematical formalism presented in General Relativity to describe the Universe.
We know that the Universe is both homogenous and isotropic on large scales due to the observations of the CMB, and is expanding from Type Ia supernovae observations.
This can be described in a mathematically succinct form known as the Friedmann-Robertson-Walker (FRW) line element \cite{Robertson:1935zz, Walker01011937}.
The line element is maximally spatially symmetric and presented in the spherical coordinates. 
It is given by
\be
\mathrm{d}s^2 = - \, \mathrm{d}t^2 + a(t)^2 \left[ \frac{ \mathrm{d}r^2}{1- \kappa r^2} + r^2( \mathrm{d} \theta^2 + \sin^2 \theta \, \mathrm{d} \phi^2) \right]~,
\ee
where $ a(t) $ is the scale factor which is a function of the cosmic time $ t $ describing the expansion of space, and $ \kappa $ is the spatial curvature.
The type of spatial curvature is assigned to a value of $ \kappa $; the values that $ \kappa $ can take are $ \kappa = {-1, 0, 1} $ describing an open, f\mbox{}lat and closed spacetime, respectively.
\newline
\\
In general, a line element can be expressed by the following
\be
\label{lineelement}
\mathrm{d}s^2 = g_{\mu \nu} \mathrm{d}x^{\mu} \mathrm{d}x^{\nu}~,
\ee
where $ g_{\mu \nu} $ is the metric of the spacetime. 
In this Thesis,  a f\mbox{}lat space is used $ \kappa = 0 $, and with an expanding spacetime where the metric is $ g_{\mu \nu} = \mathrm{diag}(-1,a^2,a^2,a^2) $, given in Cartesian coordinates.
\newline
In Einstein's 1917 paper \cite{1917SPAW.......142E}, a cosmological constant $ \Lambda $ was introduced in order to counter gravity and obtain a static Universe.
The Einstein-Hilbert action for the case with a cosmological constant and additional matter f\mbox{}ields $ S_{\mathrm{M}} $ is 
\be
S = \frac{1}{2} \int \mathrm{d}^{4} x \sqrt{-g} [  M_{\mathrm{Pl}}^2 R - 2 \Lambda ] + S_{\mathrm{M}}~,
\ee
where $ \mathrm{d}^{4} x \sqrt{-g} $ is the invariant volume element, $ g $ the determinant of the metric $ g_{\mu \nu} $, $ R $ is the Ricci scalar and $ M_{\mathrm{Pl}} $ is the reduced Planck mass with $ M_{\mathrm{Pl}}^2 = (8 \pi G)^{-1} $.
\newline
\\
By using the Principle of Least Action, this yields the Einstein f\mbox{}ield equations
\be
\label{einsteinequation}
G_{\mu \nu} + \Lambda g_{\mu \nu} = 8 \pi G \, T_{\mu \nu}~,
\ee
where $ G_{\mu \nu} $ is the Einstein tensor:
\be
\label{einsteintendef}
G_{\mu \nu} \equiv R_{\mu \nu} - \frac{1}{2} R g_{\mu \nu}~.
\ee
It is composed of the metric $ g_{\mu \nu} $, Ricci tensor $ R_{\mu \nu} $ and 
the Ricci scalar $ R = R \indices{^{\alpha}_{\alpha}} = g^{\alpha \beta} R_{\alpha \beta} $. The right hand of this equation 
contains the matter within the spacetime, represented by the matter stress-energy-momentum tensor $ T_{\mu \nu} $. 
The stress-energy momentum tensor is comprised of the energy density $ \rho $ and pressure $ p $ by the matter in question.
It is def\mbox{}ined as $ T \indices{^\mu_\nu} = \mathrm{diag}(-\rho,p,p,p) $
Eq.~\eqref{einsteinequation} describes the ef\mbox{}fect matter has on the curvature of the spacetime.
\newline
\\
In this Thesis, we will not be considering the Einstein f\mbox{}ield equations with the inclusion of the cosmological constant $ \Lambda $.
\newline
\\
A perfect f\mbox{}luid description of the system can be made. 
For this, the stress-energy momentum tensor can be written as
\be
\label{energymtensor}
T_{\mu \nu} = (\rho + p)u_{\mu} u_{\nu} + p g_{\mu \nu}~,
\ee
where $ u_{\mu} $ is the 4-velocity of a comoving observer where 
\be
u^{\mu} = \frac{d x^{\mu}}{d \lambda} = \, \delta ^{\mu}_{0}~, \qquad u_{\mu} u^{\mu} = -1~,
\ee
where $ \lambda $ is an af\mbox{}f\mbox{}ine parameter associated with a geodesic in $ g_{\mu \nu} $.
\newline
\\
Considering the FRW spacetime in Eq.~\eqref{lineelement} in general terms
\be
\label{frw}
\mathrm{d}s^2 = -\, \mathrm{d}t^2 + a^2(t) \delta_{ij} \mathrm{d}x^i \mathrm{d}x^j~,
\ee
and applying this to the Einstein f\mbox{}ield equations in Eq.~\eqref{einsteinequation} and the energy-momentum tensor in Eq.~\eqref{energymtensor}, we obtain the Friedmann and acceleration equations respectively
\begin{subequations}
\label{friedmannacceleration}
\begin{align}
3 H^2 &= 8 \pi G \rho~,	\\
\label{acceq}
-2 \dot{H} &= 8 \pi G (\rho + p)~,
\end{align}
\end{subequations}
where $ H = \dot{a}/a $ is the Hubble parameter. Note that dots represents derivatives with respect to cosmic time.
\newline
The Einstein tensor fulf\mbox{}ils the Bianchi identities and this results in $ \nabla_{\mu} G^{\mu \nu} = 0 $.
By taking the covariant derivative of the Einstein f\mbox{}ield equations, this leads to the conservation equations
of the matter f\mbox{}ields
\be
\nabla_{\mu} T \indices{^{\mu}_{\nu}} = 0~.
\ee
For a system consisting of more than one f\mbox{}luid, the energy-momentum tensor will be composed of the dif\mbox{}ferent components; the total energy density is $ \rho = \sum\limits_i \rho_i $ and the total pressure is $ p = \sum\limits_i p_i $.
By taking into account an expanding Universe (FRW spacetime) a conservation equation for each of the f\mbox{}luids can be derived
\be
\label{conservation}
\dot{\rho}_i + 3 H ( 1 + w_i ) \rho_i = 0~,
\ee
where the parameter $ w_i $ is the equation of state and is used to denote the matter in the system.
It is given by
\be
w = \frac{p}{\rho}~,
\ee
with some examples of the dif\mbox{}ferent equations of state shown below
\[ w = \left\{
  \begin{array}{l l l}
    0 			& \quad \text{matter-dominated}~,	\\
    1/3	 	& \quad \text{radiation-dominated}~,	\\
    -1 			& \quad \text{dark energy}.
  \end{array} \right.\]
We shall later study an extension of General Relativity in which the conservation equation in Eq.~\eqref{conservation} is modif\mbox{}ied; in this model, there is an interaction between the scalar f\mbox{}ield and the matter species.

\section{Inf\mbox{}lation}
In the late 1970s, cosmologists realised there were problems with cosmology in its current state.
They were encapsulated into three problems, these are the f\mbox{}latness, horizon and magnetic monopole problems.
In Guth's 1981 paper \cite{Guth:1980zm}, the f\mbox{}latness and horizon problems were stated along with a mechanism that solves all these dif\mbox{}f\mbox{}iculties --- inf\mbox{}lation.

\subsection{The problems}
\subsubsection{Flatness problem}
The f\mbox{}latness problem was stated by Dicke and Peebles in 1979 \cite{Dicke:1900mn}.
Through the observations made of the CMB, it is known that the energy density of the Universe is near critical density $ \rho_{\mathrm{crit}} $.
When the Universe is at critical density it means that the Universe is spatially f\mbox{}lat i.e. $ \kappa = 0 $.
This can be shown by rewriting the Friedmann equation with the reinstatement of the curvature $ \kappa $, in terms of the density parameter, $\Omega$
\begin{equation}
\label{densityparam}
\Omega - 1 = \frac{\kappa}{a^{2}H^{2}}~,
\end{equation}
with the parameter $ \Omega $ def\mbox{}ined as
\begin{equation}
\Omega = \frac{\rho}{\rho_{\mathrm{crit}}}~,
\end{equation}
and the critical density $\rho_{\mathrm{crit}}$ is given by
\begin{equation}
\rho_{\mathrm{crit}} = \frac{3 H^2}{8 \pi G}~.
\end{equation}
The Universe has undergone both the radiation and the matter-dominated epochs, we shall consider both.
By using the conservation equation stated in Eq.~\eqref{conservation} and the relevant equation of state, a relationship between the energy density and scale factor is established, which is used in the f\mbox{}irst of the Friedmann equations to relate the scale factor to time.
\begin{itemize}
\item{\textbf{Radiation-dominated epoch}}
\newline
Using the method outlined in the section above, we can f\mbox{}ind relationships between energy density and scale factor and hence, the scale factor and time 
\be
\rho \sim a^{-4}		\qquad \Longrightarrow	\qquad	a \sim t^{\frac{1}{2}}~,
\ee
and use this in the rewritten Friedmann equation in Eq.~\eqref{densityparam}.
For a radiation-dominated epoch where the equation of state is $ w_\gamma = 1/3 $, we f\mbox{}ind
\be
aH  \sim t^{-\frac{1}{2}}~,
\ee
and inserting this into the Friedmann equation yields
\begin{equation}
|\Omega - 1| \sim t~.
\end{equation}

\item{\textbf{Matter-dominated epoch}}
\newline
Following the same method applied to the radiation-dominated epoch, the relationships between energy density and scale factor, and therefore the scale factor and time in a matter-dominated period with $ w_{\mathrm{m}} = 0 $ are
\be
\rho \sim a^{-3}		\qquad \Longrightarrow	\qquad	a \sim t^{\frac{1}{3}}~,
\ee
and the resulting relation is formed
\be
aH \sim t^{-\frac{1}{3}}~,
\end{equation}
with the resulting Friedmann equation 
\begin{equation}
|\Omega -1| \sim t^{\frac{2}{3}}~.
\ee
\end{itemize}
In both the radiation and matter-dominated epochs, it is clear that $ |\Omega - 1| $ is an increasing function of time. 
From the results in the Planck 2013 paper \cite{Ade:2013zuv}, the total present day density parameter $ \Omega_{\mathrm{tot}} = \sum\limits_i \Omega_i + \Omega_{\Lambda} $ where $ i $ denotes the matter f\mbox{}ields such as radiation, baryons and dark matter, is
\be
\Omega_{\mathrm{tot}} = 1.0005 \pm 0.0033~.
\ee
This implies that $ | \Omega - 1 | $ must be extremely close to zero in the early Universe, indicating that the initial conditions of our Universe must be incredibly f\mbox{}ine-tuned.
However, such conditions are extremely unlikely, and cosmologists are continuously searching for dynamical reasons.

\subsubsection{Horizon problem}
The horizon problem was f\mbox{}irst posed by Misner in 1967 \cite{Misner:1967uu}.
We know from observations made of the CMB that the background radiation and therefore temperature, is approximately uniform in the Universe.
From this, we can infer that regions on opposite sides of the Universe must have been in causal contact in the past.
However, when extrapolating the light rays backward towards the surface of last-scattering, it is shown that these regions cannot have communicated. 
The surface of last-scattering is the point in time where the photons decoupled from charged matter.
We now introduce the particle horizon (or comoving horizon) $ \eta $, and the comoving Hubble radius $ (aH)^{-1} $, the particle horizon is def\mbox{}ined as
\be
d_{\mathrm{hor}} = \int^t_0 \frac{dt'}{a(t')}~,
\ee
and is the distance that light has travelled since the start of the Big Bang; a paper on the dif\mbox{}ferent horizons used in cosmology was presented in 1956 by Rindler \cite{Rindler:1956yx}.
The comoving Hubble radius is the distance that determines whether two regions in the Universe can communicate or not. 
If we consider two particles that are separated by a distance greater than the horizon, there is no possibility of these particles interacting.
However, if the distance is now greater than the comoving Hubble radius (and less than the horizon), the two particles could have possibly communicated in the past.
The horizon problem involves answering why the CMB has a roughly uniform temperature, when the Universe contains many causally disconnected regions, approximately $ \sim 10^{83} $ regions.

\subsubsection{Monopoles problem}
The problem of the magnetic monopoles was documented in 1978 by Zeldovich and Khlopov \cite{Zeldovich:1978wj}, and later in 1979 by Preskill \cite{Preskill:1979zi}.
It was restated by Guth and Tye \cite{Guth:1979bh} in 1979.
In all the papers stated above, the Grand Unif\mbox{}ied Theory (GUT) was under consideration.
The GUT is a theory that treats the electromagnetic, string and weak forces as one force at very high temperatures, similar to those during the very early stages of the Universe.
The aim is to combine this unif\mbox{}ied force with classical gravity in the hopes to understand the physics that takes place below the Planck scale.
A feature of GUTs is that they predict the existence of stable magnetic monopoles at very high temperatures.

\subsection{The solution}
As a solution to three problems in cosmology, an inf\mbox{}lationary period is required, which can be summarised in the relation
\be
\ddot{a} > 0~.
\ee
Using the def\mbox{}inition of the Hubble parameter, the condition for inf\mbox{}lation above can be rewritten in terms of the comoving Hubble radius:
\be
\frac{d}{dt} \bigg( \frac{1}{aH} \bigg) < 0~.
\ee
Inf\mbox{}lation is able to solve the f\mbox{}latness problem as it causes the term $ | \Omega - 1 | $ to drive towards zero, and hence present day density parameter $ \Omega_0 $ stays approximately at unity.
The horizon problem is solved as the comoving Hubble radius will always decrease during an inf\mbox{}lationary period, which allows more regions of the Universe to be in casual contact, and hence the CMB will be of uniform temperature.
Due to the increase in volume generated by inf\mbox{}lation, this causes any possible magnetic monopoles to be undetectable.

\subsection{Scalar f\mbox{}ields and slow-roll inf\mbox{}lation}
\label{sec:chap1scalar}
Let us now examine the second Friedmann equation. 
We can use this equation to study the accelerated expansion of the Universe; it can be written as the following
\be
\frac{\ddot{a}}{a} = - \frac{1}{6} ( 1 + 3 w ) \rho~.
\ee
In order for this type of expansion to occur, $ \ddot{a} > 0 $, which can be obtained when the equation of state is $ w < -1/3 $.
The scalar f\mbox{}ield is an example where this is possible, as we shall now show.
\newline
\\
The action for a canonical scalar f\mbox{}ield $ \phi $ with potential $ V(\phi) $ in Einstein gravity is given by
\be
S_{\phi} = \int \mathrm{d}^{4} x \sqrt{-g} \left[ \frac{R}{2} - \frac{1}{2} g^{\mu \nu} \partial_{\mu} \phi \, \partial_{\nu} \phi - V(\phi) \right]~,
\ee
and by varying this action with respect to the f\mbox{}ield, we can f\mbox{}ind its background equation of motion.
The background equation of motion for the f\mbox{}ield $ \phi $ can be calculated by using the Euler-Lagrange equation
\be
\label{eulerlagrange}
\frac{\delta \mathcal{L}}{\delta \phi} - \nabla_{\mu} \bigg( \frac{ \delta \mathcal{L} }{ \delta \partial_{\mu} \phi } \bigg) = 0~.
\ee
Strictly speaking, a scalar f\mbox{}ield is dependent on both time and space i.e. $ \phi = \phi (t, \bold{x}) $.
However, we know that the Universe is both homogeneous and isotropic on large scales and therefore, we can say that the scalar f\mbox{}ield is dependent only on time at the background level.
Using this statement and the Euler-Lagrange equation in Eq.~\eqref{eulerlagrange} yields the Klein-Gordon equation in FRW spacetime
\be
\label{kgscalar}
\ddot{\phi} + 3H \dot{\phi} + V'({\phi}) = 0~,
\ee
with primes denoting derivatives with respect to the associated f\mbox{}ield.
\newline
\\
The energy-momentum tensor of the scalar f\mbox{}ield is obtained by varying the action with respect to the metric $ g_{\mu \nu} $
\begin{subequations}
\begin{align}
T_{\mu \nu} & = - \frac{2}{\sqrt{-g}} \frac{ \delta S }{ \delta g^{\mu \nu} }~,		\\
\label{tensorem}
			 & = \partial_{\mu} \phi \, \partial_{\nu} \phi 
				- g_{\mu \nu} \left[ \frac{1}{2} g^{\alpha \beta} \partial_{\alpha} \phi \, \partial_{\beta} \phi 
								+ V(\phi) \right]~.
\end{align}
\end{subequations}
By considering the leading diagonal of this tensor $ T\indices{^\mu_\nu} = g^{\mu \alpha} T_{\alpha \nu} $, the energy density and pressure of the scalar f\mbox{}ield are
\begin{subequations}
\label{setenergypress}
\begin{align}
\rho_{\phi} & = \frac{1}{2} \dot{\phi}^2 + V(\phi)~, 		\\
p_{\phi} & = \frac{1}{2} \dot{\phi}^2 - V(\phi)~,
\end{align}
\end{subequations}
and hence the equation of state for the scalar f\mbox{}ield is
\be
w_{\phi} = \frac{\frac{1}{2} \dot{\phi}^2 - V(\phi)}{\frac{1}{2} \dot{\phi}^2 + V(\phi)}~.
\ee
The scalar f\mbox{}ield $ \phi $ is able to achieve the requirement for inf\mbox{}lation, $ w < -1/3 $, by simply through the condition $ \dot{\phi}^2 \ll V(\phi) $.
This is known as slow-roll inf\mbox{}lation.
\newline
\\
Slow-roll inf\mbox{}lation occurs when a scalar f\mbox{}ield (dominant for the case with more than one scalar f\mbox{}ield present) rolls down a potential $ V(\phi) $ \cite{Steinhardt:1984jj}.
In order to obtain this mechanism, the potential of the f\mbox{}ield must be signif\mbox{}icantly greater than its kinetic energy $ \dot{\phi}^2 \ll V(\phi) $ and very slow acceleration of the scalar f\mbox{}ield is required $ \ddot{\phi} \ll 1 $.
Under these slow-roll conditions, the Friedmann equation and the Klein-Gordon equation for the scalar f\mbox{}ield are
\begin{subequations}
\begin{align}
\label{slowroll1}
3 H \dot{\phi} & \simeq -V'({\phi})~,		\\
3 H^2 & \simeq V(\phi)~.
\end{align}
\end{subequations}
Adding to this, slow-roll inf\mbox{}lation is usually def\mbox{}ined using two parameters $ \epsilon $ and $ \eta $
\begin{align}
\epsilon & = -\frac{\dot{H}}{H^2} \simeq \frac{1}{2} \bigg( \frac{V'}{V} \bigg)^2~,		\\
\eta & = - \frac{ \ddot{\phi} }{H \dot{\phi}} \simeq \frac{V''}{V}~.
\end{align}
These slow-roll approximations hold when $ \epsilon \ll 1 $ and $ |\eta| \ll 1 $.
The formulation of the parameters were constructed by Liddle and Lyth in the early 1990s \cite{Liddle:1992wi, LiddLyth}.
A formal approach of the slow-roll parameters is presented in \cite{Liddle:1994dx}.
\newline
\\
A quantity that we shall require in the Thesis is the conformal time $ \eta $.
The relationship between conformal time and cosmic time is
\be
d \eta = \frac{dt}{a(t)}~.
\ee
During inf\mbox{}lation, this unit of time will run from large negative numbers to zero.
Another time variable which is useful in cosmology is the $ e $-fold number $ N $ given by
\be
N = \ln \bigg( \frac{a_{\mathrm{end}}}{a} \bigg)~,
\ee
where $ a_{\mathrm{end}} $ is the value of the scale factor at the end of inf\mbox{}lation.

\section{Cosmological perturbation theory}
We will now move to the subject of cosmological perturbations.
This area of cosmology has been studied extensively; see the following references for more detail \cite{Bardeen:1980kt, Kodama:1985bj, Mukhanov:1990me, Ma:1995ey, Lyth:1998xn}.
The linear perturbations of the metric $ g_{\mu \nu} $, i.e. $ \delta g_{\mu \nu} $, can be decomposed and classif\mbox{}ied depending on how they transform on spatial hypersurfaces; there are three types of metric perturbation: scalar, vector and tensor perturbations \cite{Bardeen:1980kt, Stewart:1990fm}.
In this Thesis, we shall be studying two types of perturbations: scalar and tensor perturbations.
\newline
\\
Scalar perturbations are necessary in order to obtain structures in the Universe; some examples of structures include stars and galaxies.
During the inf\mbox{}lationary epoch, there were quantum f\mbox{}ield f\mbox{}luctuations that evolved to yield small density inhomogeneities in the Universe.
This implies that there were seeds planted in the primitive Universe.
The primordial density inhomogeneities were measured to be approximately $ \delta \rho / \rho \sim 10^{-5} $ by COBE \cite{Smoot:1992td}.
Once the Universe reached the matter-dominated epoch, gravity causes these density inhomogeneities to grow in size leading to the formation of large scale structure.
The other type of perturbation, namely tensor perturbations (or primordial gravitational waves) are a prediction of inf\mbox{}lationary models, which can be used to eliminate such models.

\subsection{SVT decomposition}
The metric perturbations $ \delta g_{\mu \nu} $ can be decomposed into 4 scalar perturbations, 2 vector perturbations and 1 tensor perturbation.
We shall brief\mbox{}ly summarise the scalar-vector-tensor (SVT) decomposition of cosmological perturbations.

\subsubsection{Scalar perturbations}
The line element for the scalar metric perturbations in the FRW spacetime is
\be
\label{scalarlin}
\mathrm{d}s^2 = -(1 + 2 \Psi ) \, \mathrm{d}t^2 + 2\,a(t) \, B_i \, \mathrm{d}t \, \mathrm{d}x^i 	
					+ a^2(t)[ (1 - 2 \Phi) \delta_{ij} + 2 \, E_{ij} ] \mathrm{d}x^i \mathrm{d}x^j~,
\ee
where $ \Psi $, $ B $, $ \Phi $ and $ E $ are scalar perturbations of the metric.
The gauge transformations can be made
\begin{subequations}
\begin{align}
t \; & \rightarrow \; t + \alpha~,		\\
x^{i} \; & \rightarrow \; x^i + \delta^{ij} \, \beta_j~.
\end{align}
\end{subequations}
where $ \alpha $ is a scalar function and $ \beta_i $ is a divergence-free 3-vector.
\newline
\\
Under these transformations the 4 scalar metric perturbations transform via the following
\begin{subequations}
\label{scaltrans}
\begin{align}
\Psi 		\; & \rightarrow \; \Psi - \dot{\alpha}~,		\\
B 		\; & \rightarrow \; B + a^{-1} \alpha - a \dot{\beta}~,		\\
E 		\; & \rightarrow \; E - \beta~,		\\
\Phi 		\; & \rightarrow \; \Phi + H\alpha~.
\end{align}
\end{subequations}
The matter perturbations given by $ \delta \rho $ and $ \delta p $ --- which are the perturbations in the energy density and pressure --- and the perturbation of the momentum potential $ \delta q $ are also gauge-dependent and transform as
\begin{subequations}
\label{scalpertmetric}
\begin{align}
\delta \rho \; & \rightarrow \; \delta \rho - \dot{\rho} \alpha~,		\\
\delta p 	\; & \rightarrow \; \delta p - \dot{p} \alpha~,			\\
\delta q 	\; & \rightarrow \; \delta q + (\rho + p ) \alpha~.
\end{align}
\end{subequations}

\subsubsection{Vector perturbations}
The vector metric perturbations $ S_i $ and $ F_i $ are given by this line element
\be
\mathrm{d} s^2 = - \mathrm{d} t^2 + 2a S_i \mathrm{d}t \mathrm{d} x^i+ a^2 [ \delta_{ij} + 2 F_{(i,j)} ] \mathrm{d} x^i \mathrm{d} x^j~,
\ee
where the perturbations satisfy the two conditions: $ S_{i,i} = 0 $ and $ F_{i,i} = 0 $.
The gauge transformation for these vector type perturbations are
\be
x^i \; \rightarrow \; x^i + \beta^i~,
\ee
and vector type metric perturbations transforms as
\begin{subequations}
\begin{align}
S_i	\; & \rightarrow \; S_i + a \dot{\beta_i}~,		\\
F_i	\; & \rightarrow \; F_i - \beta_i~.
\end{align}
\end{subequations}

\subsubsection{Tensor perturbations}
The tensor metric perturbations $ h_{ij} $ are given by the line element
\be
\label{tensorelem}
\mathrm{d} s^2 = - \mathrm{d} t^2 + a^2 [ \delta_{ij} + h_{ij} ] \mathrm{d} x^i \mathrm{d} x^j~,
\ee
and $ h_{ij} $ can be written as follows
\be
h_{ij} = h_{+} e^{+}_{ij} \, + \, h_{\times} e^{\times}_{ij}~,
\ee
where $ e^{+}_{ij} $ and $ e^{\times}_{ij} $ are the two polarization tensors and $ h_{+} $ and $ h_{\times} $ are the coef\mbox{}f\mbox{}icient functions.
In addition, $ h_{ij} $ is both traceless and transverse \cite{Grishchuk:1974ny}: 
\be
\delta^{ij} h_{ij} = 0~,  \qquad \partial^i h_{ij} = 0~,
\ee
and the polarization tensors have the following properties
\be
e_{ij}(-\bold{k}, \lambda) = e^{\ast}_{ij}(\bold{k},\lambda)		\qquad \mathrm{and} \qquad \sum_{\lambda} e^{\ast}_{ij} (\bold{k}, \lambda) e^{ij} (\bold{k}, \lambda) = 4~,
\ee
where $ \lambda $ indicates the polarization i.e. $ \lambda = + , \times $.
Unlike the scalar and vector metric perturbations, the tensor metric perturbations are unaffected by gauge transformations.

\subsection{Scalar modes}
\label{sec:chap1scalmodes}
In the previous subsection, we described the formalism required to study the f\mbox{}irst-order perturbations generated in cosmology.
We shall begin with the scalar perturbations before turning our attention to tensor perturbations.
%
There are two types of scalar perturbation that are produced during inf\mbox{}lation; they are quantif\mbox{}ied as the curvature and isocurvature perturbations.
\newline
\\
Curvature (otherwise known as adiabatic) perturbations are those which are related to the perturbations in the total energy density $ \delta \rho_{\mathrm{total}} $. 
The other type of perturbation is known as the isocurvature perturbation; this perturbation is one that does not af\mbox{}fect the local curvature in such a way that it does not af\mbox{}fect the total energy density perturbation $ \delta \rho_{\mathrm{total}} = 0 $.
For this perturbation type, the individual matter species components are allowed to vary.
In the case of single-f\mbox{}ield inf\mbox{}lation, there are no isocurvature perturbations generated.
The presence of isocurvature perturbations causes growth in the curvature perturbation on superhorizon scales.
\newline
\\
In this Thesis, only the Newtonian (or longitudinal) gauge will be used.
The name arises due to the system considered reducing to Newtonian gravity when in the small-scale limit; when at small-scales, the Newtonian potential $ \Psi $ is that satisfying the Poisson equation $ \nabla^2 \Psi = 4 \pi G \rho $ for nonrelativistic matter.
In the Newtonian gauge, a coordinate transformation is chosen in such a way that the scalar metric perturbations $ B $ and $ E $ are equal to zero $ B = E = 0 $.
This results in the full scalar perturbation FRW line element in Eq.~\eqref{scalarlin} simplifying to contain only $ \Phi $ and $ \Psi $ which are now gauge invariant potentials
\be
\label{lineelfull}
\mathrm{d}s^2 = -(1 + 2 \Psi ) \mathrm{d}t^2 + a^2(t)(1 - 2 \Phi) \delta_{ij} \mathrm{d}x^i \mathrm{d}x^j~.
\ee
The perturbed Einstein f\mbox{}ield equations can be calculated by knowing the perturbed Einstein tensors which are as follows
\begin{subequations}
\begin{align}
\delta G\indices{^0_0} & = - 2 \nabla^2 \Phi + 6 H \dot{\Phi} + 6 H^2 \Psi~,		\\
\delta G\indices{^0_i} & = -2 ( \dot{\Phi} + H \Psi )_{,i}~,		\\
\delta G\indices{^i_j} & = [ 2 \ddot{\Phi} + \nabla^2 ( \Psi- \Phi ) + (4 \dot{H} + 6 H^2 ) \Psi + H (2 \dot{\Psi} + 6 \dot{\Phi} ) ] \delta^i_j 		\nonumber \\
						& \qquad + \partial^i \partial_j ( \Phi - \Psi )~.
\end{align}
\end{subequations}
As an example, we can calculate the perturbed components of the energy-momentum tensor for a scalar f\mbox{}ield.
We will assume that the scalar f\mbox{}ield is composed of two parts: the background (homogeneous) and the linear f\mbox{}irst-order perturbation
\be
\phi(t,\bold{x}) = \phi(t) + \delta \phi (t, \bold{x})~,
\ee
and we will be working with the Fourier components of the perturbations, $ \delta \phi_{\bold{k}} $
\be
\delta \phi (t, \bold{x}) = \int \frac{ \mathrm{d}^3 \bold{k} }{ (2 \pi)^{3/2} } \delta \phi_{\bold{k}} e^{i \bold{k} \cdot \bold{x} }~,
\ee
where $ \delta \varphi_\bold{k} $ satisf\mbox{}ies the Poisson equation: $ \nabla^2 \delta \varphi_\bold{k} = -k^2 \delta \varphi $.
In order to shorten the expressions, the subscript $ \bold{k} $ will be omitted.
Furthermore, it is convenient to work with the gauge-invariant Sasaki-Mukhanov variables \cite{Sasaki:1986hm, Mukhanov:1988jd}.
They are related to the f\mbox{}ield perturbations and are def\mbox{}ined as
\be
\label{muksasvar}
Q_{\phi} \equiv  \delta \phi + \frac{\dot{\phi}}{H} \Psi~.
\ee
From the perfect f\mbox{}luid description in Eq.~\eqref{energymtensor} and the energy-momentum tensor for the f\mbox{}ield $ \phi $ in Eq.~\eqref{tensorem}, the perturbed components are
\begin{subequations}
\begin{align}
\delta T\indices{^0_0} = - \delta \rho & = \dot{\phi}^2 \Psi - \dot{\phi} \delta \dot{\phi} - V' \delta \phi~,		\\
\delta T\indices{^0_i} = \delta q & = - \dot{\phi} \, \delta \phi_{,i}~,	\\
\delta T\indices{^i_j} = \delta p & =  [-\dot{\phi}^2 \Psi + \dot{\phi} \delta \dot{\phi} + V' \delta \phi \, ] \, \delta^i_j~,
\end{align}
\end{subequations}
where we have used
\begin{subequations}
\label{vartensor}
\begin{align}
\delta T\indices{^\mu_\nu} & = \delta ( g^{\mu \alpha} T_{\alpha \nu} )~,		\\
						& = \delta g^{\mu \alpha} T_{\alpha \nu} + g^{\mu \alpha} \delta T_{\alpha \nu}~.
\end{align}
\end{subequations}
In order to study the curvature and isocurvature perturbations, it is preferable to work with quantities that are gauge-invariant.
For both types of perturbations in question a def\mbox{}inition does exist for each.
Starting with the curvature perturbations, f\mbox{}irst is the curvature perturbation $ \mathcal{R} $ as def\mbox{}ined:
\be
\label{curvR}
\mathcal{R} \equiv \Phi - \frac{H}{\rho + p} \delta q~,
\ee
$ \delta q $ is associated with the 0i-component of the perturbed Einstein f\mbox{}ield equations.
We can show that this curvature perturbation is a gauge-invariant quantity.
By using the def\mbox{}inition in Eq.~\eqref{curvR} and considering the scalar metric perturbation gauge transformations in Eqs.~\eqref{scaltrans} and \eqref{scalpertmetric}
\begin{align}
\mathcal{R} & = \Phi + H \alpha - \frac{H}{ \rho + p } [ \, \delta q + ( \rho + p ) \alpha \, ]~,	\nonumber	\\
			& = \Phi + H \alpha - \frac{H}{ \rho + p } \delta q - H \alpha~,			\nonumber	\\
\therefore \qquad \mathcal{R} & = \Phi - \frac{H}{\rho + p} \delta q~.					\nonumber			
\end{align}
This type of curvature perturbation represents the gravitational potential on comoving hypersurfaces where the inf\mbox{}lation f\mbox{}luctuations are $ \delta \phi = 0 $
\be
\mathcal{R} = \Phi |_{\delta \phi = 0 }~.
\ee
For the case of a single scalar f\mbox{}ield, the comoving curvature perturbation is
\be
\label{curvsame}
\mathcal{R} = \Phi + \frac{H}{\dot{\phi}} \delta \phi~.
\ee
Second is the curvature perturbation on uniform density hypersurfaces $ \zeta $ \cite{Wands:2000dp}
\be
\label{zeta}
- \, \zeta \equiv \Phi + \frac{H}{\dot{\rho}} \delta \rho~,
\ee
and is gauge-invariant (using Eqs.~\eqref{scaltrans} and \eqref{scalpertmetric}) as shown below
\begin{align}
- \zeta & = \Phi + H \alpha + \frac{H}{ \dot{\rho} } [ \, \delta \rho - \dot{\rho} \alpha \, ]~,	\nonumber	\\
			& = \Phi + H \alpha + \frac{H}{ \dot{\rho} } \delta \rho - H \alpha~,			\nonumber	\\
\therefore \qquad - \zeta & = \Phi + \frac{H}{ \dot{\rho} } \delta \rho~.					\nonumber			
\end{align}
This curvature perturbation describes the gravitational potential on uniform energy density slices:
\be
-\zeta = \Phi |_{\delta \rho = 0 }~.
\ee
Using the conservation equation in Eq.~\eqref{conservation}, the expression for $ \zeta $ can be rewritten as
\be
-\zeta = \Phi - \frac{\delta \rho}{3 (\rho + p) }~.
\ee
During the slow-roll period in the case of a single scalar f\mbox{}ield, the following hold true
\begin{align}
\rho + p & = \dot{\phi}^2~,		\\
\delta \rho & \simeq -3 H \dot{\phi} \, \delta \phi~.
\end{align}
The last relation comes from using the Klein-Gordon equation of the scalar f\mbox{}ield; f\mbox{}irst the f\mbox{}ield perturbation $ \delta \phi $ is frozen in and under slow-roll conditions Eq.~\eqref{slowroll1} is obeyed.
This results in the following relation at superhorizon scales
\begin{subequations}
\begin{align}
-\zeta & = \Phi + \frac{3 H \dot{\phi}}{3 \dot{\phi}^2} \, \delta \phi~,		\\
		& = \Phi + \frac{H}{\dot{\phi}} \delta \phi~,
\end{align}
\end{subequations}
which is the same as Eq.~\eqref{curvsame}.
Therefore at superhorizon scales, both $- \zeta $ and $ \mathcal{R} $ are the same.
\newline
\\
The other type of perturbation that can be calculated is the isocurvature perturbation.
For a system to be adiabatic, the pressure perturbation is as follows
\be
\delta p = c_\mathrm{s}^2 \delta \rho~,
\ee
where $ c^{2}_{\mathrm{s}} $ the adiabatic sound speed of the f\mbox{}luid given by $ c^{2}_{\mathrm{s}} = \dot{p}/\dot{\rho} $.
However, in general, the total pressure perturbation is expressed containing two contributions: an adiabatic part and an entropic part also known as the nonadiabatic pressure perturbation $ \delta p_{\mathrm{nad}} $ \cite{Wands:2000dp}:
\be
\delta p = c_\mathrm{s}^2 \delta \rho + \delta p_{\mathrm{nad}}~,
\ee
where $ \delta p_{\mathrm{nad}} $ is def\mbox{}ined as $  \delta p_{\mathrm{nad}} = \dot{p} \Gamma $.
The entropy perturbation $ \Gamma $ is def\mbox{}ined as
\be
\Gamma = \frac{\delta p}{\dot{p}} - \frac{\delta \rho}{\dot{\rho}}~,
\ee
which is by construction a gauge-invariant quantity as shown below
\begin{align}
\Gamma & = \frac{ \delta p - \dot{p} \alpha }{ \dot{p} } - \frac{ \delta \rho - \dot{\rho} \alpha }{ \dot{\rho} }~,	\nonumber	\\
			& = \frac{ \delta p  }{ \dot{p} } - \alpha - \frac{ \delta \rho }{ \dot{\rho} } + \alpha~,			\nonumber	\\
\therefore \qquad \Gamma & =\frac{\delta p}{\dot{p}} - \frac{\delta \rho}{\dot{\rho}}~.					\nonumber			
\end{align}
This quantity describes the dif\mbox{}ference between hypersurfaces of uniform pressure and uniform density \cite{Malik:2008im}.
\newline
\\
An isocurvature perturbation that is regularly calculated in literature is known as the entropy perturbation $ \mathcal{S} $ which is def\mbox{}ined as \cite{Gordon:2000hv, Malik:2004tf} 
\be
\mathcal{S} = \frac{H}{\dot{p}} \, \delta p_{\mathrm{nad}}~.
\ee
Earlier studies into the entropy and nonadiabatic perturbations are shown in references \cite{Mollerach:1990ue} and \cite{Mollerach:1989hu}.
The nonadiabatic pressure perturbation also appears in the rate of change of the curvature perturbation on hypersurfaces of uniform density; the full expression is presented in \cite{Wands:2000dp}
\be
\label{zetadotfull}
\dot{\zeta} = - \frac{H}{ \rho + p} \, \delta p_{\mathrm{nad}} - \frac{1}{3} \nabla^2 ( \sigma + v + B )~,
\ee
where $ \nabla^i v $ is the perturbed 3-velocity of the f\mbox{}luid, $ \sigma $ denotes the shear given by
\be
\sigma = \dot{E} - B~,
\ee
where $ B $ and $ E $ are the metric perturbations in the line element stated in Eq.~\eqref{scalarlin}.
At large scales, the gradient term in Eq.~\eqref{zetadotfull} can be neglected resulting in
\be
\dot{\zeta} = - \frac{H}{ \rho + p} \, \delta p_{\mathrm{nad}}~.
\ee
We require the primordial power spectrum as it can be used to distinguish between various inf\mbox{}lationary models.
Some of these models produce features in the primordial power spectrum in the form of bumps and oscillations, examples of models include those with features in the ef\mbox{}fective inf\mbox{}laton potential \cite{Starobinsky:1992ts, Leach:2001zf, Adams:2001vc, Covi:2006ci, Hamann:2007pa, Lerner:2008ad} and disruption to the slow-roll evolution caused by phase transitions \cite{Adams:1997de}; for more examples see \cite{Hunt:2013bha}.
\newline
The power spectra for the curvature and entropy perturbations and their cross spectrum between the two quantities are stated below
\begin{subequations}
\begin{align}
\mathcal{P}_{\zeta} = {} &\frac{k^3}{2 \pi^2} ( \, |\zeta_1|^2 + |\zeta_2|^2 \, )~,	\\
\mathcal{P}_{\mathcal{S}} = {} & \frac{k^3}{2 \pi^2} ( \, |\mathcal{S}_1|^2 + |\mathcal{S}_2|^2 \, )~,	\\
\mathcal{P}_{\mathrm{C}} = {} & \frac{k^3}{2 \pi^2} | \, \zeta_1 \, \mathcal{S}_1 + \zeta_2 \, \mathcal{S}_2 \, |~,
\end{align}
\end{subequations}
where $ k $ is the wavenumber, this quantity arises in the perturbation equations, and the subscripts indicate the run number.
For an overview of the numerical method performed in this Thesis, see Appendix A.
\newline
\\
Another quantity that is commonly seen with the cross spectrum is the correlation $ r_\mathrm{C} $ which is def\mbox{}ined as
\be
r_\mathrm{C} = \frac{\mathcal{P}_\mathrm{C}}{\sqrt{ \mathcal{P}_{\zeta} \mathcal{P}_{\mathcal{S}} }}~,
\ee
and varies between the values zero and one. 
Once the power spectrum has been obtained, other quantities can be derived from it.
The power-law spectrum of the primordial curvature perturbations is def\mbox{}ined as
\be
\mathcal{P}_{\zeta} (k) = \mathcal{P}_{\zeta} (k_0) \bigg( \frac{k}{k_0} \bigg)^{n_{\mathrm{s}} - 1}~,
\ee
where $ \mathcal{P}_{\zeta} (k_0) $ is the amplitude of the curvature power spectrum at a pivot scale denoted by $ k_0 $ and the spectral index $ n_{\mathrm{s}} $ is 
\be
n_{\mathrm{s}} - 1 \equiv \frac{ \mathrm{d} \ln \mathcal{P}_{\zeta} }{ \mathrm{d} \ln{k} }~.
\ee
Some models predict a scale dependence, this is also known as the running in the almost power-law spectrum of curvature perturbations.
For this case where the power spectrum contains running \cite{Kosowsky:1995aa}, it is given as
\be
\label{scalarpowspec}
\mathcal{P}_{\zeta} (k) = \mathcal{P}_{\zeta} (k_0) \bigg( \frac{k}{k_0} \bigg)^{n_{\mathrm{s}} - 1 + \frac{1}{2} \ln ( k/k_0 )\alpha }~,
\ee
with $ \alpha $ as the running of the spectral index.
\be
\label{runninindex}
\alpha = \frac{ \mathrm{d} \, n_{\mathrm{s}} }{ \mathrm{d} \ln{k} }~.
\ee
In the WMAP experiment, the pivot scale is set at $ k_0 = 0.002 \, \mathrm{Mpc}^{-1} $ \cite{Komatsu:2010fb}.

\subsection{Fluctuation origins}
It was proposed in early 1980s by various theorists that scalar f\mbox{}ields are subjected to quantum f\mbox{}luctuations with these f\mbox{}luctuations stretched when on superhorizon scales \cite{Mukhanov:1981xt, Hawking:1982cz, Starobinsky:1982ee, Guth:1982ec, Bardeen:1983qw}; this was also looked at later in \cite{Fischler:1985ky}.
The quantum f\mbox{}ield f\mbox{}luctuations, which originate at very small scales, will evolve during inf\mbox{}lation and will become the initial density perturbations which will give rise to structures we see today.
\newline
\\
During the inf\mbox{}lationary period, the f\mbox{}luctuations of the inf\mbox{}laton f\mbox{}ield $ \phi $ will grow until a point where its wavelength is greater than the Hubble radius.
The amplitude of the f\mbox{}ield f\mbox{}luctuations will then freeze at a nonzero value and will remain so until the end of inf\mbox{}lation.
At some stage, inf\mbox{}lation eventually ends causing the f\mbox{}ield f\mbox{}luctuations to reenter the Hubble radius due to the Hubble radius increasing faster than the scale factor.
The reentering of the f\mbox{}luctuations does not happen immediately, but occurs after reheating during either the radiation or matter-dominated epochs.
For the f\mbox{}luctuations that exit the horizon between 50 and 60 $ e $-foldings before the end of inf\mbox{}lation, these will reenter at a scale corresponding to cosmologically relevant wavelengths.
When these modes reenter the Hubble radius, they will leave an imprint in the curvature spectrum which can be measured and compared to various models. 
\newline
\\
The f\mbox{}luctuations of the scalar f\mbox{}ield can be studied by using its second-order action; the second-order action of the scalar f\mbox{}ield $ \phi $ written in terms of the canonically-normalised Mukhanov variable $ u $ in conformal time $ \eta $ is given by
\be
\label{uaction}
S^{(2)} = \frac{1}{2} \int \mathrm{d} \eta \, \mathrm{d}^{3} \bold{x} \, \bigg[ \, (u')^2 - (\partial_i u)^2 + \frac{z''}{z} u^2 \, \bigg]~,
\ee
where the parameters $ u $ and $ z $ are as follows
\begin{subequations}
\begin{align}
\label{udef}
u & = z \mathcal{R}~,		\\
\label{zmukh}
z^2 & = \frac{a^2 \dot{\phi}^2 }{H^2}~,
\end{align}
\end{subequations}
where $ \mathcal{R} $ is the comoving curvature perturbation.
This calculation is shown explicitly in the following references: \cite{Sasaki:1986hm, Mukhanov:1988jd, Mukhanov:1990me}.
Varying the action in Eq.~\eqref{uaction} yields the Sasaki-Mukhanov equation
\be
\label{muksaseq}
u''_{\boldsymbol{\mathrm{k}}} + \omega^2_{\boldsymbol{\mathrm{k}}} (\eta) \, u_{\boldsymbol{\mathrm{k}}} = 0~,		\qquad	\omega^2_{\boldsymbol{\mathrm{k}}} = \bigg( k^2 - \frac{z''}{z} \bigg)~,
\ee
with the following Fourier modes def\mbox{}ined as
\be
u_{\boldsymbol{\mathrm{k}}} (\eta) = \int \mathrm{d}^{3} \bold{x} \, e^{-i \bold{k} \cdot \bold{x} } u( \eta , \bold{x} )~.
\ee
In a de Sitter spacetime which is a spatially f\mbox{}lat spacetime containing a positive cosmological constant, the scale factor is
\be
a(\eta) = - \frac{1}{H \eta}~,
\ee
which yields an ef\mbox{}fective oscillator frequency of
\be
\omega^2_{\bold{k}} = k^2 - \frac{2}{\eta^2}~.
\ee
This example in de Sitter spacetime can also be studied in two dif\mbox{}ferent regimes:
\begin{itemize}
\item{\textbf{Subhorizon scales}}
\newline
Subhorizon scales occur when the wavelengths of the f\mbox{}luctuations are much smaller than the Hubble radius $ \lambda \ll H^{-1} $ and hence $ k^2 \gg | z''/z | $, leading to a frequency of 
\be
\omega^2_{\bold{k}} = k^2	~,
\ee
and yielding the oscillatory solutions to the Sasaki-Mukhanov equation
\be
u_{\bold{k}} \propto e^{\pm ik \eta}~.
\ee
\item{\textbf{Superhorizon scales}}
\newline
Superhorizon scales are when the wavelengths are much greater than the Hubble radius $ \lambda \gg H^{-1} $ and therefore $ k^2 \ll | z''/z | $
\be
\omega^2_{\bold{k}} = - \frac{2}{\eta^2}~.
\ee
The resulting solution for this regime is
\be
u_{\bold{k}} \propto \frac{1}{\eta} \qquad \mathrm{and} \qquad u_{\bold{k}} \propto \eta^2~.
\ee
\end{itemize}
Using the fact that during inf\mbox{}lation conformal time will run from large negative values to zero, we have here the f\mbox{}irst of these solutions is a growing mode and the second is a decaying mode.
From using Eq.~\eqref{udef} and that $ z \sim \eta^{-1} $, it can be shown that the comoving curvature perturbation $ \mathcal{R} $ is constant on superhorizon scales.
\newline
\\
A vacuum state will need to be chosen and the standard choice is to use the Bunch-Davies vacuum, named after the authors whose work was published in 1978 \cite{Bunch:1978yq}.
At very early times, the wavelengths of all f\mbox{}luctuations existed inside the Hubble radius (subhorizon scales) and therefore the following equation holds
\be
\label{redequation}
u''_{\boldsymbol{\mathrm{k}}} + k^2  u_{\boldsymbol{\mathrm{k}}} = 0~,
\ee
and through the quantization of the f\mbox{}ield $ u_{\bold{k}} $ the solution to the equation above is
\be
\lim_{\eta \to - \infty} \, u_{\bold{k}} (\eta) = \frac{1}{\sqrt{2k}} e^{-i k \eta }~.
\ee
The power spectrum for the scalar modes is
\begin{subequations}
\begin{align}
\mathcal{P}_{\mathcal{R}} & = \frac{k^3}{2 \pi^2} \bigg| \frac{ u_{\bold{k}} (\eta) }{ z (\eta) } \bigg|^2~,			\\
				& =  \frac{k^3}{2 \pi^2} | \mathcal{R}_{\bold{k}} (\eta) |^2~.
\end{align}
\end{subequations}

\subsection{Tensor modes}
\label{sec:chap1tensor}
Inf\mbox{}lation also predicts the existence of gravitational waves which are tensor modes.
The line element for tensor perturbations is given in Eq.~\eqref{tensorelem}
We would like to study the f\mbox{}luctuations of the tensor models and in order to do this, we will need to f\mbox{}ind the second-order action of the tensor modes.
This is done by expanding the gravitational sector of the action (Einstein-Hilbert action) to second-order; the second-order action for the tensor perturbations is
\be
S^{(2)} = \int \mathrm{d} \eta \, \mathrm{d}^{3} \bold{x} \frac{a^2}{2} \bigg[ (h')^2 - ( \partial_i h )^2 \bigg]~,
\ee
which will lead to the equation of motion for the quantity $ h_{\bold{k}} $
\be
\label{tensoreqmotion}
h_{\bold{k}}'' + 2 \frac{a'}{a} h_{\bold{k}} + k^2 h_{\bold{k}} = 0~.
\ee
This equation can be rewritten by def\mbox{}ining a new variable $ v_{\bold{k}} $ where 
\be
v_{\bold{k}} = a h_{\bold{k}}~,
\ee
thus yielding the equation
\be
\label{tensmuksaseq}
v_{\bold{k}}'' + \bigg( k^2 - \frac{a''}{a} \bigg) \, v_{\bold{k}} = 0~.
\ee
It is important to note that this equation is of the same form as the Sasaki-Mukhanov equation for scalar modes in Eq.~\eqref{muksaseq}, with the exception of the scale factor replacing the parameter $ z $.
\newline
\\
The same arguments used on the scalar modes can be applied to tensor modes --- all wavelengths in the very early Universe existed within subhorizon scales.
Due to this, the term $ a''/a $ can be neglected and Eq.~\eqref{tensmuksaseq} is reduced to that in Eq.~\eqref{redequation}.
The initial conditions for the gravitational waves are the same as for the scalar modes
\be
v_{\bold{k}} = \frac{1}{\sqrt{2k}} e^{-i k \eta}~.
\ee
The power spectrum generated for the tensor modes is
\begin{subequations}
\begin{align}
\mathcal{P}_{h} & = \frac{k^3}{2 \pi^2} \bigg| \frac{v_{\bold{k}}(\eta)}{a(\eta)} \bigg|^2~,		\\
					& = \frac{k^3}{2 \pi^2} | h_{\bold{k}} (\eta) |^2~.
\end{align}
\end{subequations}
In order to relate this to the gravitational waves (tensor) power spectrum, we will need to sum over all polarizations (appears as the additional factor of 8):
\begin{align}
\mathcal{P}_{\mathrm{T}} & = 4 \times 2 \frac{k^3}{2 \pi^2} | h_{\bold{k}} (\eta) |^2~,		\nonumber	\\
					& = 8 \mathcal{P}_{h}~,
\end{align}
A quantity that is commonly used to relate the tensor power spectrum $ \mathcal{P}_{\mathrm{T}} $ to the scalar curvature power spectrum $ \mathcal{P}_{\zeta} $ is the tensor-to-scalar ratio, $ r $ and is def\mbox{}ined as
\be
r = \frac{\mathcal{P}_{\mathrm{T}}}{\mathcal{P}_{\zeta}} = \frac{8 \mathcal{P}_{h}}{\mathcal{P}_{\zeta}}~.
\ee

\section{Alternative models}
\label{sec:chap1altmodel}
Most inf\mbox{}lationary models are based upon scalar f\mbox{}ields, both single and multif\mbox{}ield, in General Relativity.
However, modif\mbox{}ications to Einstein's Theory of General Relativity have also been studied, for example through scalar-tensor theories, such as Brans-Dicke theory \cite{Brans:1961sx} and $ f(R) $ theories \cite{Starobinsky:1980te, Buchdahl:1983zz}.
For further reading on $ f(R) $ theories see the following reviews \cite{Sotiriou:2008rp, DeFelice:2010aj}, and a more general overview of the various extended models of gravity see the review \cite{Capozziello:2011et}.
\newline
The action of a scalar-tensor theory is usually displayed as
\be
\label{actionjordan}
S = \int \mathrm{d}^{4}x \sqrt{-g} \bigg[ \frac{1}{2}F(\phi)R - \frac{1}{2} g^{\mu \nu} \partial_{\mu} \phi \, \partial_{\nu} \phi - U(\phi) \bigg]
		+ S_{\chi} [ g_{\mu \nu} ; \chi ]~,
\ee
where $ F(\phi) $ is the coupling between the scalar f\mbox{}ield and the gravity sector. 
The action $ S_{\chi}[g_{\mu \nu} ; \chi ]  $ is that of the matter f\mbox{}ields $ \chi $; this action is only composed of the metric $ g_{\mu \nu} $ and the f\mbox{}ield $ \chi $ and is independent of the f\mbox{}ield $ \phi $.
By rewriting the action, the Einstein-Hilbert term can be recovered.
The most common practice is to perform a conformal transformation of the metric given by
\be
\tilde{g}_{\mu \nu} = F(\phi) g_{\mu \nu}~,
\ee
where $ \tilde{g}_{\mu \nu} $ is the redef\mbox{}ined metric in the new representation, and a scalar f\mbox{}ield redef\mbox{}inition detailed below
\begin{align}
\label{transfield}
\bigg( \frac{d \tilde{\phi}}{d \phi} \bigg)^2 &= \frac{3}{2} \bigg( \frac{F_{,\phi}}{F} \bigg)^2 + \frac{1}{F}~,		\\
A(\tilde{\phi}) &= F^{-\frac{1}{2}}(\phi)~,			\\
V(\tilde{\phi}) &= U(\phi) F^{-2}(\phi)~.
\end{align}
By using the new def\mbox{}initions stated above, the action in Eq.~\eqref{actionjordan} is rewritten as follows
\be
\label{actioneins}
\tilde{S} = \int \mathrm{d}^{4}x \sqrt{-\tilde{g}} \bigg[ \frac{1}{2}\tilde{R} - \frac{1}{2} \tilde{g}^{\mu \nu} \partial_{\mu} \tilde{\phi} \, \partial_{\nu} \tilde{\phi} - V(\tilde{\phi}) \bigg]
			+ S_{\chi} [ A^2 (\tilde{\phi}) \tilde{g}_{\mu \nu}; \chi ]~,		\\
\ee
where $ \tilde{g} $ is the determinant of the metric $ \tilde{g}_{\mu \nu} $ and $ \tilde{R} $ is its Ricci scalar obtained by using the tilde metric.
The usage of the conformal transformation in scalar-tensor theories are presented in the following references:\cite{Maeda:1988ab, Barrow:1990nv}.
\newline
\\
In the presence of a second scalar f\mbox{}ield, these theories will provide the system with nonstandard kinetic terms, such as kinetic couplings between scalar f\mbox{}ields.
\newline
The idea of a Lagrangian containing nonstandard kinetic terms coined as ``$k$-inf\mbox{}lation'' was f\mbox{}irst introduced in the papers \cite{ArmendarizPicon:1999rj, Garriga:1999vw} published in 1999 as a possible solution to inf\mbox{}lation, using inspiration from models in string theory.
From string theory based models, an inf\mbox{}laton candidate known as a moduli f\mbox{}ield which is a weakly coupled scalar f\mbox{}ield, naturally arises.
However, it was shown that the potentials generated from these models do not allow for slow roll to occur.
It was later shown that these nonstandard kinetic terms are able to produce interesting models in the context of dark energy; the class of models called ``$k$-essence'' 
\cite{ArmendarizPicon:2000ah, Malquarti:2003nn, Scherrer:2004au}; for an extensive review of dark energy see \cite{Copeland:2006wr}.
\newline
\\
The kinetic couplings between scalar f\mbox{}ields are obtained by performing the conformal transformation on the action stated in Eq.~\eqref{actionjordan}; works in this area include \cite{Berkin:1991nm, GarciaBellido:1995fz, Starobinsky:2001xq, DiMarco:2002eb, Lalak:2007vi}.

\section{Thesis outline}
This Thesis is divided up into f\mbox{}ive chapters, with the f\mbox{}irst chapter dedicated to the history of cosmology and the mathematical formalism required.
The following three chapters will each discuss models involving coupled scalar f\mbox{}ields under a dif\mbox{}ferent setting; Chapters 2 and 3 will consider the inf\mbox{}lationary scenario and Chapter 4 in the context of dark energy.
\newline
\\
Chapter 2 considers three models of multif\mbox{}ield inf\mbox{}lation each dif\mbox{}fering in kinetic terms, and their ef\mbox{}fects upon the isocurvature (nonadiabatic pressure) perturbations generated during the inf\mbox{}lationary epoch.
Each of the models considered contains at least one minimally coupled scalar f\mbox{}ield, with two of these models classif\mbox{}ied under $ k $-inf\mbox{}lation.
The models studied are as follows: the f\mbox{}irst model containing two minimally coupled scalar f\mbox{}ields, second model consisting of two f\mbox{}ields, one minimally coupled and the other containing a kinetic coupling, and lastly, the third model composing of a minimally coupled f\mbox{}ield and a Dirac-Born-Infeld f\mbox{}ield.
\newline
\\
In Chapter 3, a study of the ef\mbox{}fects induced by sharp transitions in the ef\mbox{}fective Planck mass upon primordial power spectra during inf\mbox{}lation is presented.
A single f\mbox{}ield model of scalar-tensor form is initially studied before extending to a two-f\mbox{}ield model with the inclusion of a minimally coupled auxiliary f\mbox{}ield.
In both cases, the scalar and tensor power spectra are generated numerically.
It was shown that the sharp transitions af\mbox{}fected slow-roll evolution, and this was ref\mbox{}lected through a feature in the scalar power spectrum.
This feature allowed the easing of tension between observations from the two experiments, Planck \cite{Ade:2013zuv} and BICEP2 \cite{Ade:2014xna}.
\newline
\\
The move to more later times in cosmic history occurs in Chapter 4 where a noncanonical transformation is studied in the context of dark energy.
In this chapter, a model of scalar-tensor theory form is presented in which the scalar and matter f\mbox{}ields are disformally coupled.
The intention is to study the ef\mbox{}fects disformal coupling has upon observables, such as the temperature and spectral distortions of the CMB.
In order to check the consistency of the disformal theory, a f\mbox{}luid description and a kinetic theory was verif\mbox{}ied.
Through the verif\mbox{}ication, two important traits of the disformal theory were discovered: the equation of state and the distribution function are not frame-independent quantities.
\newline
\\
The Thesis concludes with a f\mbox{}inal chapter recalling and summarising the main points from the three studies contained in the preceding chapters.
Suggestions to extend the work performed in this Thesis and concluding statements are also included.

\subsubsection{Notation}
We will not be considering the cosmological constant $ \Lambda $.
For the most part of the Thesis, the reduced Planck mass $ M_{\mathrm{Pl}} $ will be set to one $ M_{\mathrm{Pl}} = 1 $, except in fundamental equations such as the action of a theory.
All the calculations in this Thesis will be performed in the Newtonian gauge.
\newline
\\
The metric signature used throughout this Thesis is $ (-,+,+,+) $.
For mathematical objects such as the metric $ g_{\mu \nu} $, Greek indices i.e. $ ( \alpha, \beta, \ldots ) $ run from 0 to 3, and denote over all dimensions, and lower case Roman indices i.e. $ ( a, b, \ldots ) $ run from 1 to 3, and represent the spatial coordinates.

\chapter{Noncanonical multif\mbox{}ield inf\mbox{}lation}

The principles of inflation were laid out in the Introduction, alongside a brief note of alternative models in its self-titled section.
There are numerous theories beyond the standard model of physics that provide motivation for inflationary models; some examples of these theories include those from supergravity \cite{Nanopoulos:1982bv, Goncharov:1983mw, Jensen:1985yt} and string theory \cite{Sen:2002an, Lambert:2002hk}.
As a result, these models predict many scalar fields that could drive the inflationary epoch.
In addition, these scalar fields couple through either the potential or kinetic terms.
We shall be concentrating on the details of the latter.
\newline
\\
Many of these extension of standard physics models give rise to f\mbox{}ields with noncanonical kinetic terms; this group of models are known as $ k $-inf\mbox{}lation.
Some examples of work in this field include \cite{ArmendarizPicon:1999rj, Garriga:1999vw, Helmer:2006tz, Panotopoulos:2007ky, Yue:2009wq, Bose:2009kc, Ringeval:2009jd, Brax:2011si, Ohashi:2013pca}.
One example of $ k $-inf\mbox{}lation is one driven by the Dirac-Born-Infeld (DBI) f\mbox{}ield, which has been studied in \cite{Silverstein:2003hf, Alishahiha:2004eh}.
\newline
\\
Inf\mbox{}lation in the DBI model is caused by a D3-brane travelling along a warped region called a throat of a compactif\mbox{}ied space.
Additionally, a speed limit naturally arises on the brane restricting the speed at which it travels, and is dependent on the speed of the brane and the warping of the throat.
With this speed limit, a parameter is introduced which is the analogue of the Lorentz factor in special relativity and is allowed to grow until the speed limit is reached.
This boost factor also has an effect on the perturbations generated. 
Unlike f\mbox{}luctuations of standard scalar f\mbox{}ields which travel at the speed of light, the f\mbox{}luctuations in the DBI f\mbox{}ield travel at a sound speed which is related to the analogous Lorentz factor.
\newline
\\
Features from these noncanonical models with multiple fields can be used to distinguish and eliminate models that do not agree with current constraints set by observations of the Cosmic Microwave Background (CMB) radiation \cite{Ohashi:2011na, Devi:2011qm, Li:2012vta}.
One such constraint is through using isocurvature perturbations;
the amount of isocurvature perturbations is restricted to approximately of order 10\% of the curvature perturbations \cite{Komatsu:2010fb}.
\newline
\\
Recently, the nonadiabatic pressure perturbation has been identified as an important ingredient in the study of cosmological perturbation theory.
The perturbation can source vorticity perturbations at second order in cosmological perturbation theory \cite{Christopherson:2009bt}.
In fact, the second order vorticity is shown to evolve as
\be
\omega_{2ij}' - 3{\cal H} c_{\mathrm{s}}^2 \omega_{2ij}= \frac{2a}{\rho+p}\left[3{\cal H} V_{1[i}\delta p_{{\rm nad}1,j]} +  \frac{\delta\rho_{1,[j}\delta p_{{\rm nad}1,i]}}{\rho + p} \right]~,
\ee
where $ \delta \rho_1 $ and $ \delta p_{{\rm nad}1} $ are the f\mbox{}irst order energy density and nonadiabatic pressure perturbations.
It was found that if the f\mbox{}irst order nonadiabatic pressure perturbation vanishes, it causes the vorticity to decay at both f\mbox{}irst and second orders.
In contrast, under the context of multi-f\mbox{}ield inf\mbox{}lation, the nonadiabatic pressure perturbation is usually nonzero and therefore sources vorticity at second order \cite{Christopherson:2009bt, Christopherson:2010dw, Christopherson:2010ek}.
Vorticity can possibly be detected by searching for B-mode polarization of the CMB radiation \cite{Hu:1997hv}, hence creating further tests of inf\mbox{}lation.
\newline
\\
The evolution of the nonadiabatic pressure perturbation has been studied in detail, specif\mbox{}ically for the theory involving two f\mbox{}ields with canonical kinetic terms and a variety of potentials by Huston and Christopherson in \cite{Huston:2011fr}.
The potential choices included the double quadratic \cite{Langlois:1999dw}, quartic \cite{Avgoustidis:2011em, Kodama:2011vs} and product exponential \cite{Byrnes:2008wi}.
In their paper, the authors stated and used two known def\mbox{}initions of the entropy perturbation: one def\mbox{}inition in which directly uses the nonadiabatic pressure perturbation, and the other more commonly used def\mbox{}inition of utilising f\mbox{}ield decomposition as proposed by \cite{Gordon:2000hv}.
Both these entropy perturbation def\mbox{}initions were used for each potential considered.
With each potential, the evolution of the power spectra with respect to the $e$-fold number and wavenumber for the curvature and entropy perturbations were plotted.
From this, they found that the entropy perturbation which uses the nonadiabatic pressure perturbation evolves differently compared to that using the regularly used isocurvature mode through f\mbox{}ield decomposition.
\newline
\\
In this chapter, we extend the works by Huston and Christopherson in 2012 \cite{Huston:2011fr}, by studying various models with noncanonical kinetic terms, concentrating on the method which incorporates the nonadiabatic pressure perturbation.
In particular, we will consider two theories in which both will contain a canonical kinetic f\mbox{}ield: one theory with the inclusion of a kinetic coupling term between the two f\mbox{}ields \cite{DiMarco:2002eb, GarciaBellido:1995qq, Choi:2007su}, and the other with a DBI f\mbox{}ield, see \cite{Silverstein:2003hf, Alishahiha:2004eh, Peiris:2007gz, Baumann:2014nda} and references therein.
This chapter is presented as follows: the actions for the three models considered are stated in the following section along with their background and perturbation equations in Section~\ref{sec:chap2sec1}. 
Section~\ref{sec:results} presents the results for the three models considered.
For each model apart from double inf\mbox{}lation, there are two scenarios: one in which the canonical f\mbox{}ield dominates the inf\mbox{}lationary epoch, and the other, vice versa.
A summary of the f\mbox{}indings are presented in Section~\ref{sec:chap2conclusion}.

\section{The models}
\label{sec:chap2sec1}
We will consider three models: one model containing two canonical scalar f\mbox{}ields, the second containing two scalar f\mbox{}ields with a kinetic coupling between them and lastly, a model containing a canonical scalar f\mbox{}ield and a DBI f\mbox{}ield.
Their actions are given as follows:
\begin{enumerate}
	\item The f\mbox{}irst two models are described by actions of the form 
		\begin{align}
		\label{kineticcouplingaction}
			S & =  \int \mathrm{d}^{4} x\sqrt{-g} \bigg[ \frac{M_{\mathrm{Pl}}^2}{2}R 
						- \frac{1}{2} g^{\mu \nu} \partial_\mu \phi \, \partial_\nu \phi 	\nonumber \\
					& \qquad \qquad \qquad \qquad - \frac{1}{2} e^{2b(\phi)} g^{\mu \nu} \partial_\mu \chi \, \partial_\nu \chi - V(\phi,\chi) \bigg]~,
		\end{align}
with $ b(\phi) $ denoting the kinetic coupling between the two f\mbox{}ields $ \phi $ and $ \chi $. 
In the f\mbox{}irst model, there is no kinetic coupling and so $b(\phi)=0$, whereas in the second model $ b $ is def\mbox{}ined as
\be
b(\phi) = \beta \phi 
\ee
where $ \beta $ is a constant. 

\item The third model we examine contains a scalar f\mbox{}ield with a canonical kinetic term $ \phi $, and one DBI f\mbox{}ield $ \chi $. 
The action is given by
\be
\label{dbiaction}
S =  \int \mathrm{d}^{4} x\sqrt{-g} \bigg[ \frac{M_{\mathrm{Pl}}^2}{2}R 
									- \frac{1}{2} g^{\mu \nu} \partial_\mu \phi \, \partial_\nu \phi 
									- \frac{1}{f(\chi)} ( 1 - \gamma^{-1} ) - V(\phi,\chi) \bigg]~,
\ee
where the two parameters exclusive to DBI f\mbox{}ield are def\mbox{}ined by the following
\begin{subequations}
\begin{align}
\gamma = {} & \frac{1}{ \sqrt{1 + f(\chi) g^{\mu \nu} \partial_\mu \chi \, \partial_\nu \chi  } }~, 		\\
f(\chi) = {} & \frac{\lambda}{ (\chi^2 + \mu^2)^2 }~.
\end{align}
\end{subequations}
$ \gamma $ is a parameter which is analogous to the Lorentz contraction factor in special relativity and is related to the sound speed of the DBI f\mbox{}ield by
\be
c_{\mathrm{s}} = \frac{1}{\gamma}~.
\ee
It is important to note that $ c_{\mathrm{s}} $ is the speed at which the fluctuations of the DBI field $ \chi $ propagate.
$ f(\chi) $ is the warp factor which describes the shape of the extra dimensions. 
The two parameters $ \lambda $ and $ \mu $ in the warp factor are constants.
\end{enumerate}
In all the actions given, $ M_{\mathrm{Pl}} $ is the reduced Planck mass where $ M_{\mathrm{Pl}}^2 = (8 \pi G)^{-1} $, $R$ is the Ricci scalar and $ V(\phi, \chi) $ is the potential.
We shall set the reduced Planck units to one, $ M_{\mathrm{Pl}} = 1 $.

\subsection{Background}
We assume a spatially f\mbox{}lat expanding spacetime which is both homogeneous and isotropic.
This spacetime is given by the FRW line element as given in Eq.~\eqref{frw} in Section~\ref{sec:chap1prelim}.
There are no modif\mbox{}ications to the Einstein-Hilbert part of the three actions and hence, all the actions obey the Einstein f\mbox{}ield equations
\be
\label{fieldeqs}
G_{\mu \nu} = T_{\mu \nu}~,
\ee
where $ G_{\mu \nu} $ is the Einstein tensor and $ T_{\mu \nu} $ is the energy-momentum tensor.
We assume that the energy-momentum tensors of the models considered are of a perfect fluid form, $ T \indices{^{\mu}_{\nu}} = \mathrm{diag}( -\rho, p, p, p) $ where $ \rho $ and $ p $ are the energy density and pressure.
\newline
\\
The background Klein-Gordon equations are obtained by varying the actions with respect to the f\mbox{}ields $ \phi $ and $ \chi $.
Due to no change in the form of the Einstein f\mbox{}ield equations, the Friedmann and acceleration equations are the same as those in Eq.~\eqref{friedmannacceleration}.
\newline
\\
For the actions stated in Eqs.~\eqref{kineticcouplingaction} and \eqref{dbiaction}, the Klein-Gordon equation, energy density and pressure are summarised in the following:
\begin{itemize}
\item{\textbf{Kinetic coupling}}
\newline
The Klein-Gordon equations are found by varying the action with respect to the two fields and are as follows:
\begin{subequations}
\label{generalkinetic}
\begin{align}
\frac{1}{\sqrt{-g}} \nabla_{\mu} ( \sqrt{-g} g^{\mu \nu} \nabla_{\nu} \phi ) & = - b_{,\phi} g^{\mu \nu} \nabla_{\mu} \chi \nabla_{\nu} \chi - V_{,\phi}~,	\\
\frac{1}{\sqrt{-g}} \nabla_{\mu} ( \sqrt{-g} g^{\mu \nu} e^{2b} \nabla_{\nu} \chi ) & = V_{,\chi}~.
\end{align}
\end{subequations}
The FRW metric is applied to the equations above and yield the equations of motion in cosmic time.
In addition, the energy density and pressure, and hence Friedmann equations are also stated here; they are given by \cite{DiMarco:2002eb, Lalak:2007vi}
	\begin{subequations}
	\begin{align}
		\ddot{\phi} + 3H \dot{\phi} + V_{\phi} = {} & b_{\phi} e^{2b} \dot{\chi}^2~, \\
		\ddot{\chi} + ( 3H + 2b_{\phi} \dot{\phi} ) \dot{\chi} + e^{-2b} V_{\chi} = {} &  0~.
	\end{align}
	\end{subequations}
The energy-momentum tensor is given by
\begin{align}
\label{tensor1}
T_{\mu \nu} & = \phi_{,\mu} \phi_{,\nu} - g_{\mu \nu} \bigg( \frac{1}{2} g^{\alpha \beta} \phi_{,\alpha} \phi_{,\beta} + V \bigg)		\nonumber \\
			& \qquad + e^{2b} \chi_{,\mu} \chi_{,\nu} - g_{\mu \nu} \bigg( \frac{1}{2} g^{\alpha \beta} e^{2b} \chi_{,\alpha} \chi_{,\beta} \bigg)~.
\end{align}
Both the energy density and pressure are obtained by considering the leading diagonal of the energy-momentum tensor (the same method as employed in Section~\ref{sec:chap1scalar})
\begin{subequations}
\begin{align}
		\rho = {} & \frac{1}{2} ( \dot{\phi}^2 + e^{2b} \dot{\chi}^2 ) + V~, \\
		p = {} & \frac{1}{2} ( \dot{\phi}^2 + e^{2b} \dot{\chi}^2 ) - V~,
\end{align}
\end{subequations}	
where $ b_{\phi} = db(\phi)/d\phi $; subscripts denote differentiation with respect to the subscript.
For all the equations stated above, they will reduce to the two standard f\mbox{}ield case when the kinetic coupling is set to zero.

\item{\textbf{DBI f\mbox{}ield}}
\newline
The Klein-Gordon equations for the two f\mbox{}ields in the DBI model are
\begin{subequations}
\begin{align}
\label{generalDBI}
\frac{1}{\sqrt{-g}} \nabla_{\mu} ( \sqrt{-g} g^{\mu \nu} \nabla_{\nu} \phi ) & = V_{,\phi}~,	\\
\frac{1}{\sqrt{-g}} \nabla_{\mu} ( \sqrt{-g} g^{\mu \nu} \gamma \nabla_{\nu} \chi) & = \frac{f'}{f} \bigg( 1 - \frac{1}{2} \gamma - \frac{1}{2} \frac{1}{\gamma} \bigg) + V_{,\chi}~,
\end{align}
\end{subequations}
and when the FRW metric is applied it yields
\begin{subequations}
	\begin{align}
		\ddot{\phi} + 3H \dot{\phi} + V_{\phi} = {} & 0~, \\
		\ddot{\chi} + 3H \gamma^{-2} \dot{\chi} + \frac{1}{2} \frac{f_{\chi}}{f^2} ( 1 - 3 \gamma^{-2} + 2 \gamma^{-3} ) + \gamma^{-3} V_{\chi} = {} & 0~,
	\end{align}
\end{subequations}
where $ f_{\chi} = df/d\chi $.
The energy-momentum tensor for this model is
\begin{align}
T_{\mu \nu} & = \phi_{,\mu} \phi_{,\nu} - g_{\mu \nu} \bigg( \frac{1}{2} g^{\alpha \beta} \phi_{,\alpha} \phi_{,\beta} + V \bigg)		\nonumber \\
			& \qquad + \gamma \chi_{,\mu} \chi_{,\nu} + g_{\mu \nu} \bigg[ \frac{1}{f} \bigg( 1 - \frac{1}{\gamma} \bigg) \bigg]~.
\end{align}
From this, the energy density and pressure are
\begin{subequations}
	\begin{align}
		\rho = {} & \frac{1}{2} \dot{\phi}^2 + \frac{1}{f}( \gamma - 1 ) + V~, \\
		p = {} & \frac{1}{2} \dot{\phi}^2 + \frac{1}{f} \bigg( 1 - \frac{1}{\gamma} \bigg) - V~.
	\end{align}
\end{subequations}
From the def\mbox{}initions above, the Friedmann equations are 
\begin{subequations}
	\begin{align}
		3 H^2 = {} & \frac{1}{2} \dot{\phi}^2 + \frac{1}{f}( \gamma - 1 ) + V~, \\
		-2 \dot{H} = {} & \dot{\phi}^2 + \gamma \dot{\chi}^2~,
	\end{align}
\end{subequations}
where the following relation
\be
1 - \frac{1}{\gamma^2} = f \dot{\chi}^2~
\ee
has been utilised.
Likewise to the case containing the kinetic coupling, when the boost factor is set to one, the equations reduce to those of the two canonical f\mbox{}ields.
\end{itemize}

\subsection{Perturbations}
We turn now our attention to the f\mbox{}irst order perturbation equations.
First, we choose a gauge to perform this calculation in; for this work, we will perform the calculation in the longitudinal gauge, which was discussed in Section~\ref{sec:chap1scalmodes}.
In this gauge and without the presence of anisotropic stress, the two scalar metric perturbations $ \Psi $ and $ \Phi $ are equal.
The perturbed FRW line element is essentially the same as that given in Eq.~\eqref{lineelfull} but now including $ \Phi = \Psi $, we will repeat it here for convenience
\be
\mathrm{d}s^2 = -\,(1-2\Psi) \mathrm{d}t^2 + a^2(1+2\Psi) \delta_{ij} \mathrm{d}x^i \mathrm{d}x^j~.
\ee
The perturbed Einstein f\mbox{}ield equations in the longitudinal gauge are
\begin{subequations}
\begin{align}
3H( H\Psi + \dot{\Psi} ) + \frac{k^2}{a^2} \Psi = {} & - \frac{1}{2} \delta \rho~,		\\
\label{deltaqeq}
\dot{\Psi} + H \Psi = {} & - \frac{1}{2} \delta q~, 							\\
\ddot{\Psi} + 4H\dot{\Psi} + ( 2\dot{H} + 3H^2 )\Psi = {} & \frac{1}{2} \delta p~,
\end{align}
\end{subequations}
where the perturbations in the energy density, momentum potential and pressure are $ \delta \rho $, $ \delta q $ and $ \delta p $, respectively, as stated in Section~\ref{sec:chap1scalmodes}.
We will now state the perturbation equations --- both the Klein-Gordon and Einstein equations --- for the models containing the noncanonical kinetic terms.
\newline
\\
We begin with the model involving the kinetic coupling between the two f\mbox{}ields.
The components, $ \delta \rho, \delta q $ and $ \delta p $, of the perturbed energy-momentum tensor are calculated using Eqs.~\eqref{vartensor} and \eqref{tensor1}, and are as follows
\begin{subequations}
\begin{align}
\delta \rho = {} & \dot{\phi} \delta \dot{\phi} 
			+ e^{2b} \dot{\chi} \delta \dot{\chi}
			+ b_{\phi} e^{2b} \dot{\chi}^2 \delta \phi
			+ V_{\phi} \delta \phi
			+ V_{\chi} \delta \chi
			- \dot{\phi}^2 \Psi
			- e^{2b} \dot{\chi}^2 \Psi~,			\\
\delta q 	= {} & \dot{\phi} \delta \phi
			+ e^{2b} \dot{\chi} \delta \chi~,		\\
\delta p	= {} & \dot{\phi} \delta \dot{\phi} 
			+ e^{2b} \dot{\chi} \delta \dot{\chi}
			+ b_{\phi} e^{2b} \dot{\chi}^2 \delta \phi
			- V_{\phi} \delta \phi
			- V_{\chi} \delta \chi
			- \dot{\phi}^2 \Psi
			- e^{2b} \dot{\chi}^2 \Psi~.
\end{align}
\end{subequations}
The perturbation equations for the two fields are given by perturbing the general form of their equations of motion given in Eqs.~\eqref{generalkinetic}, and are given as
\begin{subequations}
\label{kintpert}
\begin{align}
\delta \ddot{\phi} {} & + 3 H \delta \dot{\phi} 
				+ \bigg[ \frac{k^2}{a^2} + V_{\phi \phi} - ( b_{\phi \phi} + 2 \, b_{\phi}^2 ) \dot{\phi}^2 e^{2b} \bigg] \delta \phi 		
				- 2 \, b_{\phi} e^{2b} \dot{\chi} \delta \dot{\chi} 		\nonumber		\\
				& \qquad + V_{\phi \chi} \delta \chi
				- 4 \dot{\phi} \dot{\Psi} 
				+ 2 V_{\phi} \Psi
				= 0~,			\\
\delta \ddot{\chi} {} & + ( 3H + 2b_{\phi} \dot{\phi} ) \delta \dot{\chi}
				+ \bigg[ \frac{k^2}{a^2} + e^{-2b} V_{\chi \chi} \bigg] \delta \chi
				+ 2b_{\phi} \dot{\chi} \delta \dot{\phi}				\nonumber		\\
				& \qquad + \bigg[ e^{-2b} ( V_{\chi \phi} - 2b_{\phi} V_{\chi} ) + 2b_{\phi \phi} \dot{\phi} \dot{\chi} \bigg] \delta \phi
				- 4 \dot{\chi} \dot{\Psi} 
				+ 2e^{-2b} V_{\chi} \Psi
				= 0~.
\end{align}
\end{subequations}
We can rewrite these perturbation equations in Eq.~\eqref{kintpert} in a gauge-invariant form using the Sasaki-Mukhanov variables stated earlier in Eq.~\eqref{muksasvar} in Section~\ref{sec:chap1scalmodes}; they are given as follows
\begin{subequations}
\begin{align}
\ddot{Q}_{\phi} & + 3H \dot{Q}_{\phi}
			- 2 e^{2b} b_{\phi} \dot{\chi} \dot{Q}_{\chi}
			+ \bigg( \frac{k^2}{a^2} + C_{\phi \phi} \bigg) Q_{\phi}
			+ C_{\phi \chi} Q_{\chi}
			= 0~,			\\\
\ddot{Q}_{\chi} & + 3H \dot{Q}_{\chi}
			+ 2 b_{\phi}\dot{\phi} \dot{Q}_{\chi}
			+ 2 b_{\phi} \dot{\chi} \dot{Q}_{\phi}
			+ \bigg( \frac{k^2}{a^2} + C_{\chi \chi} \bigg) Q_{\chi}
			+ C_{\chi \phi} Q_{\phi} 
			= 0~,
\end{align}
\end{subequations}
with the coefficients $ C_{\phi \phi}, C_{\phi \chi}, C_{\chi \chi} $ and $ C_{\chi \phi} $ as
\begin{subequations}
\begin{align}
C_{\phi \phi} = {} & -2 e^{2b} b_{\phi}^2 \dot{\chi}^2
				+ 3 \dot{\phi}^2
				- \frac{e^{2b} \dot{\phi}^2 \dot{\chi}^2}{2 H^2}
				- \frac{\dot{\phi}^4}{2 H^2}
				- e^{2b} b_{\phi \phi} \dot{\chi}^2
				+ \frac{2 \dot{\phi} V_{\phi} }{H}
				+ V_{\phi \phi}~,		\\
C_{\phi \chi} = {} & 3 e^{2b} \dot{\phi} \dot{\chi}
				- \frac{e^{4b} \dot{\phi} \dot{\chi}^3}{2 H^2}
				- \frac{e^{2b} \dot{\phi}^3 \dot{\chi} }{2 H^2}
				+ \frac{\dot{\phi} V_{\chi} }{H}
				+ \frac{e^{2b} \dot{\chi} V_{\phi} }{H}
				+ V_{\phi \chi}~,		\\
C_{\chi \chi} = {} & 3 e^{2b} \dot{\chi}^2
				- \frac{e^{4b} \dot{\chi}^4}{2 H^2}
				- \frac{e^{2b} \dot{\phi}^2 \dot{\chi}^2}{2 H^2}
				+ \frac{2 \dot{\chi} V_{\chi} }{H}
				+ e^{-2b} V_{\chi \chi}~,	\\
C_{\chi \phi} = {} & 3 \dot{\phi} \dot{\chi}
				- \frac{e^{2b} \dot{\phi} \dot{\chi}^3 }{2 H^2}
				- \frac{ \dot{\phi}^3 \dot{\chi} }{2 H^2}
				+ 2 b_{\phi \phi} \dot{\phi} \dot{\chi}
				- 2 e^{-2b} b_{\phi} V_{\chi}
				+ \frac{e^{-2b} \dot{\phi} V_{\chi} }{H}
				+ \frac{\dot{\chi} V_{\phi} }{H}			\nonumber	\\
				& \qquad + e^{-2b} V_{\phi \chi}~.
\end{align}
\end{subequations}
The perturbation equations (Klein-Gordon and Einstein equations) for the model involving the kinetic coupling between the two f\mbox{}ields can be found in the papers \cite{DiMarco:2002eb, Lalak:2007vi, vandeBruck:2014ata}.
\newline
\\
We now move to the perturbation equations for the model concerning the canonical scalar and DBI f\mbox{}ields.
In this model, the energy density, momentum potential and pressure perturbations are as follows
\begin{subequations}
\begin{align}
\delta \rho = {} & \dot{\phi} \delta \dot{\phi} 
				- \dot{\phi}^2 \Psi 
				+ V_{\phi} \delta \phi 
				+ V_{\chi} \delta \chi 	
				+  \frac{1}{2} \frac{f_{\chi}}{f^2} \bigg( 2 - 3\gamma + \gamma^3 \bigg) \delta \chi 			\nonumber	\\
				& \qquad - \gamma^3( \dot{\chi} \delta \dot{\chi} + \dot{\chi}^2 \Psi )~,	\\
\delta q = {} & - ( \dot{\phi} \delta \phi + \gamma \dot{\chi} \delta \chi )~,		\\
\delta p = {} & \dot{\phi} \delta \dot{\phi} 
				- \dot{\phi}^2 \Psi 
				- V_{\phi} \delta \phi 
				- V_{\chi} \delta \chi 		
				- \frac{1}{2} \frac{f_{\chi}}{f^2} \bigg( 2 - \frac{1}{\gamma} - \gamma \bigg) \delta \chi 			\nonumber	\\
				& \qquad + \gamma( \dot{\chi} \delta \dot{\chi} - \dot{\chi}^2 \Psi )~.
\end{align}
\end{subequations}
We f\mbox{}ind the Klein-Gordon equations for the f\mbox{}ield perturbations are
\begin{subequations}
\label{dbipert}
\begin{align}
\delta \ddot{\phi} {} & + 3H \delta \dot{\phi} + \bigg( \frac{k^2}{a^2} + V_{\phi \phi} \!\bigg) \delta \phi 
						+ V_{\phi \chi} \delta \chi - 4 \dot{\phi} \dot{\Psi} + 2 V_{\phi} \Psi = 0~, \\
\delta \ddot{\chi} {} & +  3\bigg( H + \frac{\dot{\gamma}}{\gamma} \bigg) \delta \dot{\chi} 
					+  \bigg[ \frac{k^2}{a^2 \gamma^2} + \frac{V_{\chi \chi}}{\gamma^3} 
					+ \frac{f_{\chi} \dot{\gamma}}{f \gamma} \dot{\chi} 
					- \frac{1}{2} \frac{f_{\chi}}{\gamma} \dot{\chi}^2 V_{\chi} 		\nonumber \\
					& + \frac{1}{2} \bigg(1 - \frac{1}{\gamma} \bigg)^2 \left\{ \frac{1}{\gamma} \bigg( \frac{f_{\chi}}{f^2} \bigg)_{\!,\chi} 
					+ \bigg(1 + \frac{1}{\gamma} \bigg)\frac{1}{f} \bigg( \frac{f_{\chi}}{f} \bigg)_{\!,\chi} \right\} \bigg] \delta \chi			\nonumber \\
					& + \frac{V_{\chi \phi}}{\gamma^3} \delta \phi 
					- \bigg( \frac{3}{\gamma^2} + 1 \bigg) \dot{\chi}\dot{\Psi} 				\nonumber	\\
					& + \bigg[ \frac{f_{\chi}}{f^2 \gamma^3} (1 - \gamma )^2 
							+ \frac{V_{\chi}}{\gamma^3}(1 + \gamma^2) 
							- 2\frac{\dot{\gamma}}{\gamma}\dot{\chi} \bigg] \Psi = 0~.
\end{align}
\end{subequations}
By rewriting the field perturbation equations in Eq.~\eqref{dbipert} using the Sasaki-Mukhanov variables as stated in Eq.~\eqref{muksasvar}, the gauge-invariant form of the perturbation equations are
\begin{subequations}
\begin{align}
\label{perturbationdbi1}
\ddot{Q}_{\phi} {} & + 3H\dot{Q}_{\phi} +B_{\phi}\dot{Q}_{\chi} + \bigg(\frac{k^{2}}{a^2} + C_{\phi \phi} \bigg)Q_{\phi} + C_{\phi \chi}Q_{\chi} = 0~, \\
\label{perturbationdbi2}
\ddot{Q}_{\chi} {} & + \bigg{(}3H + 3\frac{\dot{\gamma}}{\gamma} \bigg{)} \dot{Q}_{\chi} + B_{\chi}\dot{Q}_{\phi} + \bigg{(}\frac{k^{2}}{a^2 \gamma^{2}} + C_{\chi \chi}\bigg{)}Q_{\chi} + C_{\chi \phi}Q_{\phi} = 0~,
\end{align}
\end{subequations}
with the coeff\mbox{}icients $B_{\phi}, B_{\chi}, C_{\phi \phi}, C_{\phi \chi}, C_{\chi \chi}, C_{\chi \phi} $, in the equations are as follows
\begin{subequations}
\begin{align}
B_{\phi} = {} & \frac{\dot{\phi}}{2H}\gamma^{3}\bigg{(}1-\frac{1}{\gamma^{2}}\bigg{)}\dot{\chi}~, \\
B_{\chi} = {} &-\frac{\dot{\phi}}{2H}\bigg{(}1 - \frac{1}{\gamma^{2}}\bigg{)}\dot{\chi}~, \\
C_{\phi \phi} = {} & 3\dot{\phi}^{2} 
				- \gamma^{3}\bigg{(}1 + \frac{1}{\gamma^{2}} \bigg{)}\frac{\dot{\phi}^{2}\dot{\chi}^{2}}{4H^{2}} 
				- \frac{\dot{\phi}^{4}}{2H^{2}} 
				+ 2\frac{\dot{\phi}}{H}V_{\phi} 
				+ V_{\phi \phi}~, \\
C_{\phi \chi} = {} &\frac{\dot{\phi}}{4H}\frac{f_{\chi}}{f^{2}}\frac{1}{\gamma}(1-\gamma)^{2}(\gamma^{2} + 2\gamma -1) 
				+ 3\gamma\dot{\chi}\dot{\phi} 			\nonumber		\\
				& - \frac{\gamma^{4}}{4H^{2}}\bigg{(}1+\frac{1}{\gamma^{2}}\bigg{)}\dot{\phi}\dot{\chi}^{3} 
				- \frac{\gamma\dot{\phi}^{3}\dot{\chi}}{2H^{2}} + \frac{\dot{\phi}}{H}V_{\chi} + \frac{\gamma\dot{\chi}}{H} V_{\phi} + V_{\phi \chi}~, \\
C_{\chi \chi} = {} &\frac{1}{H}\frac{f_{\chi}}{f^{2}}\bigg{(}1 - \frac{1}{\gamma}\bigg{)}^{2}\dot{\chi} + \bigg{(}\frac{f_{\chi}}{f} 
					- \frac{\gamma\dot{\chi}}{H}\bigg{)}\frac{\dot{\gamma}}{\gamma}\dot{\chi} 
					-\frac{1}{2}\frac{f_{\chi}}{\gamma}\dot{\chi}^{2}V_{\chi}				\nonumber \\
			& + \frac{1}{2}\bigg{(}1 - \frac{1}{\gamma}\bigg{)}^{2}\bigg{[}\frac{1}{\gamma}\bigg{(}\frac{f_{\chi}}{f^{2}}\bigg{)}_{\!\!,\chi}		
					+ \bigg{(}1+\frac{1}{\gamma}\bigg{)}\frac{1}{f}\bigg{(}\frac{f_{\chi}}{f} \bigg{)}_{\!\!,\chi}\bigg{]}		\nonumber \\
			& - \frac{\gamma^{2} \dot{\chi}^{4}}{2H^{2}} - \frac{\gamma}{4H^{2}}\bigg{(}1+ \frac{1}{\gamma^{2}}\bigg{)}\dot{\chi}^{2}\dot{\phi}^{2} 
					+ \frac{1}{H}\bigg{(}1+\frac{1}{\gamma^{2}}\bigg{)}\dot{\chi}V_{\chi} 	\nonumber \\
			& + \frac{V_{\chi \chi}}{\gamma^{3}} + \frac{3}{2}\gamma\bigg{(}1 + \frac{1}{\gamma^{2}}\bigg{)}\dot{\chi}^{2}~, \\
C_{\chi \phi} = {} &\frac{\dot{\phi}}{H}\bigg{[}\frac{1}{2}\frac{f_{\chi}}{f^{2}}\frac{1}{\gamma}\bigg{(}1 - \frac{1}{\gamma}\bigg{)}^{2} 
					- \frac{\dot{\gamma}}{\gamma}\dot{\chi}\bigg{]} - \frac{\gamma\dot{\phi}\dot{\chi}^{3}}{2H^{2}} 		\nonumber \\
			& + \frac{1}{2}\bigg{(}1+\frac{1}{\gamma^{2}}\bigg{)} \bigg{[} 3\dot{\phi}\dot{\chi} - \frac{\dot{\chi}\dot{\phi}^{3}}{2H^{2}} 
					+ \frac{\dot{\phi}}{\gamma H}V_{\chi} 
					+ \frac{\dot{\chi}}{H}V_{\phi}\bigg{]}~.
\end{align}
\end{subequations}
Further on in this work, we shall calculate the power spectra for both the curvature and isocurvature perturbations.
We will now recall elements of the different perturbations from Section~\ref{sec:chap1scalmodes}.
\newline
\\
The curvature perturbation that we have chosen to study is the curvature perturbation $ \mathcal{R} $ in Eq.~\eqref{curvR}.
It can be rewritten in a different form by using the acceleration equation in Eq.~\eqref{acceq} and the perturbed Einstein equation in Eq.~\eqref{deltaqeq}
\be
\mathcal{R} = -\frac{H^2}{\dot{H}} \bigg( \Psi + \frac{\dot{\Psi}}{H} \bigg) + \Psi~.
\ee
For the isocurvature perturbation, we shall calculate the entropy perturbation $ \mathcal{S} $ and this is a gauge-invariant quantity \cite{Gordon:2000hv, Malik:2004tf}, and it is related to the nonadiabatic pressure perturbation $ \delta p_{\mathrm{nad}} $.
The gauge-invariant entropy perturbation and the nonadiabatic pressure perturbation are related through the def\mbox{}inition:
\be
\mathcal{S} = \frac{H}{\dot{p}} \delta p_{\rm nad}~.
\ee
The nonadiabatic pressure perturbation is a constituent of the pressure perturbation $ \delta p $; the pressure perturbation can be written as
\begin{align}
\label{pressure}
\delta p & = c_{\mathrm{s,a}}^{\,2} \delta \rho + \delta p_{\rm nad}~,		\\
\label{adiabspeed}
c_{\mathrm{s,a}}^{\,2} & = \frac{\dot{p}}{\dot{\rho}}~,
\end{align}
with $ c_{\mathrm{s,a}}^{\,2} $ denoting the adiabatic sound speed.
\newline
\\
The adiabatic sound speed for each of the three models can be calculated by using the time derivatives of the Friedmann and acceleration equations, and the Klein-Gordon equations for the two f\mbox{}ields.
We now state the adiabatic sound speeds for the three models considered in this work. 
This speed should not be confused with the sound speed $ c_{\mathrm{s}}^{\,2} $ of the DBI field fluctuation.
\newline
\\
There are two speeds that we use in this work: the adiabatic sound speed $ c_{\mathrm{s,a}}^{\,2} $ and the DBI field perturbation sound speed $ c_{\mathrm{s}}^{\,2} $.
To be clear, we now make the distinction between the two.
The first of these two speeds, the adiabatic sound speed, is given by the standard definition stated in Eq.~\eqref{adiabspeed},
and the second speed is introduced when considering the speed at which a field perturbation propagates.
This is the field perturbation sound speed, otherwise known as the phase speed; for further discussion on this topic, see \cite{Christopherson:2008ry}.
\begin{enumerate}
\item Two canonical scalar f\mbox{}ields
\begin{equation}
c_{\mathrm{s,a}}^{\,2} = 1 + \frac{2 ( V_{\phi} \dot{\phi} + V_{\chi} \dot{\chi} )}{ 3H ( \dot{\phi}^2 + \dot{\chi}^2 )}~.
\end{equation}
\item One canonical scalar f\mbox{}ield and one scalar f\mbox{}ield with a kinetic coupling
\begin{align}
c_{\mathrm{s,a}}^{\,2} = {} & 1 + \frac{2 ( V_{\phi} \dot{\phi} + V_{\chi} \dot{\chi} )}{ 3H ( \dot{\phi}^2 + e^{2b} \dot{\chi}^2 )}~.
\end{align}
\item One canonical scalar f\mbox{}ield and one DBI f\mbox{}ield
\begin{align}
c_{\mathrm{s,a}}^{\,2} & =  \frac{2 \dot{\phi}^2 + ( 1 + \gamma^2 ) \gamma^{-1} \dot{\chi}^2}{ \dot{\phi}^2 + \gamma \dot{\chi}^2 }  
							+ \frac{2V_{\phi}\dot{\phi} 
							+ ( 1+ \gamma^2 ) \gamma^{-2} V_{\chi} \dot{\chi}}{ 3H( \dot{\phi}^2 + \gamma \dot{\chi}^2 ) }		\nonumber \\
		& \qquad + \frac{1}{ 3H( \dot{\phi}^2 + \gamma \dot{\chi}^2 ) } \frac{ f_\chi}{f^2} \bigg( \frac{1}{\gamma} - 1\bigg)^2 \dot{\chi} - 1~.
\end{align}
\end{enumerate}
For the f\mbox{}inal two models that contain noncanonical kinetic terms, the adiabatic sound speed reduces to the two canonical scalar f\mbox{}ield model when the kinetic coupling and boost factor are $ b = 0 $ and $ \gamma = 1 $.
\newline
\\
Due to the complexity of these equations, for all models considered, we will solve them numerically following the method outlined in \cite{Lalak:2007vi, Liddle:2003as, vandeBruck:2010yw}.

\section{Results}
\label{sec:results}
We will now describe the results of our numerical calculations. 
To be concrete, we consider all models presented in 
Section~\ref{sec:chap2sec1} with the double quadratic potential \cite{Langlois:1999dw} which will be rewritten in the form
\begin{equation}
V(\phi,\chi) = \frac{1}{2} m_{\phi}^2 ( \phi^2 + \Gamma^{\,2} \chi^2)~,
\end{equation}
with $ \Gamma $ as the ratio between the two f\mbox{}ield masses
\begin{equation}
\Gamma = \frac{m_{\chi}}{m_{\phi}}~,
\end{equation}
where $ m_\phi $ and $ m_\chi $ are the masses of the f\mbox{}ields $ \phi $ and $ \chi $, respectively.
\newline
\\
All the plots in this chapter are shown at the Wilkinson Microwave Anisotropy Probe (WMAP) pivot scale where $ k_{\mathrm{WMAP}} = 0.002 \, \mathrm{Mpc}^{-1} $ \cite{Komatsu:2010fb} and at plotted against the number of $ e $-folds.
For all models considered, the parameter values have been chosen so that the final amplitude of the curvature power spectrum is $ \mathcal{P}_{\mathcal{R}} \sim 2 \times 10^{-9} $ at the WMAP pivot scale.

\subsection{Two canonical scalar f\mbox{}ields}
\label{sec:twostandard}
\begin{figure}
\begin{center}
\scalebox{0.55}{\includegraphics{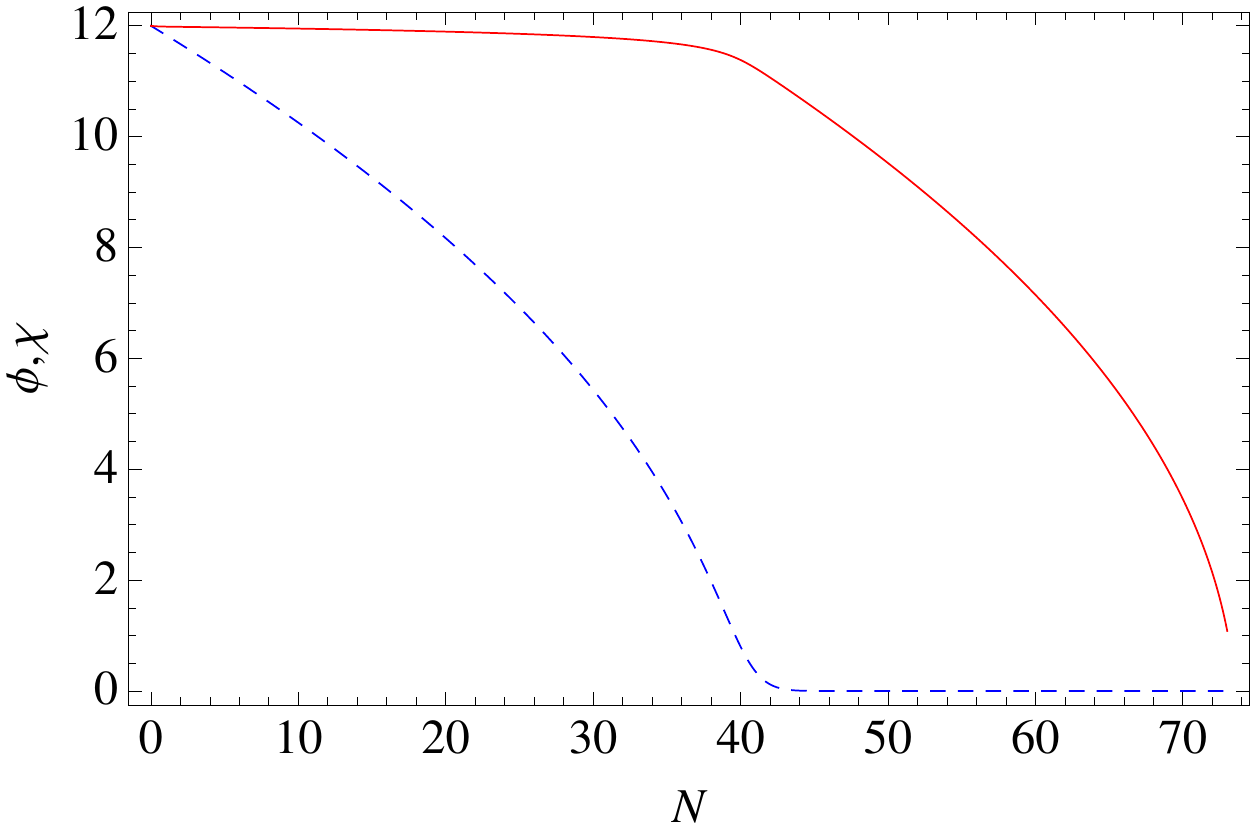}}
\caption{Background f\mbox{}ield evolution for the model containing two canonical scalar f\mbox{}ields: $ \phi $ (solid red) and $ \chi $ (blue dashed).
The initial f\mbox{}ield values are $ \phi_{\mathrm{ini}} = \chi_{\mathrm{ini}} = 12.0 $ and masses of the two f\mbox{}ields are $ \Gamma = 7.0 $ and $ m_{\phi} = 1.395 \times 10^{-6} $.
}
\label{standphichi}
\end{center}
\end{figure}

\begin{figure}
\begin{center}
\scalebox{1.1}{\includegraphics{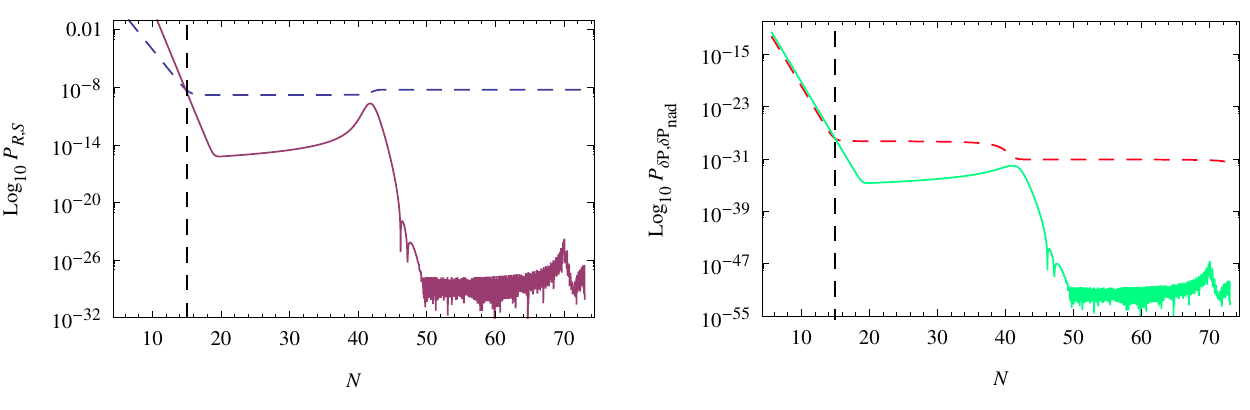}}
\caption{
Comparison of the various power spectra evolution with respect to $e$-fold number generated for the two canonical scalar f\mbox{}ield model.
The initial f\mbox{}ield values are $ \phi_{\mathrm{ini}} = \chi_{\mathrm{ini}} = 12.0 $ and masses of the two f\mbox{}ields are $ \Gamma = 7.0 $ and $ m_{\phi} = 1.395 \times 10^{-6} $.
\newline
\emph{Left panel}: Power spectra evolution for the curvature ($ \mathcal{P}_{\mathcal{R}} $, blue dashed) and entropy ($ \mathcal{P}_{\mathcal{S}} $, solid purple) perturbations. 
\emph{Right panel}: Power spectra evolution for the pressure ($ \mathcal{P}_{\delta p} $, dashed red) and nonadiabatic pressure ($ \mathcal{P}_{\delta p_{\mathrm{nad}} } $, solid green) perturbations. In both plots, the dashed black line indicates horizon crossing corresponding to $  k_{\mathrm{WMAP}} = 0.002 \, \mathrm{Mpc}^{-1} $.
}
\label{standpow}
\end{center}
\end{figure}

In this model, we consider the case where $ \Gamma = 7.0 $ as studied in \cite{Lalak:2007vi, Avgoustidis:2011em}.
To further this work, we impose a further condition.
This condition is to match the curvature power spectrum amplitude of approximately $ \mathcal{P}_{\mathcal{R}} \sim 2 \times 10^{-9} $ at the WMAP pivot scale.
In order to do this, the mass required by the f\mbox{}ield $ \phi $ is $ m_{\phi} = 1.395 \times 10^{-6} $.
The initial f\mbox{}ield values for the two f\mbox{}ields are $\chi_{\mathrm{ini}} = \phi_{\mathrm{ini}} = 12.0 $ as in \cite{Huston:2011fr}.
In addition, the two f\mbox{}ield derivatives at the start of inf\mbox{}lation are those in slow-roll.
The background dynamics for this model are presented in Figure~\ref{standphichi} and the comparison of the various power spectra evolutions, including the curvature, entropy, pressure and nonadiabatic pressure perturbations, are displayed in Figure~\ref{standpow}.
\newline
\\
We see in Figure~\ref{standphichi} that the f\mbox{}ield $ \chi $ reaches the minimum of the potential, resulting in the $ \phi $-f\mbox{}ield dominating for the last 30 $e$-folds of inf\mbox{}lation. 
This is ref\mbox{}lected in Figure~\ref{standpow}, which at this same point in time, there is a rise in the power spectrum of the curvature perturbation. 
The amplitude of the entropy perturbation power spectrum $ \mathcal{P}_{\mathcal{S}} $ reduces dramatically when the exchange in f\mbox{}ield dominance occurs at 40 $e$-folds. 
It starts to gradually increase until a peak is reached at $ N = 70 $, at which a drop is then experienced. 
\newline
\\
At the end of inf\mbox{}lation, we see the magnitude of the entropy power spectrum is many orders smaller than the curvature power spectrum.
To be specif\mbox{}ic when inf\mbox{}lation has ended, the curvature power spectrum is $ \mathcal{P}_{\zeta} \sim \times 10^{-9} $ whereas the entropy power spectrum is $ \mathcal{P}_{\mathcal{S}} \sim 10^{-31} $.
Other f\mbox{}inal power spectra amplitudes are those for the pressure and nonadiabatic pressure perturbations; they are
$ \mathcal{P}_{\delta p} \sim 10^{-32} $ and $ \mathcal{P}_{\delta p_{\mathrm{nad}} } \sim 10^{-54} $.
\newline
\\
We see that the behaviour of the power spectra evolution of the nonadiabatic pressure and entropy perturbation are generally following the same trend from horizon crossing, which occurs at $ N = 15 $ and onwards.
In addition to this, we f\mbox{}ind that Figure~\ref{standpow} is in agreement with Figures.~1 and 2 in the works of Huston and Christopherson \cite{Huston:2011fr}.

\subsection{Kinetic coupling}
\label{sec:kinetic}
\begin{figure}
\begin{center}
\scalebox{0.55}{\includegraphics{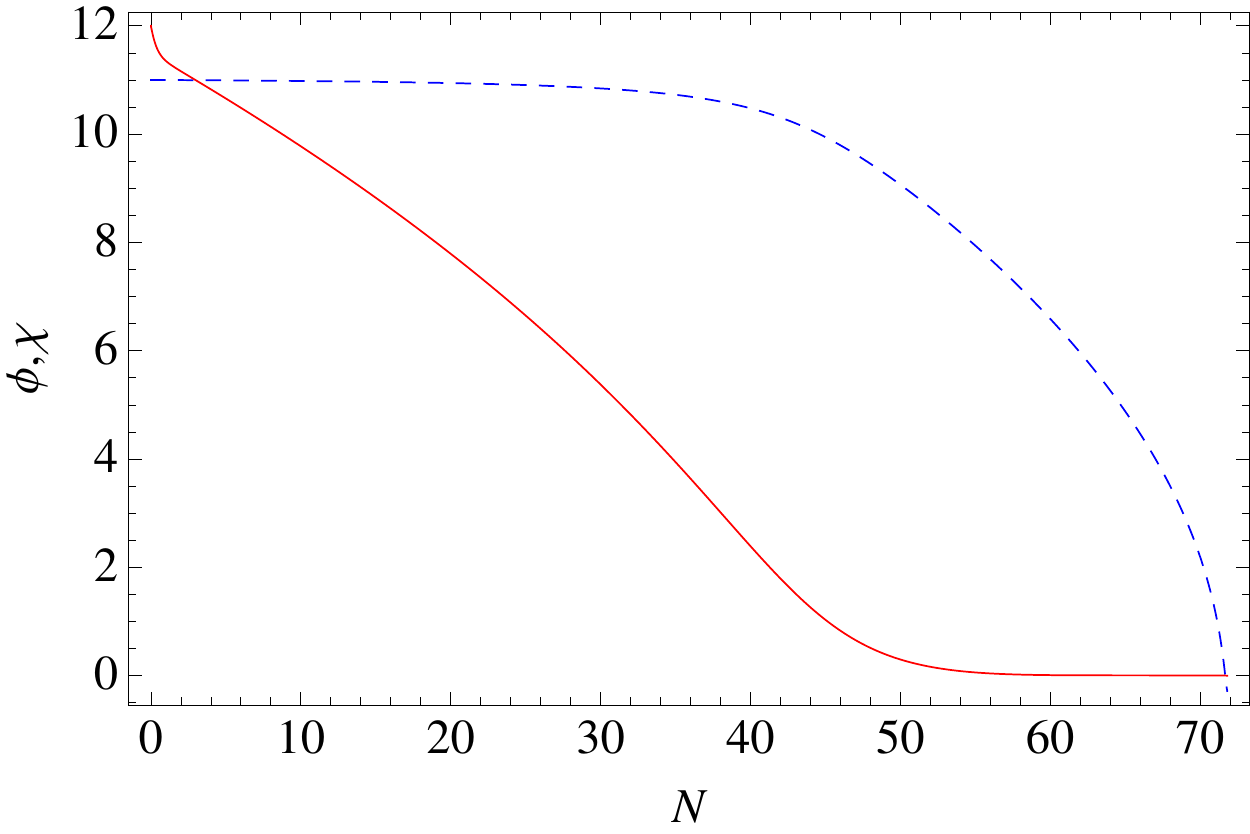}}
\caption{
Field evolution for the model with one canonical scalar f\mbox{}ield $ \phi $ (solid red) and one f\mbox{}ield containing a kinetic coupling $ \chi $ (blue dashed), with $ \phi $ dominating approximately the f\mbox{}irst 50 $ e $-folds of inf\mbox{}lation.
The initial f\mbox{}ield values are $ \phi_{\mathrm{ini}} = 11.0 $ and $ \chi_{\mathrm{ini}} = 12.0 $ and masses of the two f\mbox{}ields are $ \Gamma = 0.3 $ and $ m_{\phi} = 6.395 \times 10^{-6} $. The kinetic coupling is $ \beta = 0.1 $.
}
\label{kinetphiphichi}
\end{center}
\end{figure}

\begin{figure}
\begin{center}
\scalebox{0.55}{\includegraphics{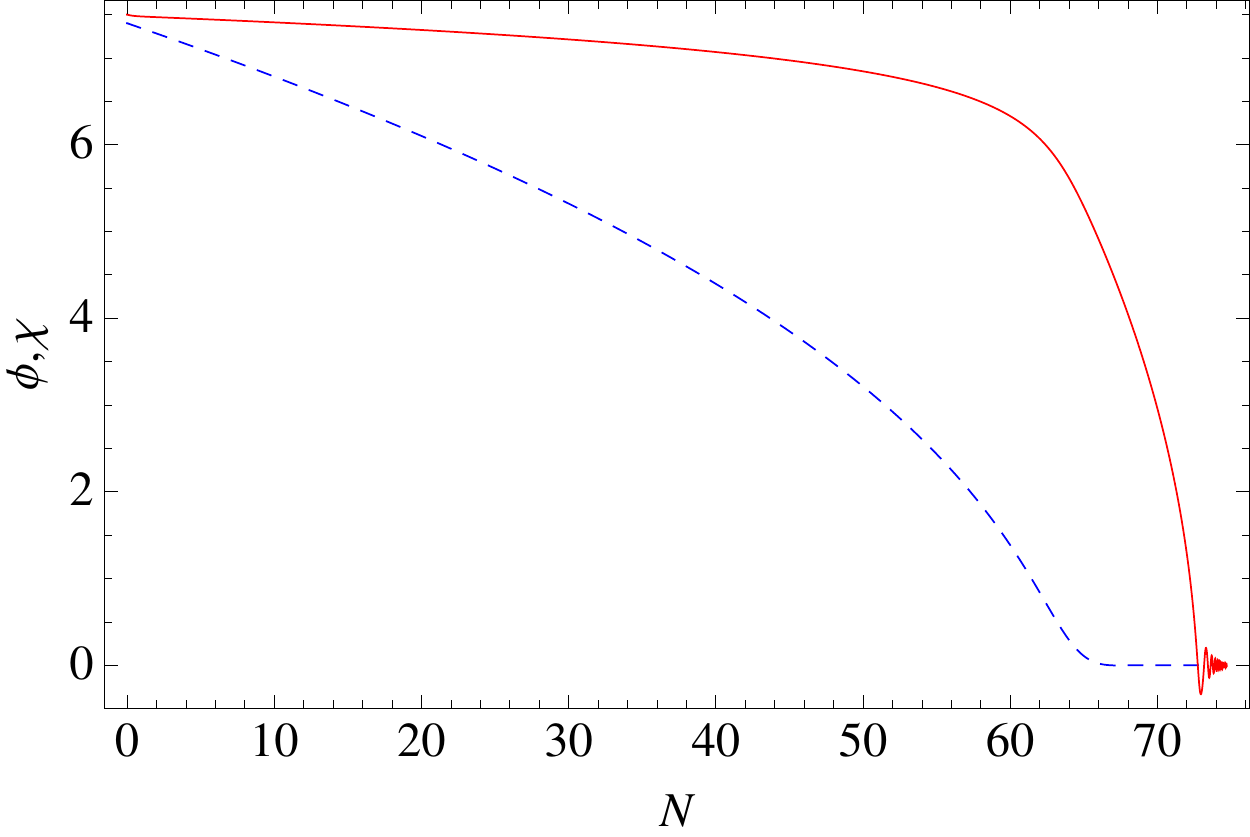}}
\caption{
Background dynamics for the model with one canonical scalar f\mbox{}ield $ \phi $ (solid red) and one f\mbox{}ield containing a kinetic coupling $ \chi $ (blue dashed), with $ \chi $ dominating inf\mbox{}lation during the f\mbox{}irst 67 $e$-foldings.
The initial f\mbox{}ield values are $ \phi_{\mathrm{ini}} = 7.5 $ and $ \chi_{\mathrm{ini}} = 7.4 $ and masses of the two f\mbox{}ields are $ \Gamma = 6.0 $ and $ m_{\phi} = 1.005 \times 10^{-6} $. The kinetic coupling is $ \beta = 0.1 $.
}
\label{kinetchiphichi}
\end{center}
\end{figure}

We have two possible scenarios that will arise in the background dynamics; one in which the f\mbox{}ield $\phi $ reaches the minimum of its potential before the f\mbox{}ield $\chi$, and vice versa. 
We will begin by considering the latter.
\newline
\\
For this scenario, the parameters in the potential are chosen as follows: the ratio of the f\mbox{}ield masses is $ \Gamma = 0.3 $ with $ m_{\phi} = 6.395 \times 10^{-6} $.
The initial conditions of the two f\mbox{}ields are $ \chi_{\mathrm{ini}} = 12.0 $ and $ \phi_{\mathrm{ini}} = 11.0 $ with the f\mbox{}ield derivatives set to slow-roll.
In addition, the kinetic coupling between the two f\mbox{}ields is $ \beta = 0.1$.
The background dynamics for this set of conditions are displayed in Figure~\ref{kinetphiphichi}.
As shown in Figure~\ref{kinetphiphichi}, the f\mbox{}irst 50 $e$-folds are dominated by the f\mbox{}ield $ \phi $ until at which point, there is an exchange in the f\mbox{}ield contributions, leaving the remaining $\chi$-f\mbox{}ield until the end of inf\mbox{}lation.
\begin{figure}
\begin{center}
\scalebox{1.1}{\includegraphics{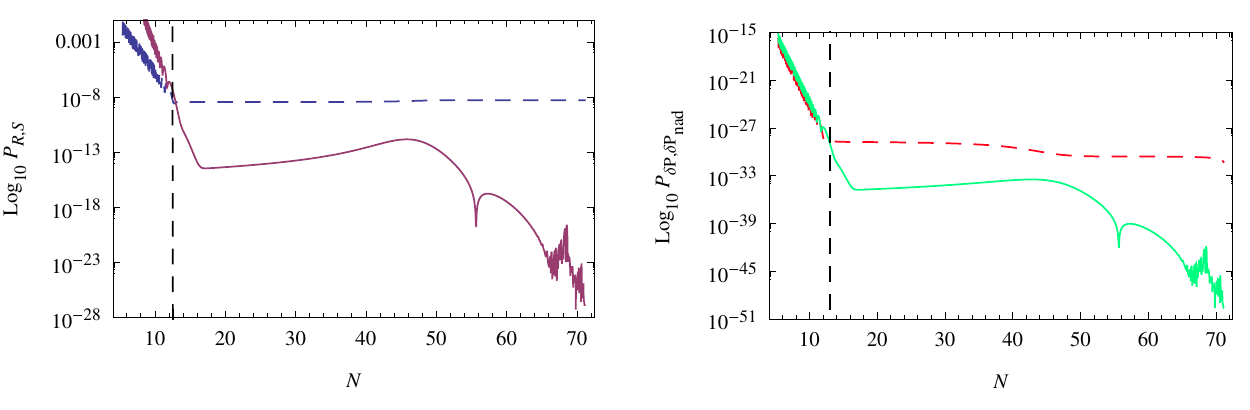}}
\caption{
Comparison of the power spectra evolution against the number of $e$-foldings for the model with one canonical scalar f\mbox{}ield and one f\mbox{}ield with a kinetic coupling. For this model, the f\mbox{}ield $ \phi $ dominates the inf\mbox{}lationary epoch for approximately 50 $e$-folds.
The initial f\mbox{}ield values are $ \phi_{\mathrm{ini}} = 11.0 $ and $ \chi_{\mathrm{ini}} = 12.0 $ and masses of the two f\mbox{}ields are $ \Gamma = 0.3 $ and $ m_{\phi} = 6.395 \times 10^{-6} $. The kinetic coupling is $ \beta = 0.1 $.
\newline
\emph{Left panel}: Power spectra evolution for the curvature ($ \mathcal{P}_{\mathcal{R}} $, blue dashed) and entropy ($ \mathcal{P}_{\mathcal{S}} $, solid purple) perturbations.
\emph{Right panel}: Power spectra evolution for the pressure ($ \mathcal{P}_{\delta p} $, dashed red) and nonadiabatic pressure ($ \mathcal{P}_{\delta p_{\mathrm{nad}} } $, solid green) perturbations. In both plots, the dashed black line indicates horizon crossing corresponding to $  k_{\mathrm{WMAP}} = 0.002 \, \mathrm{Mpc}^{-1} $.
}
\label{kinetphipow}
\end{center}
\end{figure}
\newline
\\
The evolution of the various power spectra is displayed in Figure~\ref{kinetphipow}: the left-hand panel shows a comparison of the curvature and entropy power spectra evolution, and the right-hand panel shows the comparison between the power spectra evolution of the pressure and nonadiabatic pressure perturbations.
\newline
The behaviour of the entropy and nonadiabatic component of the pressure perturbations is signif\mbox{}icantly different to the model previously considered in Section~\ref{sec:twostandard}. 
However, there is a similarity between the two Figures (\ref{standpow} and \ref{kinetphipow}).
For example, a rise and fall is experienced in the evolution of the entropy power spectrum, $ \mathcal{P}_\mathcal{S} $, during the last 6 $e$-folds before the end of inf\mbox{}lation.
This feature can also be seen in Figure~\ref{standpow}. 
Like the two scalar f\mbox{}ield model previously studied, the amplitude of entropy power spectrum at the end of inf\mbox{}lation is many orders smaller than of the curvature power spectrum. 
At the end of inf\mbox{}lation, the f\mbox{}inal amplitudes for the entropy and nonadiabatic pressure perturbations are $ \mathcal{P}_\mathcal{S} \sim 10^{-26} $ and $ \mathcal{P}_{\delta p_{\mathrm{nad}}} \sim 10^{-50} $.
Furthermore, the f\mbox{}inal amplitude of the pressure perturbation power spectrum is $ \mathcal{P}_{ \delta p } \sim 10^{-33} $.
\newline
\\
We f\mbox{}ind that the entropy power spectrum amplitude at the end of inf\mbox{}lation for this model is $ 10^{5} $ times larger than found in the model considering two canonical scalar f\mbox{}ields.
\newline
\\
We will now consider the other possible scenario which can arise in this model where the f\mbox{}ield $ \chi $ initially dominates the inf\mbox{}lationary epoch.
The kinetic coupling acting between the two scalar f\mbox{}ields is kept at the same value $ \beta = 0.1 $, as used in the previous possible set up.
However, the parameters used in the potential are different with the f\mbox{}ield mass ratio being $ \Gamma = 6.0 $ and $ m_{\phi} = 1.005 \times 10^{-6} $. 
In addition to this, the starting f\mbox{}ield values for the two scalar f\mbox{}ields are $ \chi_{\mathrm{ini}} = 7.4 $ and $\phi_{\mathrm{ini}} = 7.5 $.
The background dynamics that are produced by this choice of conditions is displayed in Figure~\ref{kinetchiphichi}.
From Figure~\ref{kinetchiphichi}, it can be seen that the f\mbox{}inal 8 $e$-folds are dominated by the canonical scalar f\mbox{}ield.
\begin{figure}
\begin{center}
\scalebox{1.1}{\includegraphics{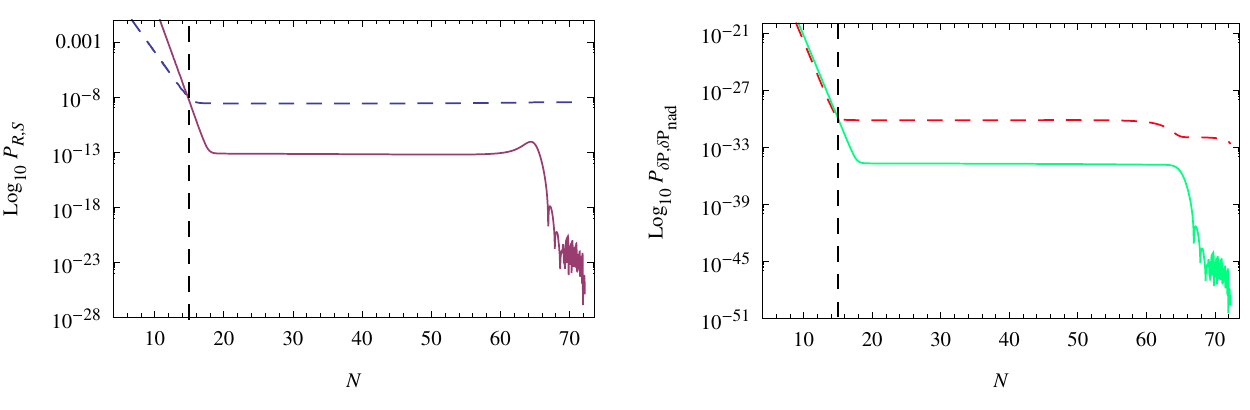}}
\caption{
Comparison of the power spectra evolution against $e$-fold number for the case with one canonical scalar f\mbox{}ield and one f\mbox{}ield with a kinetic coupling. 
Here, $ \chi $ dominates the inf\mbox{}lationary epoch for approximately 67 $e$-folds.
Starting f\mbox{}ield values are $ \phi_{\mathrm{ini}} = 7.5 $ and $ \chi_{\mathrm{ini}} = 7.4 $ and masses of the two f\mbox{}ields are $ \Gamma = 6.0 $ and $ m_{\phi} = 1.005 \times 10^{-6} $ whilst the kinetic coupling is $ \beta = 0.1 $.
\newline
\emph{Left panel}: Power spectra evolution of the curvature ($ \mathcal{P}_{\mathcal{R}} $, blue dashed) and entropy ($ \mathcal{P}_{\mathcal{S}} $, solid purple) perturbations. 
\emph{Right panel}: Power spectra evolution for the pressure ($ \mathcal{P}_{\delta p} $, dashed red) and nonadiabatic pressure ($ \mathcal{P}_{\delta p_{\mathrm{nad}} } $, solid green) perturbations. In both plots, the dashed black line indicates horizon crossing corresponding to $  k_{\mathrm{WMAP}} = 0.002 \, \mathrm{Mpc}^{-1} $.
}
\label{kinetchipow}
\end{center}
\end{figure}
\newline
As a consequence of the scalar f\mbox{}ield containing the kinetic coupling reaching its potential minimum earlier during inf\mbox{}lation, this has resulted in a one canonical scalar f\mbox{}ield scenario at the end of inf\mbox{}lation, which is similar to the two scalar f\mbox{}ields case studied in Section~\ref{sec:twostandard}.
Due to this, the shape of the power spectra evolution for all considered quantities will be similar.
The power spectra evolution for the curvature, entropy, pressure and nonadiabatic pressure perturbations are displayed and compared in Figure~\ref{kinetchipow}.
Adding to the statement about the similarities in the power spectra evolution between this model and that of the two canonical f\mbox{}ields, there is a slight difference in the last few $e$-folds in $ \mathcal{P}_{\mathcal{S}} $ (and $ \mathcal{P}_{ \delta p_{\mathrm{nad}} } $) which relates to the behaviour of the remaining canonical scalar f\mbox{}ield. 
\newline
\\
The evolution of the entropy power spectrum is seen initially rising and then falling during the remaining $e$-folds of inf\mbox{}lation in Figure~\ref{standpow} and Figure~\ref{kinetphipow}. 
Instead we see in Figure~\ref{kinetchipow} that the entropy power spectrum continues to decrease over time until slow-roll is no longer satisf\mbox{}ied. 
We f\mbox{}ind the f\mbox{}inal amplitudes for the entropy and nonadiabatic pressure perturbations are signif\mbox{}icantly larger than for the two standard scalar f\mbox{}ields case.
However, the values for the entropy and nonadiabatic pressure perturbation power spectra amplitudes are the similar, if not the same, as those found in the previous scenario, where the f\mbox{}ield $ \chi $ reaches its potential minimum before the f\mbox{}ield $ \phi $.
To be specif\mbox{}ic, the amplitudes for the entropy and nonadiabatic pressure perturbations power spectra are $ \mathcal{P}_{\mathcal{S}} \sim 10^{-27} $ and $ \mathcal{P}_{ \delta p_{\mathrm{nad}} } \sim 10^{-50} $.
Adding to this, the power spectrum amplitude of the pressure perturbation at the end of inf\mbox{}lation is $ \mathcal{P}_{\delta p} \sim 10^{-33} $.
\newline
\\
Furthermore, there appears to be no difference in the curvature and entropy power spectra amplitudes between the two scenarios studied here in Section~\ref{sec:kinetic}.

\subsection{A scalar f\mbox{}ield and DBI f\mbox{}ield}
\begin{figure}
\begin{center}
\scalebox{1.1}{\includegraphics{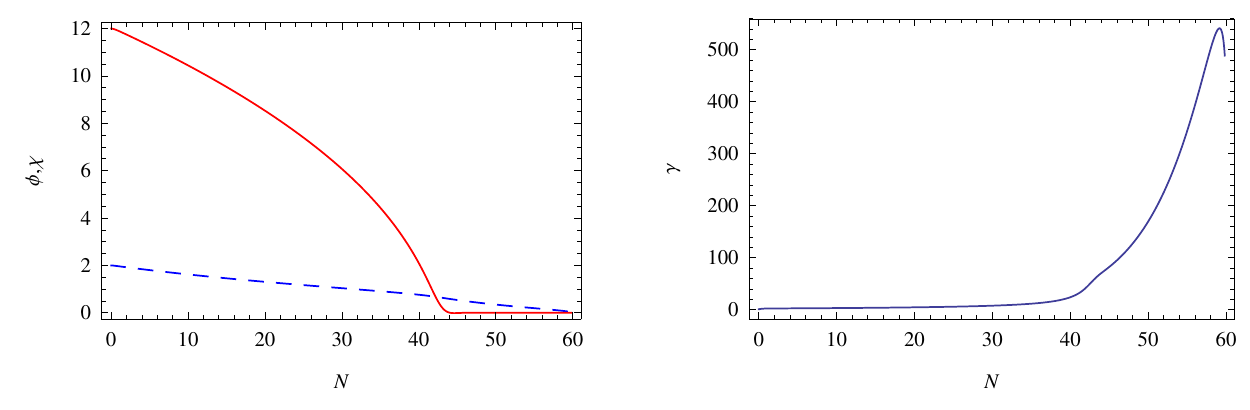}}
\caption{
Background dynamics for the model with a canonical scalar f\mbox{}ield $ \phi $ and DBI f\mbox{}ield $ \chi $ where the scalar f\mbox{}ield $ \phi $ dominates the inf\mbox{}lationary period.
The initial f\mbox{}ield values are $ \phi_{\mathrm{ini}} = 12.0 $ and $ \chi_{\mathrm{ini}} = 2.0 $ and masses of the two f\mbox{}ields are $ \Gamma = 2.0 $ and $ m_{\phi} = 1.156 \times 10^{-5} $.
The parameters required in the DBI model are $ \lambda = 2.0 \times 10^{12} $ and $ \mu = 0.2 $.
\newline
\emph{Left panel}: Field evolution of the canonical scalar f\mbox{}ield $ \phi $ (solid red) and DBI f\mbox{}ield $ \chi $ (blue dashed).
\emph{Right panel}: Evolution of the boost factor $ \gamma $ during inf\mbox{}lation.
}
\label{dbiphibggam}
\end{center}
\end{figure}

\begin{figure}
\begin{center}
\scalebox{1.1}{\includegraphics{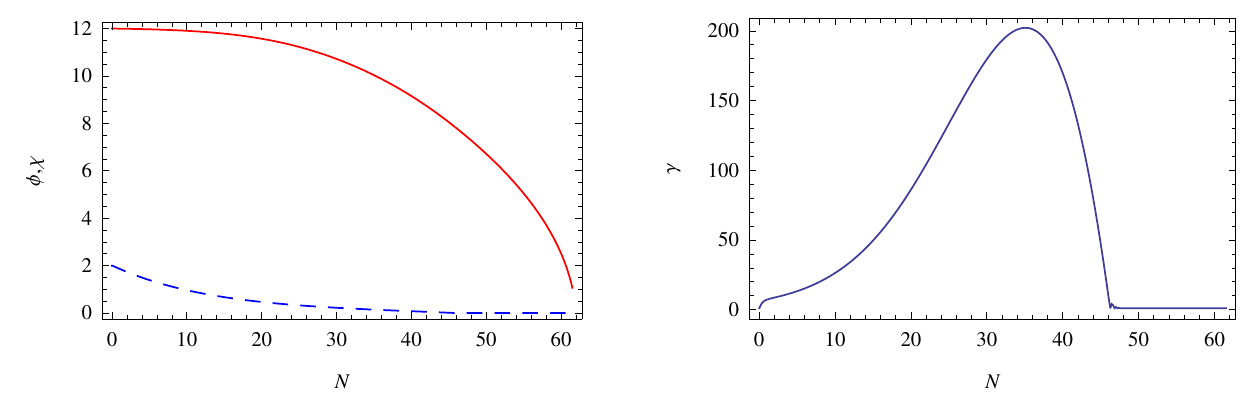}}
\caption{
Background dynamics of the model containing a canonical scalar f\mbox{}ield $ \phi $ and DBI f\mbox{}ield $ \chi $.
In this scenario, the mass of the DBI f\mbox{}ield $ \chi $ is much heavier than the canonical f\mbox{}ield $ \phi $, and therefore decays earlier.
The initial f\mbox{}ield values are the same as those in the previous case: $ \phi_{\mathrm{ini}} = 12.0 $ and $ \chi_{\mathrm{ini}} = 2.0 $.
Masses of the two f\mbox{}ields are now $ \Gamma = 35.1 $ and $ m_{\phi} = 6.50 \times 10^{-7} $.
The parameters required in the DBI model are $ \lambda = 2.0 \times 10^{12} $ and $ \mu = 0.2 $.
\newline
\emph{Left panel}: Field evolution of the canonical scalar f\mbox{}ield $ \phi $ (solid red) and DBI f\mbox{}ield $ \chi $ (blue dashed).
\emph{Right panel}: Evolution of the boost factor $ \gamma $ during inf\mbox{}lation.
}
\label{dbichibggam}
\end{center}
\end{figure}

\begin{figure}
\begin{center}
\scalebox{1.1}{\includegraphics{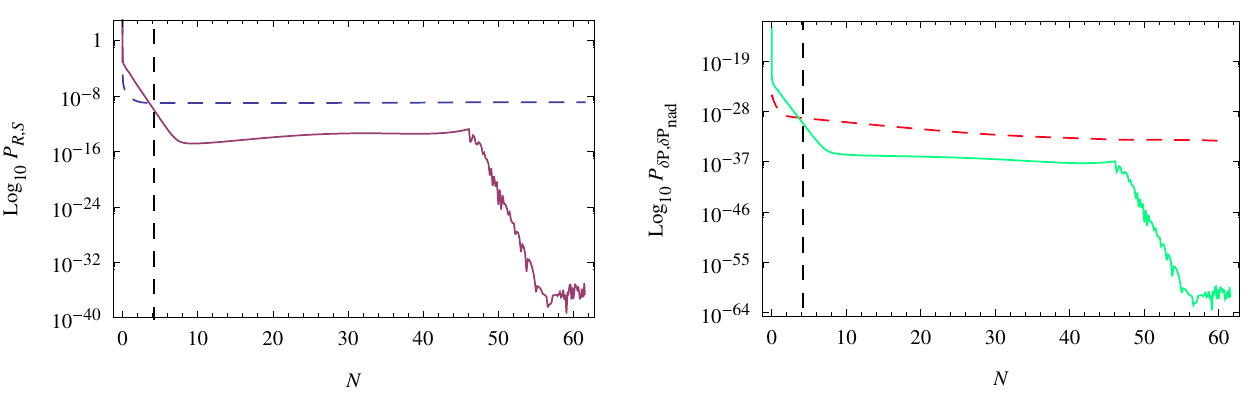}}
\caption{
Comparison of the power spectra evolution versus $e$-fold number for the model containing a canonical scalar f\mbox{}ield $ \phi $ and a DBI f\mbox{}ield $ \chi $.
The DBI f\mbox{}ield $ \chi $ decays earlier during inf\mbox{}lation as it is much heavier than the canonical f\mbox{}ield $ \phi $.
The initial f\mbox{}ield values are $ \phi_{\mathrm{ini}} = 12.0 $ and $ \chi_{\mathrm{ini}} = 2.0 $ and masses of the two f\mbox{}ields are $ \Gamma = 35.1 $ and $ m_{\phi} = 6.50 \times 10^{-7} $.
Parameters needed in the DBI model are $ \lambda = 2.0 \times 10^{12} $ and $ \mu = 0.2 $.
\newline
\emph{Left panel}: Power spectra evolution of the curvature ($ \mathcal{P}_{\mathcal{R}} $, blue dashed) and entropy ($ \mathcal{P}_{\mathcal{S}} $, solid purple) perturbations. 
\emph{Right panel}: Power spectra evolution of the pressure ($ \mathcal{P}_{\delta p} $, red dashed) and nonadiabatic pressure ($ \mathcal{P}_{ \delta p_{\mathrm{nad}} } $, solid green) perturbations. 
In both plots, the dashed black line indicates horizon crossing corresponding to $ k_{\mathrm{WMAP}} = 0.002 \, \mathrm{Mpc}^{-1} $.
}
\label{dbichipow}
\end{center}
\end{figure}

We will now study the model containing a canonical scalar f\mbox{}ield and DBI f\mbox{}ield.
In a similar fashion to the previous model containing the kinetic coupling, this DBI model will also be associated with the same two scenarios: one in which the DBI f\mbox{}ield decays before the scalar f\mbox{}ield and the reverse. 
For all the cases considered, the parameters within the DBI model will hold the following values: $ \lambda = 2.0 \times 10^{12} $ and $ \mu = 0.2 $ \cite{Silverstein:2003hf}.
\newline
\\
First we consider the case where inf\mbox{}lation is initially dominated by the scalar f\mbox{}ield $ \phi $. 
In this scenario, the f\mbox{}ield mass ratio is $ \Gamma = 2.0 $ with the mass of the f\mbox{}ield $ \phi $ as $ m_{\phi} = 1.156 \times 10^{-5} $.
The initial values for the two f\mbox{}ields are $ \chi_{\mathrm{ini}} = 2.0 $ and $ \phi_{\mathrm{ini}} = 12.0 $.
We display the background dynamics and the boost factor $ \gamma $ against the number of $e$-foldings in Figure~\ref{dbiphibggam}.
\begin{figure}
\begin{center}
\scalebox{1.1}{\includegraphics{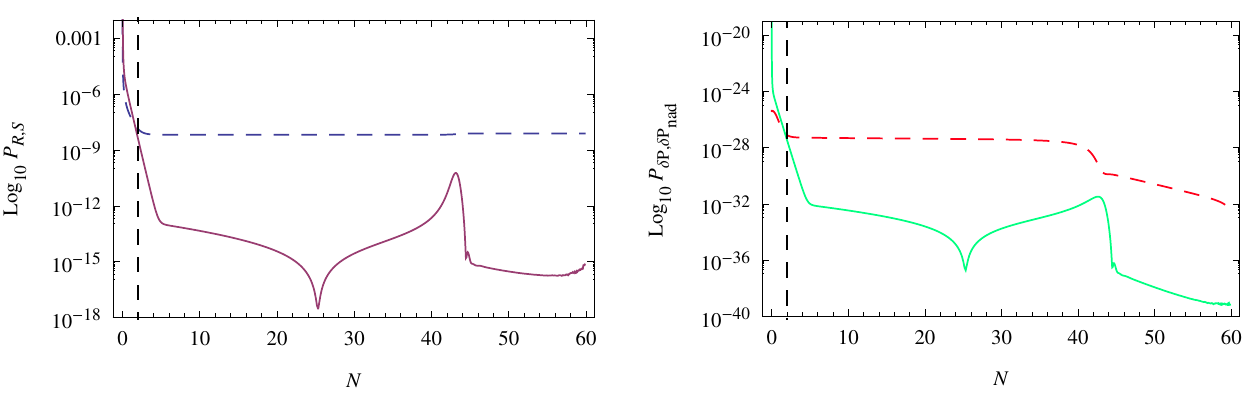}}
\caption{
Comparison of the power spectra evolution against the number of $e$-foldings for the model containing a canonical scalar f\mbox{}ield $ \phi $ and a DBI f\mbox{}ield $ \chi $, where $ \phi $ dominates the inf\mbox{}lationary epoch.
The starting f\mbox{}ield values are $ \phi_{\mathrm{ini}} = 12.0 $ and $ \chi_{\mathrm{ini}} = 2.0 $ and masses of the two f\mbox{}ields are $ \Gamma = 2.0 $ and $ m_{\phi} = 1.156 \times 10^{-5} $.
Parameters needed in the DBI model are $ \lambda = 2.0 \times 10^{12} $ and $ \mu = 0.2 $.
\newline
\emph{Left panel}: Power spectra evolution of the curvature ($ \mathcal{P}_{\mathcal{R}} $, blue dashed) and entropy ($ \mathcal{P}_{\mathcal{S}} $, solid purple) perturbations. 
\emph{Right panel}: Power spectra evolution of the pressure ($ \mathcal{P}_{\delta p} $, red dashed) and nonadiabatic pressure ($ \mathcal{P}_{ \delta p_{\mathrm{nad}} } $, solid green) perturbations. 
In both plots, the dashed black line indicates horizon crossing corresponding to $  k_{\mathrm{WMAP}} = 0.002 \, \mathrm{Mpc}^{-1} $.
}
\label{dbiphipow}
\end{center}
\end{figure}
\newline
\\
In Figure~\ref{dbiphibggam}, we see that the f\mbox{}ield $ \phi $ reaches its potential minimum when $ N = 44 $ leaving the DBI f\mbox{}ield $ \chi $ driving inf\mbox{}lation for the remaining $ e $-foldings.
From this point onwards, the boost factor $ \gamma $ starts to increase reaching a peak value of $ \gamma = 540 $ when $ N = 59 $, 
before it drops slightly to $ \gamma = 490 $ at the end of inf\mbox{}lation.
\newline
\\
In addition to this, the exchange in the f\mbox{}ield contributions at $ N = 44 $ affects the power spectra evolution; this is displayed in Figure~\ref{dbiphipow} through the fall in the power spectrum amplitude of the entropy perturbation. 
We see in this model that the evolution of the entropy and nonadiabatic pressure power spectra are significantly different to the other models previously considered.
\newline
\\
Furthermore, the f\mbox{}inal value of the entropy power spectrum amplitude, $ \mathcal{P}_\mathcal{S} \sim 10^{-16} $ is markedly greater than found in the two canonical f\mbox{}ields model studied in Section~\ref{sec:twostandard}. 
Similarly, this increase in amplitude is also found in the power spectrum evolution of the nonadiabatic pressure perturbation.
For example, in this model,  the nonadiabatic pressure perturbation power spectrum amplitude at the end of inf\mbox{}lation is $ \mathcal{P}_{\delta p_{\mathrm{nad}}} \sim 10^{-40} $ whereas for the two scalar f\mbox{}ield case $ \mathcal{P}_{ \delta p_{\mathrm{nad}}} \sim 10^{-54} $, which is a difference of 14 orders of magnitude.
The power spectrum amplitude of the pressure perturbation in this model is $ \mathcal{P}_{\delta p} \sim 10^{-33} $.
\newline
\\
We will now study the other case where the DBI f\mbox{}ield decays before the scalar f\mbox{}ield.
In the potential, the parameters are chosen as $ \Gamma = 35.1 $ with the mass of the $ \phi$-f\mbox{}ield as $6.50 \times 10^{-7} $. 
The f\mbox{}ields will have the same starting values as in the previously considered DBI f\mbox{}ield case.
The evolution of the canonical and DBI f\mbox{}ields alongside the boost factor are presented in Figure~\ref{dbichibggam}, and the resulting power spectra evolution for the curvature, entropy, pressure and nonadiabatic pressure perturbations are displayed in Figure~\ref{dbichipow}.
\newline
\\
At f\mbox{}irst, the DBI f\mbox{}ield dominates the inf\mbox{}lationary period until it reaches the minimum of the potential well at $ N = 47 $, at which point the f\mbox{}ield $ \phi $ will become dominant.
The DBI f\mbox{}ield will now proceed to oscillate about its minimum.
During these 45 $e$-foldings of inf\mbox{}lation, the boost factor is affected.
Specif\mbox{}ically, the boost factor increases to a maximum value of $ \gamma = 200 $ at $ N = 35 $ before decreasing rapidly to $ \gamma = 1 $ at $ N = 47 $.
This is ref\mbox{}lected in Figure~\ref{dbichipow} through the drop in the power spectrum amplitude of the entropy perturbation 12 $ e $-foldings before the end of inf\mbox{}lation.
\newline
\\
The behaviour of all power spectra is similar to those seen in the two canonical scalar f\mbox{}ields in Section~\ref{sec:twostandard} and kinetic coupling models, where the $ \chi $-f\mbox{}ield is the f\mbox{}irst to reach the potential's minimum in Section~\ref{sec:kinetic}.
This is due to the presence of the remaining f\mbox{}ield being a canonical f\mbox{}ield experiencing slow-roll.
From this, the drop in the entropy power spectrum at $ N = 47 $ is due to the DBI f\mbox{}ield reaching the minimum of its potential. 
The f\mbox{}inal power spectra amplitudes for the entropy and nonadiabatic pressure perturbations are $ \mathcal{P}_{\mathcal{S}} \sim 10^{-37} $ and $ \mathcal{P}_{\delta p_{\mathrm{nad}}} \sim 10^{-62} $. 
For completeness, the pressure perturbation power spectrum amplitude is $ \mathcal{P}_{\delta p} \sim 10^{-34} $.
\newline
\\
The f\mbox{}inal amplitudes for the entropy and nonadiabatic pressure perturbations, $ \mathcal{P}_{\mathcal{S}} $ and $ \mathcal{P}_{\delta p_{\mathrm{nad}}} $ for this scenario are much smaller than in the two scalar f\mbox{}ield case in Section~\ref{sec:twostandard}.
\newline
\\
We f\mbox{}ind in the model containing the kinetic coupling that regardless of which f\mbox{}ield reaches the minimum of its potential f\mbox{}irst, it has no effect on the power spectra generated.
However, in the DBI model considered in this section, we f\mbox{}ind that the ordering of f\mbox{}ield decay has an effect on the power spectra produced, in particular, the power spectra of the nonadiabatic pressure perturbation and ultimately, the entropy perturbation.
We f\mbox{}ind that the difference between nonadiabatic pressure power spectrum amplitude of the two scenarios considered here is incredibly large.
To be precise, in the scenario where the canonical f\mbox{}ield is the f\mbox{}irst to reach the potential minimum, the power spectrum amplitude of the nonadiabatic pressure perturbation is $ \mathcal{P}_{\delta p_{\mathrm{nad}} } \sim 10^{-40} $, and for the alternative scenario, the amplitude is $ \mathcal{P}_{\delta p_{\mathrm{nad}} } \sim 10^{-62} $.
Naturally, the amplitude of the entropy power spectrum is also greatly affected; when the canonical f\mbox{}ield reaches the minimum f\mbox{}irst, the entropy power spectrum amplitude is $ \mathcal{P}_{\mathcal{S}} \sim 10^{-16} $, and vice versa, the amplitude is $ \mathcal{P}_{\mathcal{S}} \sim 10^{-37} $.
There is no difference in the pressure perturbation power spectrum amplitude between the two DBI scenarios.

\section{Conclusion}
\label{sec:chap2conclusion}
We have studied the evolution of the nonadiabatic pressure perturbation $ \delta p_{\mathrm{nad}} $ in the context of multi-f\mbox{}ield inf\mbox{}lation, furthering the work of Huston and Christopherson \cite{Huston:2011fr}.
In this work, we have examined three models where each model is composed of two massive scalar f\mbox{}ields.
However, they all differ in kinetic terms: f\mbox{}irst model consisting on two canonical scalar f\mbox{}ields, second model with one canonical scalar f\mbox{}ield and one f\mbox{}ield with the presence of a kinetic coupling, and f\mbox{}inally, the third model containing a canonical scalar f\mbox{}ield and a DBI f\mbox{}ield.
For all models considered, inf\mbox{}lation is driven by both f\mbox{}ields until one f\mbox{}ield reaches the minimum of its potential, at which point, inf\mbox{}lation is then driven solely by the remaining f\mbox{}ield.
\newline
\\
In the process of this numerical study, we conf\mbox{}irm our results for the model composed of two canonical scalar f\mbox{}ields are in agreement with Huston and Christopherson \cite{Huston:2011fr}.
In addition to this, we found that the nature of the kinetic term affects the evolution and ultimately, the f\mbox{}inal value of the nonadiabatic pressure perturbation.
For example, in the case involving the kinetic coupling, we f\mbox{}ind that the f\mbox{}inal amplitude of $ \mathcal{P}_{\delta p_{\mathrm{nad}}} $ is independent of which f\mbox{}ield primarily decays. 
Yet, an increase in the kinetic coupling generates an increase in $ \mathcal{P}_{\delta p_{\mathrm{nad}}} $. 
In particular, the difference between no and a nonzero, specif\mbox{}ically when $ \beta = 0.1 $, kinetic coupling yields an increase in the f\mbox{}inal amplitude of $ \mathcal{P}_{\delta p_{\mathrm{nad}}} $ of roughly f\mbox{}ive orders of magnitude.
\newline
\\
The background dynamics, specif\mbox{}ically the speed at which the f\mbox{}irst f\mbox{}ield reaches the minimum of its potential, greatly inf\mbox{}luences the evolution of $ \mathcal{P}_{\mathcal{S}} $.
At the point where the f\mbox{}irst f\mbox{}ield reaches its potential minimum, it corresponds to an increase in $ \mathcal{P}_{\mathcal{S}} $, and in turn affects $ \mathcal{P}_{\mathcal{R}} $.
We show that this effect is most apparent when considering the case of two canonical scalar f\mbox{}ields.
For all other models considered here, the decay of the f\mbox{}irst f\mbox{}ield is not as dramatic as that of the canonical case, and hence, the changes in $ \mathcal{P}_{\mathcal{S}} $ are less visible.
\newline
\\
In the case of the canonical scalar and DBI f\mbox{}ields, we see that if the DBI f\mbox{}ield drives the remaining $e$-folds of inf\mbox{}lation, it causes the f\mbox{}inal amplitude of $ \mathcal{P}_{\delta p_{\mathrm{nad}}} $ to be signif\mbox{}icantly larger than other models examined.
On the contrary, if the DBI f\mbox{}ield has reached its potential minimum earlier on during the inf\mbox{}lationary epoch, and therefore is not signif\mbox{}icant for last few $e$-folds of inf\mbox{}lation, we see that $ \mathcal{P}_{\delta p_{\mathrm{nad}}} $ is similar to the canonical case.
Furthermore, we show that the DBI f\mbox{}ield driving the last few inf\mbox{}lationary $e$-foldings produces the largest amplitude of $\mathcal{P}_{\cal S}$.
\newline
\\
The nonadiabatic pressure perturbation can be shown to source vorticity at second order in perturbation theory, and in turn will impact CMB B-mode polarization predictions.
We have shown that a model containing a DBI f\mbox{}ield driving the remaining $e$-folds of inf\mbox{}lation will provide the largest f\mbox{}inal 
amplitude for $ \mathcal{P}_{\delta p_{\mathrm{nad}}} $, and hence supply a larger vorticity source when compared to other models considered here.
Further study into the mechanism of both pre- and reheating in these models is imperative in order to estimate the amount of entropy perturbations in the radiation dominated epoch; from \cite{Brown:2011dn}, it was shown that in this epoch, the nonadiabatic pressure perturbation can be generated.
The results will depend on the model considered due to the particular features within, for instance on couplings between the inf\mbox{}laton f\mbox{}ield(s) and matter f\mbox{}ields.

\chapter{Implications of sharp transitions on the ef\mbox{}fective Planck mass during inf\mbox{}lation}
\chaptermark{Transitions in the effective Planck mass}
Cosmological inflation has been widely accepted as the process which gives rise to the formation of structure in the Universe.
However, there are many inflationary models which satisfy the vast quantity of observational data that have been obtained through experiments.
This makes determining the definitive model of inflation incredibly dif\mbox{}ficult.
One area of study which has been active are scalar-tensor theories.
\newline
\\
Scalar-tensor theories are the most natural extensions to Einstein's Theory of General Relativity and were first studied by Jordan in 1959 \cite{Jordan1959}.
In 1961, Brans and Dicke published a paper suggesting an alternative to Einstein's Theory of General Relativity \cite{Brans:1961sx} 
as a way to describe classical gravity.
Within this theory, there are gravity-matter couplings and from this, two frames arise: the Jordan frame and the Einstein frame. 
The Jordan frame is a frame in which the Einstein-Hilbert action has been modified to include a coupling to a scalar field.
By performing a conformal transformation of the metric \cite{Faraoni:1998qx, Kaiser:2010ps}, the Jordan frame can be related to the Einstein frame.
In the Einstein frame, the Einstein-Hilbert action is as usual with the additional of a scalar field.
The action written in these two frames can be seen in Chapter 1 in Section~\ref{sec:chap1altmodel} entitled ``Alternative models'': Eq.~\eqref{actionjordan} for the Jordan frame and Eq.~\eqref{actioneins} for the Einstein frame.
Scalar-tensor theories provide interesting phenomenological models; they can be used to study the variation of Newton's constant $ G $.
\newline
\\
In 1991, Accetta and Steinhardt stated that if the gravitational constant oscillates at high frequencies when compared to the Hubble expansion rate, it can af\mbox{}fect cosmological measurements \cite{Accetta:1990yb}.
The authors state the only requirement is that the system contains a massive scalar field which is coupled nonminimally to gravity.
Oscillations in the gravitational coupling are obtained through the scalar field oscillating about its ground state value.
The frequency of the scalar field oscillations are related to the mass of the field.
In their 1991 paper, the authors focus on epochs after inflation, i.e. the radiation and matter dominated epochs.
This work was extended in 1994 by Steinhardt and Will who considered the variation of the the gravitational coupling after inflation \cite{Steinhardt:1994vs}.
The authors constructed a model with oscillations in Newton's constant by using a Brans-Dicke model with a nonminimally coupled massive scalar field $ \psi $, which has the ability to be displaced from the minimum of its ef\mbox{}fective potential during inflation.
Inflation in this model is not driven by the field $ \psi $, but instead by an inflaton field within the matter sector.
During the inflationary epoch, the coupling between the gravity and scalar field sectors generates a finite displacement of the field $ \psi $ from its potential minimum, creating a potential minimum which has a false vacuum energy density.
Once inflation has ended, reheating starts, which causes the inflaton energy density to be converted into radiation.
Due to this, the potential minimum now occurs at $ \psi = 0 $ resulting in the field having two options: to oscillate or to slowly roll.
The choice is determined whether the period of oscillation is greater than the expansion time scale.
For the case when the period of oscillation is greater than the expansion time scale, the field begins by slow-rolling down the potential eventually leading to oscillations. For the reverse case, the field immediately starts with oscillatory behaviour.
\newline
\\
It is known that the gravitational constant could not have varied too much during the Universe's lifetime.
In fact, the gravitational constant could have only varied by \cite{Uzan:2002vq, Uzan:2010pm}
\be
\frac{\dot{G_N}}{G_N} < 10^{-12} \, \mathrm{yr}^{-1}~.
\ee
There are many experiments that measure the time variation of the gravitational constant: lunar ranging observations \cite{Williams:2004qba, Muller:2007zzb}, Big Bang Nucleosynthesis (BBN) \cite{Santiago:1997mu, Copi:2003xd, Bambi:2005fi, Coc:2006rt}, gravitational waves \cite{Yunes:2009bv}, and more recently WiggleZ \cite{Nesseris:2011pc}.
Other experiments that constraint the varying Newton's constant involve stellar objects; some examples include comparing the ages of individual Globular Clusters to the age of the Universe \cite{Degl'Innocenti:1995nt}, observations of Type Ia supernovae \cite{GarciaBerro:2005yw, Nesseris:2006jc}, pulsating white dwarfs \cite{GarciaBerro:2011wc, Corsico:2013ida}, pulsars \cite{BisnovatyiKogan:2005ap, Jofre:2006ug, Verbiest:2008gy,  Lazaridis:2009kq} and neutron star surface temperatures \cite{Krastev:2007en}.
\newline
As we will see in this chapter, varying the gravitational coupling can af\mbox{}fect the scalar-to-tensor ratio.
In the first quarter of 2014, the BICEP2 team reported a detection of B-mode polarization of the CMB; this was published later in 2014 \cite{Ade:2014xna}.
When this detection is interpreted as gravitational waves produced during the inflationary epoch, it yields a tensor-to-scalar ratio $ r $ of 
\be
r = 0.20^{+0.07}_{-0.05}~,
\ee
at around $ \ell \simeq  80 $ which corresponds to a scale of $ k_{\mathrm{BICEP2}} \simeq 0.005 \, \mathrm{Mpc}^{-1} $.
However, this is in contention with the Planck results from 2013 \cite{Ade:2013zuv} stating an upper limit of $ r < 0.11 $ (95\% C.L.) at the scale $ k_{\mathrm{Planck}} \simeq 0.002 \, \mathrm{Mpc}^{-1} $ i.e. $ \ell \simeq 28 $.
These values suggest that the scale of inflation is at the GUT scale and furthermore, a tensor-to-scalar ratio that is $ r > 0.11 $ will rule out a significant number of inflationary models such as those where the field displacement is less than the Planck mass $ \Delta \phi \ll M_{\mathrm{Pl}} $ \cite{Lyth:1996im, Antusch:2014cpa}; models where $ r > 0.1 $ is obtainable include those inspired by string theory: assisted inflation \cite{Liddle:1998jc, Malik:1998gy, Copeland:1999cs, Kanti:1999vt}, N-flation \cite{Dimopoulos:2005ac, Kim:2006ys}, a model similar to the latter \cite{Ashoorioon:2009wa, Ashoorioon:2009sr, Postma:2010wd} and monodromy \cite{Silverstein:2008sg, McAllister:2008hb, McAllister:2014mpa}.
\newline
\\
Since the announcement of the BICEP2 results, many authors have proposed solutions to resolve the discrepancies between these results and from Planck on both theoretical \cite{Ashoorioon:2014nta, Mukohyama:2014gba, Xu:2014laa, Kawasaki:2014fwa, McDonald:2014kia, Wang:2014kqa, Bastero-Gil:2014oga, Cai:2014hja} and experimental grounds \cite{Liu:2014mpa, Mortonson:2014bja, Flauger:2014qra}.
With the upcoming Planck 2014 results and those from experiments scheduled for years to come, they will continue to test existing models of inflation; other searches for features in the power spectra include \cite{Covi:2006ci, Hamann:2007pa, Meerburg:2014kna, Meerburg:2013dla, Martin:2003sg, Aslanyan:2014mqa}.
\newline
\\
In this chapter, we consider an action of scalar-tensor form, specifically studying the ef\mbox{}fects of a sudden transition in Newton's gravitational constant during inflation and its consequences on the scalar curvature and tensor perturbations.
The transition made is smooth step in nature and not violent, and so the variations of $ M_{\mathrm{Pl}} $ are not of order one, but rather of order a percent or less.
Two scenarios will be considered in this chapter: one in which the Brans-Dicke field will play the role of the inflaton, and a two-field model where this role will be assigned to the second auxiliary field.
We begin by presenting the action, background and perturbation equations for the single-field case in Section~\ref{sec:chap3sec1} and the model in Section~\ref{sec:chap3sec2}, which contains three benchmark models in Section~\ref{sec:chap3singlefield}.
The addition of a minimally-coupled field to the action is considered in Section~\ref{sec:chap3twofield}.
In this section, the relevant equations are presented alongside one case.
We conclude with Section~\ref{sec:chap3sec4} containing a summary of our findings.

\section{Field equations}
\label{sec:chap3sec1}
\subsection{The background}
The action that we consider is of the form
\be
\label{action}
S^{(\varphi)} =  \int \mathrm{d}^{4} x \sqrt{-g} \bigg[ \frac{M_{\mathrm{Pl}}^2}{2} F(\varphi) R 
						- \frac{1}{2} g^{\mu \nu} \partial_{\mu} \varphi \, \partial_{\nu} \varphi - U(\varphi) \bigg]~,
\ee
where $ M_{\mathrm{Pl}} $ is the bare reduced Planck mass with $ \mathrm{M_{Pl}}^2 = (8 \pi G_{\ast})^{-1} $, and contained is the bare gravitational constant $ G_{\ast} $, $ R $ is the Ricci scalar and $ U(\varphi) $ is the potential of the scalar field $ \varphi $.
The gravitational sector is coupled to the scalar field via the ef\mbox{}fective Planck mass $ F(\varphi) $, i.e. $ M_{\mathrm{Pl}}^2 \times F(\varphi) $.
From now on, the bare reduced Planck mass is set to one $ M_{\mathrm{Pl}} = 1 $.
\newline
\\
All calculations will be performed in this frame known as the Jordan frame; this will allow us to model the variations in the Planck mass with ease.
Physical quantities such as the power spectra are the same in both the Jordan frame and the Einstein frame (see \cite{Kaiser:1994vs, Kaiser:1995nv, Prokopec:2013zya} and the references therein).
\newline
\\
By varying the action in Eq.~\eqref{action} with respect to the metric $ g_{\mu \nu} $ yields the Einstein equations
\begin{align}
\label{einsteinstab}
G_{\mu \nu} & = \frac{1}{F(\varphi)} \bigg[ \partial_{\mu} \varphi \, \partial_{\nu} \varphi
								- \frac{1}{2} g_{\mu \nu} g^{\alpha \beta} \partial_{\alpha} \varphi \, \partial_{\beta} \varphi		\nonumber 	\\
					& \qquad \qquad + \nabla_{\mu} \nabla_{\nu}F(\varphi) 
								- g_{\mu \nu} \Box F(\varphi) 
								- g_{\mu \nu} U(\varphi) \bigg]~,
\end{align}
where $ G_{\mu \nu} $ is the Einstein tensor and $ \Box = g^{\mu \nu} \nabla_\mu \nabla_\nu $ is the covariant d'Alembertian operator.
\newline
\\
The equation of motion of a scalar field is obtained by varying the action in Eq.~\eqref{action} with respect to such field; for the field $ \varphi $, it is as follows
\be
\label{chap3eom}
\Box \varphi = U_{,\varphi} (\varphi) - \frac{1}{2} F_{,\varphi} (\varphi) R~,
\ee
with $ F_{,\varphi} = \mathrm{d}F/\mathrm{d}\varphi $, noting that $ _{,\varphi} $ denotes partial derivatives with respect to the field $ \varphi $.
This background equation can be rewritten in a dif\mbox{}ferent form known as the Brans-Dicke equation.
We first require the Ricci scalar $ R $.
By taking the trace of the Einstein field equations in Eq.~\eqref{einsteinstab} and using the definition of the Einstein tensor as stated in Eq.~\eqref{einsteintendef}, the Ricci scalar is
\be
R = \frac{1}{F(\varphi)} \bigg[ g^{\mu \nu} \, \partial_{\mu} \varphi \, \partial_{\nu} \varphi + 3 \, \Box F(\varphi) + 4 \, U(\varphi) \bigg]~,
\ee
with
\be
\Box F(\varphi) = F_{,\varphi} (\varphi) \Box \varphi + F_{,\varphi \varphi} (\varphi) g^{\mu \nu} \partial_{\mu} \varphi \, \partial_{\nu} \varphi~.
\ee
Substituting this for $ R $ in Eq.~\eqref{chap3eom} yields the Brans-Dicke equation
\be
\label{Brans-Dicke}
2 \varpi \Box \varphi = - \varpi_{,\varphi} g^{\mu \nu} \partial_{\mu} \varphi \, \partial_{\nu} \varphi 
					- 4 F_{,\varphi} (\varphi) U(\varphi) 
					+ 2F(\varphi) U_{,\varphi}(\varphi)~,
\ee
where $ \varpi = F(\varphi) + \frac{3}{2} F_{,\varphi} (\varphi)^2 $. 
\newline
\\
We assume that the Universe is both homogeneous and isotropic on large scales, and therefore we use the FRW line element as stated in Eq.~\eqref{frw} in Section~\ref{sec:chap1prelim}.
Using this, the Brans-Dicke equation in Eq.~\eqref{Brans-Dicke} is
\be
\ddot{\varphi} + 3H \dot{\varphi} = \frac{1}{2 \varpi} [- \varpi_{,\varphi} \dot{\varphi}^2 + 4 F_{,\varphi} U - 2 F U_{,\varphi} ]~.
\ee
where dots represent derivatives with respect to cosmic time.
Furthermore, due to the form of the Einstein field equations, the standard Friedmann and acceleration equations are obeyed --- see Eqs.~\eqref{friedmannacceleration} in Section~\ref{sec:chap1prelim} --- with the ef\mbox{}fective energy density and ef\mbox{}fective pressure given by
\begin{subequations}
\begin{align}
\rho &= \frac{1}{F} \left[ \, \frac{1}{2} \, \dot{\varphi}^2 + U - 3H \dot{F} \, \right]~,		\\
p &= \frac{1}{F} \left[ \, \frac{1}{2} \, \dot{\varphi}^2 - U + \ddot{F} + 2H \dot{F} \, \right]~.
\end{align}
\end{subequations}
It is important to note that these are ef\mbox{}fective quantities and that the corresponding energy-momentum tensor $ T^{(\varphi)}_{\mu \nu} $ behaves as a perfect fluid and hence obeys $ \nabla_{\mu} T^{\mu \nu} = 0 $, and the leading diagonal is $ T \indices{^{\mu}_{\nu}} = \mathrm{diag}( -\rho, p, p, p) $, where $ \rho $ and $ p $ denotes the energy density and pressure of the perfect fluid.
\newline
\\
In order to extend and test the generalities of this single-field theory, we later consider a two-field model.
The two-field model will contain a second minimally-coupled field $ \chi $ in conjunction to the single-field case.
The action of the auxiliary field is
\be
\label{actionchi}
S^{(\chi)} = - \frac{1}{2} \int \mathrm{d}^{4} x \sqrt{-g} \bigg{[} g^{\mu \nu} \partial_{\mu} \chi \, \partial_{\nu} \chi + 2 V(\chi) \bigg{]}~,
\ee
with $ V(\chi) $ denoting its potential, which is of the quadratic form:
\be
V(\chi) = \frac{1}{2} m_{\chi}^2 \chi^2~.
\ee
We shall first proceed with the single-field model and later in the chapter, specifically in Section~\ref{sec:chap3twofield}, introduce the second scalar field into the model.

\subsection{Perturbations}
We will now focus our attention to the first-order perturbation equations, which will be studied in the Newtonian gauge as presented in Section~\ref{sec:chap1scalmodes}.
For convenience, we now remind the reader of the line element in the Newtonian gauge as first stated in Chapter 1
\be
\mathrm{d} s^2 =  -( 1 + 2 \Psi ) \mathrm{d} t^2 + a(t)^2(1 - 2 \Phi) \delta_{ij} \mathrm{d}x^i \mathrm{d}x^j~,
\ee
where $ \Psi $ and $ \Phi $ are the scalar metric perturbations.
In addition, we will use the field decomposition, where we split the field into its background and perturbed parts, and work with Fourier modes.
\newline
\\
By perturbing the metric, the perturbation equations of the two fields are
\begin{align}
\delta \ddot{\varphi} &+ \bigg[ 3H + \frac{\varpi_{,\varphi}}{\varpi} \dot{\varphi} \bigg] \delta \dot{\varphi}		\nonumber	\\
						&+ \bigg[ \frac{1}{2} \bigg( \frac{\varpi_{,\varphi}}{\varpi} \bigg)_{\!\!,\varphi}\! \dot{\varphi}^2
								- \frac{1}{2} \bigg( \frac{1}{\varpi} ( 4F_{,\varphi}U 
								- 2FU_{,\varphi}  ) \bigg)_{\!\!,\varphi} + \frac{k^2}{a^2} \bigg] \delta \varphi		\nonumber \\
				 &- ( \dot{\Psi} + 3 \dot{\Phi} ) \dot{\varphi}
						+ \frac{1}{\varpi} \bigg[ -4F_{,\varphi} U  + 2F U_{,\varphi}  \bigg] \Psi = 0~,
\end{align}
In the Newtonian gauge, the perturbed Einstein equations are given by the following
\begin{subequations}
\begin{align}
\label{Einseq}
3H( \dot{\Phi} + H \Psi) + \frac{k^2}{a^2} \Phi &= - \frac{1}{2} \delta \rho~,		\\
\dot{\Phi} + H \Psi &= - \frac{1}{2} \delta q~,		\\
\ddot{\Phi} + (2 \dot{H} + 3H^2 ) \Psi + H( \dot{\Psi} + 3 \dot{\Phi} ) &= \frac{1}{2} \delta p~.
\end{align}
\end{subequations}
The right-hand side of these equations are related to perturbations of the ef\mbox{}fective energy-momentum tensor $ T_{\mu \nu} $, and 
are equal to perturbations of the energy density, momentum potential and pressure, respectively
\begin{subequations}
\begin{align}
\delta \rho &= \frac{1}{F} \bigg[ \dot{\varphi} \delta \dot{\varphi} - \dot{\varphi}^2 \Psi
							+  U_{,\varphi} \delta \varphi   	
							+ 3 \dot{F} ( \dot{\Phi} + 2H \Psi )
							- 3H ( \delta \dot{F} + H \delta F ) 		\nonumber	\\
							& \qquad - \frac{k^2}{a^2} \delta F \bigg]~,	\\
\delta q &= - \frac{1}{F} \bigg[ \dot{\varphi} \delta \varphi + \delta \dot{F} - \dot{F} \Psi - H \delta F \bigg]~,		\\
\delta p &= \frac{1}{F} \bigg[ \dot{\varphi} \delta \dot{\varphi} - \dot{\varphi}^2 \Psi 
							- U_{,\varphi} \delta \varphi   		
							- p \, \delta F + \delta \ddot{F} + 2H \delta \dot{F} - \dot{F} \dot{\Psi} - 2\dot{F} \dot{\Phi}		\nonumber	\\
							& \qquad - 2( \ddot{F} + 2H \dot{F} ) \Psi + \frac{k^2}{a^2} \delta F \bigg]~.
\end{align}
\end{subequations}
By using the $ij$-component of the Einstein equations, we find that anisotropic stress is present in the Jordan frame
\be
\Phi - \Psi = \frac{ \delta F }{F} = \frac{F_{,\varphi} \delta \varphi}{F}~.
\ee
Observables are required to relate theories to experiments and examples of such observables include the spectral index $ n_\mathrm{s} $ and its running $ \alpha $.
For convenience we will now recall parts of Section~\ref{sec:chap1scalmodes} entitled ``Scalar modes''.
Both the spectral and running indices are related to the scalar perturbations through the power law, which we recall from Eq.~\eqref{scalarpowspec}
\be
\label{spectralindex}
\mathcal{P}_{\zeta} (k) = \mathcal{P}_{\zeta} (k_0) \bigg( \frac{k}{k_0} \bigg)^{n_\mathrm{s}(k_0) \,- \, 1 + \frac{1}{2} \ln ( k/k_0 ) \, \alpha }~,
\ee
where $ k_0 $ is a pivot scale and $ \mathcal{P}_{\zeta} (k_0) $ is the amplitude of the scalar power spectrum at such pivot point. 
\newline
\\
The scalar perturbations are provided by the curvature perturbation on constant hypersurfaces $ \zeta $ as defined in Eq.~\eqref{zeta}
\be
\zeta = - \, \Phi - \frac{H}{\dot{\rho}} \delta \rho~,
\ee
and the resulting power spectra generated is 
\be
\mathcal{P}_{\zeta} = \frac{k^3}{2 \pi^2} | \zeta |^2~.
\ee
The pivot scale chosen in this work is that used by the Planck experiment which is $ k_0 = 0.002 \, \mathrm{Mpc}^{-1} $, and the amplitude of the scalar curvature power spectrum at this pivot scale is $ \mathcal{P}_{\zeta} (k_0) \sim 2.15 \times 10^{-9} $ \cite{Ade:2013uln}.
Furthermore, the running index $ \alpha $ is defined as in Eq.~\eqref{runninindex}
\be
\alpha = \frac{\mathrm{d} \, n_\mathrm{s}}{\mathrm{d} \, \ln{k}}~.
\ee
From the Planck experiment \cite{Ade:2013uln}, the current best fit values for both the spectral index and its running are 
\be
n_\mathrm{s} = 0.9603 \pm 0.0073~, \quad \alpha = -\,0.0134 \pm 0.0090~.
\ee
In addition to considering scalar perturbations, we shall also study tensor perturbations.
\newline
\\
We will briefly recapitulate material on tensor perturbations from Section~\ref{sec:chap1tensor} for practical purposes.
The equation of motion for tensor perturbations is given in Eq.~\eqref{tensoreqmotion}
\be
h'' + 2 \frac{a'}{a} h' + k^2 h = 0~,
\ee
with the tensor power spectra given by
\be
\mathcal{P}_{\mathrm{T}} = \frac{k^3}{2 \pi^2} | h |^2~.
\ee
Both these equations will have the standard oscillatory initial conditions \cite{Adams:2001vc}.
\newline
\\
The tensor-to-scalar ratio $ r $ is used to relate the tensor and scalar power spectra.
It is defined as follows \cite{Uzan}
\be
r = \frac{ 8 \, \mathcal{P}_{\mathrm{T}} }{ \mathcal{P}_{\zeta} }~.
\ee

\section{Model with step variation in the Planck mass}
\label{sec:chap3sec2}
We will now examine the smooth step transition in the ef\mbox{}fective Planck mass $ F(\varphi) $; the ef\mbox{}fective Planck mass and the potential considered are
\begin{align}
\label{coupling}
F(\varphi) &= 1.0 - \beta \, \{\,  1.0 + \tanh[ (\, \varphi - \varphi_\ast )/ \gamma ] \}~,	\\
\label{potential}
U(\varphi) &= \frac{1}{2} m_{\varphi}^2 \varphi^2~,
\end{align}
where $ m_{\varphi} $ is the mass of the scalar f\mbox{}ield, $ \beta $ is a dimensionless constant, and both $ \gamma $ and $ \varphi_\ast $ are constants of mass dimension.
The parameters $ \beta $ and $ \gamma $ determines the sharpness and amplitude of the transition and the parameter $ \varphi_{\ast} $ determines the f\mbox{}ield value at which the transition occurs. 
We have chosen the quadratic potential for concreteness, but we expect the features observed due to the sharp transition to also appear in other choices of potential.
\newline
\\
In the next section, we will f\mbox{}irstly consider a single-f\mbox{}ield model, in which the Brans-Dicke f\mbox{}ield drives inflation.
Following this, we will then consider a two-f\mbox{}ield model where we introduce a second minimally-coupled scalar f\mbox{}ield which will act as the inflaton.
For the two cases and their benchmark models studied here, the values of the parameters are chosen so that successful inflation is obtained with the inflationary period lasting 66 $ e $-folds.
\newline
\\
Before proceeding, we will now show that the Jordan frame model that we are studying is not equivalent to the Einstein frame model of an inflaton potential with step as described in \cite{Adams:2001vc}.
Through the conformal transformation as outlined in Section~\ref{sec:chap1altmodel}, we can use this to transform the model considered in Eqs.~\eqref{coupling} and \eqref{potential} to the Einstein frame; the potential in the Einstein frame is
\be
\label{modpotential}
\tilde{ U }(\varphi) = \frac{ U(\varphi) }{ F^2(\varphi) } = \frac{ m_{\varphi}^2 \varphi^2 }{2 \left\{ 1.0 - \beta \, \{\,  1.0 + \tanh[ (\, \varphi - \varphi_\ast )/ \gamma ] \} \right\}^2 }~.
\ee
It is important to note that the potential above and that used in \cite{Adams:2001vc} are the same when $ \beta \ll 1 $.
However, there will be dif\mbox{}ferent dynamics due to the dif\mbox{}ferences in the kinetic terms.
When the model considered in this chapter is transformed to the Einstein frame, it will have noncanonical kinetic terms unlike that studied in \cite{Adams:2001vc}.
The canonical f\mbox{}ield $ \tilde{\varphi} $ is related to the noncanonical f\mbox{}ield $ \varphi $ via the relation --- see Eq.~\eqref{transfield}
\be
\tilde{\varphi} = \int \sqrt{\frac{ 2 F(\varphi)^2 + 3 F_{,\varphi} (\varphi)^2 }{ 2 F(\varphi)^2 }} \mathrm{d} \varphi
\ee
By rewriting the equation above, the f\mbox{}ield $ \varphi $ can be expressed in terms of $ \tilde{\varphi} $ at least implicitly.
The potential $ \tilde{U}(\varphi) $ can now be rewritten in terms of the canonical f\mbox{}ield $ \varphi $.
However, the form of the potential will not longer be a step potential as that stated in Eq.~\eqref{modpotential}.
As a result, the Einstein frame model containing a step potential is not the equivalent to a Jordan frame model containing a step transition in the ef\mbox{}fective Planck mass; the model considered in this chapter is dif\mbox{}ferent to those previously studied in the literature, and hence the results from this study will provide signif\mbox{}icantly dif\mbox{}ferent predictions for the scalar and tensor power spectra.
\newline
\\
The dynamics of the f\mbox{}ields will be numerically solved, following the method outlined in \cite{Tsujikawa:2002qx}; the derivative of the background f\mbox{}ields are given their slow-roll values, and the initial f\mbox{}ield perturbations will have the standard oscillatory Bunch-Davies initial conditions \cite{Adams:2001vc}. 
In order to calculate the tensor perturbations generated by the system, we employ the methods described in \cite{Adams:2001vc, Habib:2005mh}.
\newline
\\
For the ef\mbox{}fects of this type of transition to be observable, they must occur between 60 and 50 $ e $-folds before the end of inflation \cite{Liddle:2003as, Turner:1993xz}.
This will translate to a range of scales observable using experiments such as Planck.
These step changes in the Planck mass could result from a f\mbox{}irst-order phase transition in the vacuum expectation value (VEV) of a Brans-Dicke field \cite{Abolhasani:2012tz}.
A similar step transition in the inflaton's potential was considered in \cite{Adams:2001vc} (see \cite{Joy:2007na} for a sharp transition in the ef\mbox{}fective Planck mass) and was shown to result in oscillation in the primordial scalar curvature power spectrum.
However, no oscillations in the scalar power spectrum were seen when considering a nonminimally coupled Brans-Dicke field with a transition in the potential.

\subsection{Single-field model}
\label{sec:chap3singlefield}
\subsubsection{Minimally-coupled limit}
\begin{figure}[h]
\begin{center}
\scalebox{0.75}{\includegraphics{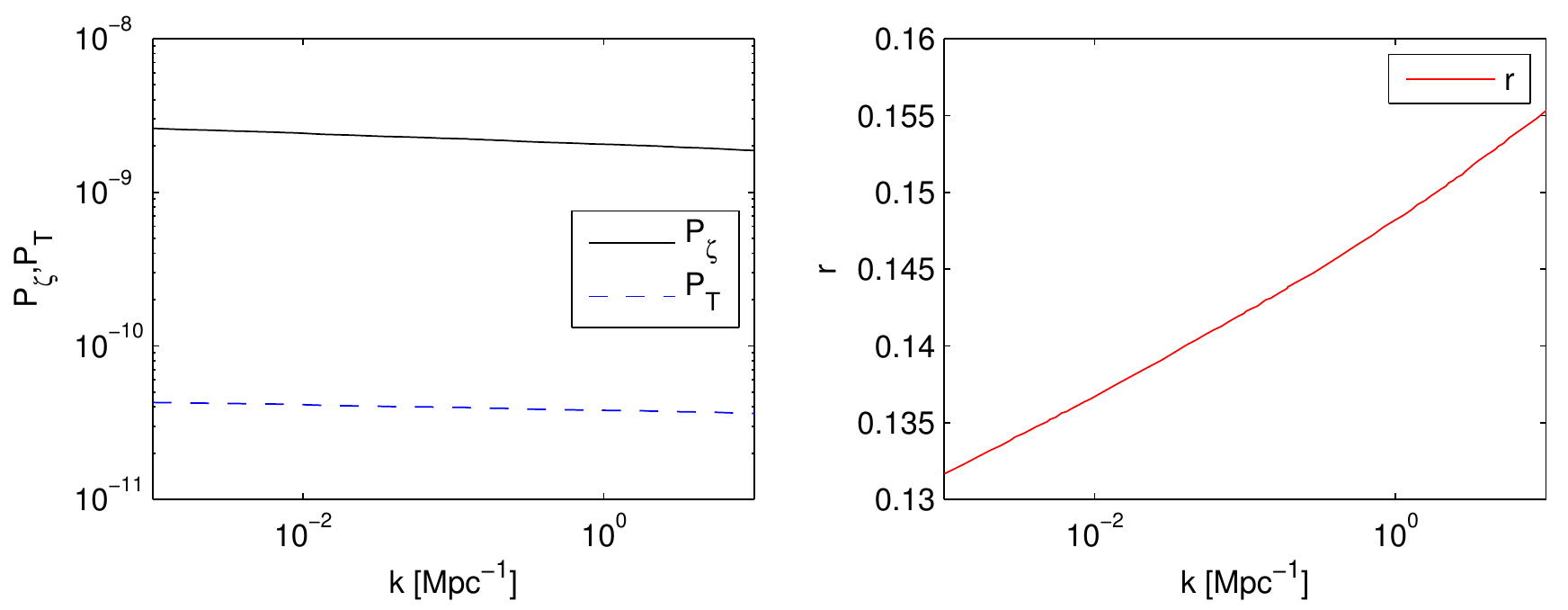}}
\caption{
The power spectra (\emph{left panel}) of the scalar ($ \mathcal{P}_{\zeta} $, solid black) and tensor ($ \mathcal{P}_{\mathrm{T}} $, blue dashed) perturbations and the associated tensor-to-scalar ratio (\emph{right panel}) against wavenumber for the minimally-coupled model, i.e. $ \beta = 0 $.
The parameters used are the initial field value, $ \varphi_{\mathrm{ini}} = 16.179 $ and field mass $ m_{\varphi} = 6.5 \times 10^{-6} $.
}
\label{standpowstensca}
\end{center}
\end{figure}

We shall begin with the minimally-coupled case; in this case, the parameter $ \beta $ is zero and hence $ F = 1 $.
The starting value for the $ \varphi $ field is $ \varphi_{\mathrm{ini}} = 16.179 $ with mass $ m_{\varphi} = 6.5 \times 10^{-6} $.
The mass is chosen so that the power spectrum for the scalar perturbations at the Planck pivot scale is approximately $ 2.15 \times 10^{-9} $.
The power spectra for the scalar and tensor perturbations are given in Figure~\ref{standpowstensca}.
Notice that we have defined the number of $e$-folds N such that N = 0 at the start of inflation.
With the coupling between the gravitational sector and scalar field set to one, there are no features generated in this model, as expected. 
For this case, the spectral and running indices are calculated to be
\be
n_{\mathrm{s}} = 0.968865~, \quad \alpha = 0.00107427~.
\ee
In addition to this, the tensor-to-scalar ratio at the Planck pivot scale is
\be
r(k_0) = 0.133205~.
\ee
For this case, a quadratic potential was chosen for simplicity.
The model is slightly under pressure from the Planck experiment \cite{Ade:2013uln} (cf. \cite{Ashoorioon:2013eia} as an attempt to reconcile the model with the Planck data), although it is still within the 68\% C.L. in the $ n_\mathrm{s} - r $ plane. 

\subsubsection{Benchmark 1}
\begin{figure}[h]
\begin{center}
\scalebox{0.75}{\includegraphics{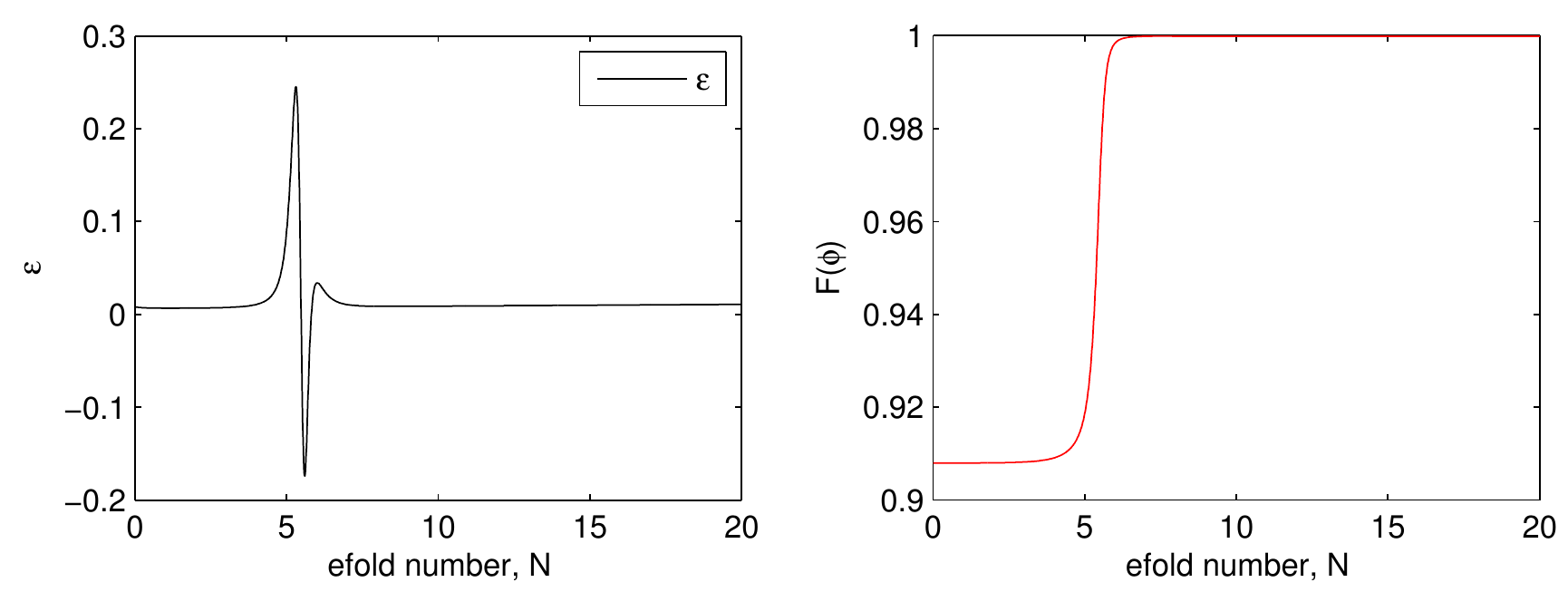}}
\caption{
The evolution of the slow-roll parameter $ \varepsilon $ (\emph{left panel}) and the nonminimal coupling $ F(\varphi) $ (\emph{right panel}) for the first 20 $e$-folds of inflation for benchmark 1.
The initial field value is $ \varphi_{\mathrm{ini}} = 16.5783 $ and mass is $ m_\varphi = 2.1\times 10^{-5} $.
Model parameters required in the nonminimal coupling $ F(\varphi) $ are $ \beta = 0.046 $, $ \gamma = 0.145 $ and $ \varphi_\ast = 15.8 $.
}
\label{violentbg}
\end{center}
\end{figure}

\begin{figure}
\begin{center}
\scalebox{0.75}{\includegraphics{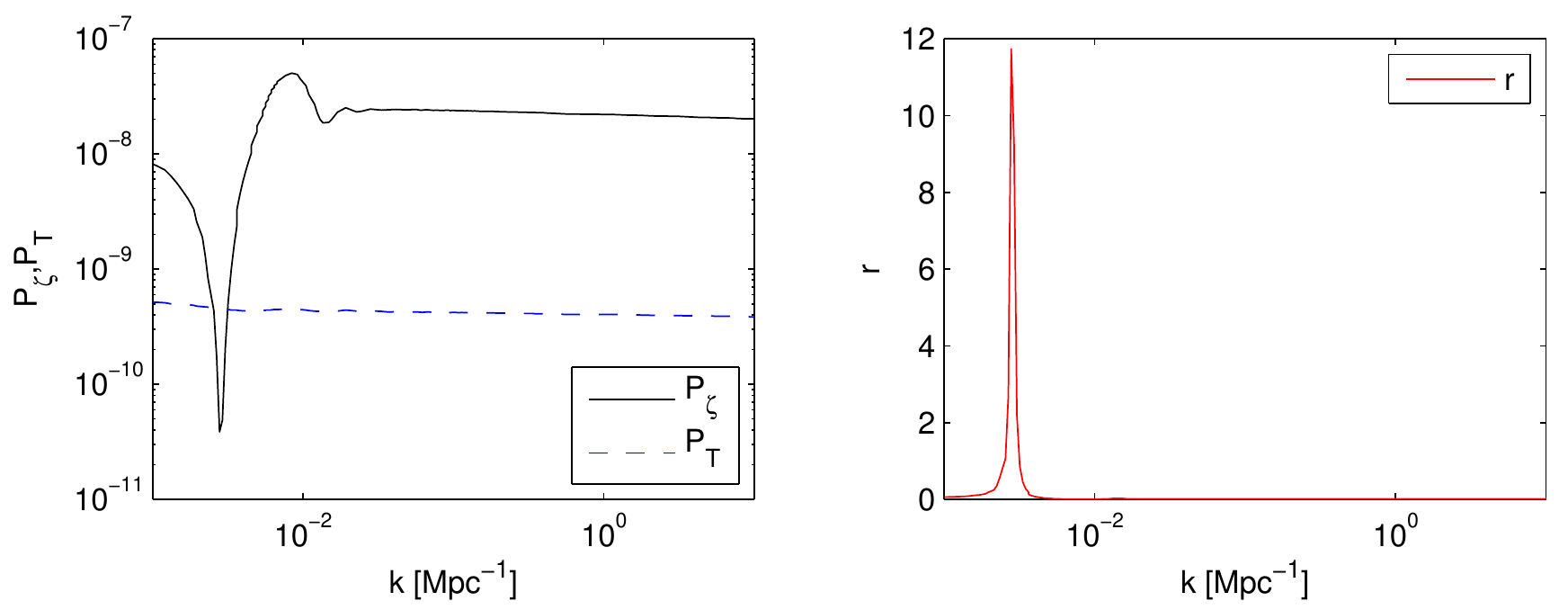}}
\caption{
The power spectra (\emph{left panel}) for the scalar ($ \mathcal{P}_{\zeta} $, solid black) and tensor ($ \mathcal{P}_{\mathrm{T}} $, blue dashed) perturbations and the associated tensor-to-scalar ratio $ r $ (\emph{right panel}) against wavenumber, for benchmark 1.
The initial field value is $ \varphi_{\mathrm{ini}} = 16.5783 $ and field mass is $ m_\varphi = 2.1 \times 10^{-5} $.
Parameters used in the nonminimal coupling $ F( \varphi ) $ are $ \beta = 0.046 $, $ \gamma = 0.145 $ and $ \varphi_\ast = 15.8 $.
}
\label{violentpowstensca}
\end{center}
\end{figure}

\begin{figure}[h]
\begin{center}
\scalebox{0.5}{\includegraphics{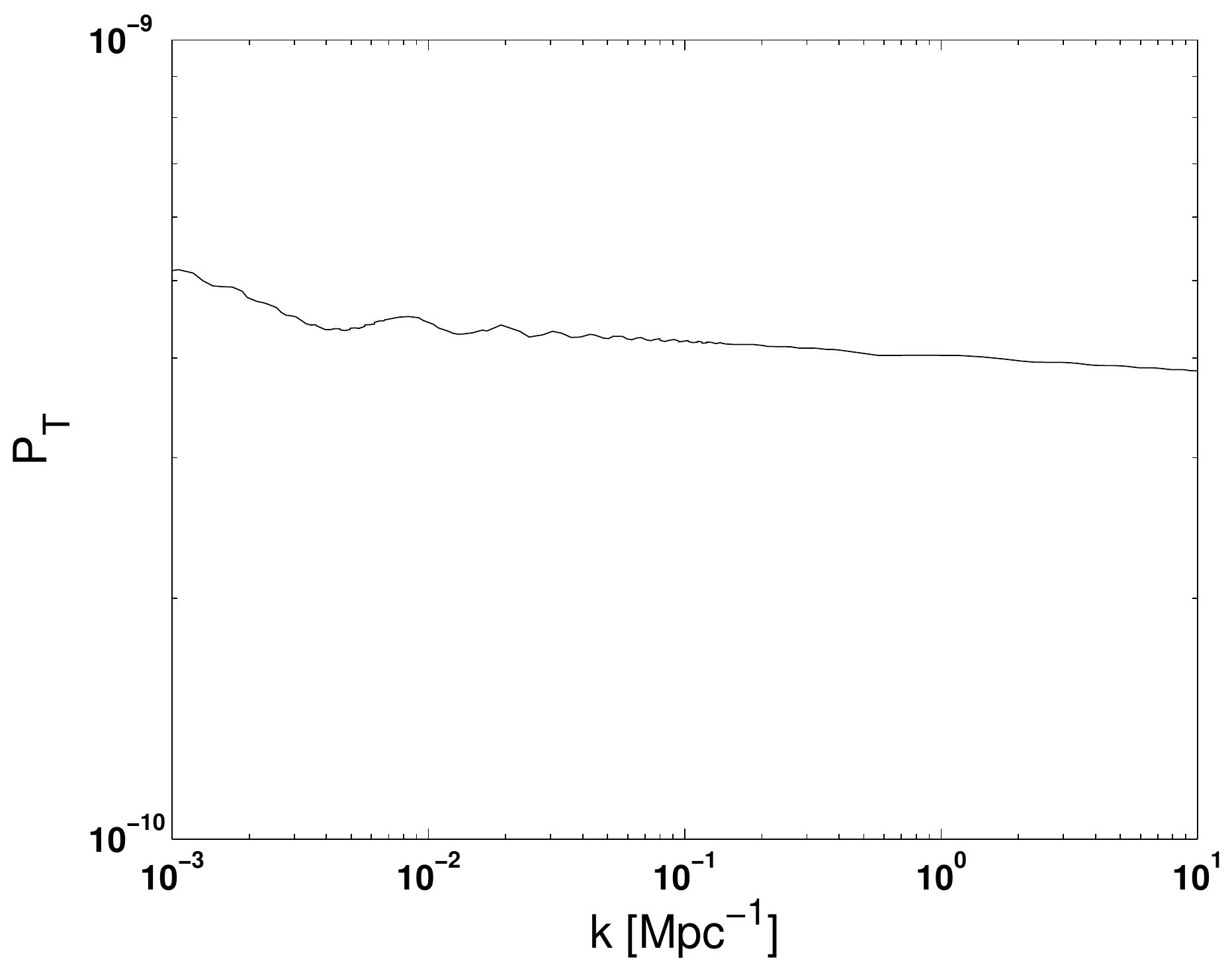}}
\caption{
The tensor power spectrum $ \mathcal{P}_{\mathrm{T}} $ of benchmark 1.
The starting field value is $ \varphi_{\mathrm{ini}} = 16.5783 $ and field mass is $ m_\varphi = 2.1 \times 10^{-5} $ and in the coupling $ F(\varphi) $ are $ \beta = 0.046 $, $ \gamma = 0.145 $ and $ \varphi_\ast = 15.8 $.}
\label{violentpowtensor}
\end{center}
\end{figure}

In the first of three benchmark models, we will consider a steep transition in the ef\mbox{}fective Planck mass, which is reflected in a violent feature in the slow-roll parameter $ \varepsilon $; this evolution in the slow-roll parameter and of the ef\mbox{}fective Planck mass against the number of $ e $-folds is displayed in Figure~\ref{violentbg}.
The parameters chosen are $ \beta = 0.0460 $, $ \gamma = 0.145 $ and $ \varphi_\ast = 15.8 $, with the mass of the field as $ m_\varphi = 2.1 \times 10^{-5} $ and the initial field value is $ \varphi_{\mathrm{ini}} = 16.5783 $.
\newline
\\
The resulting scalar and tensor power spectra are displayed in the left panel of Figure~\ref{violentpowstensca}. 
We see an extremely sharp dip in the power spectrum of the scalar perturbation at $ k \sim 0.003 \, \mathrm{Mpc}^{-1} $, which is followed by damped oscillatory behaviour for smaller wavelengths. 
For this set of parameters, there is a noticeable feature in the tensor power spectrum, as demonstrated in Figure~\ref{violentpowtensor}. 
Fading oscillatory features in the primordial scalar power spectrum can also occur from jumps in the potential \cite{Adams:2001vc, Ashoorioon:2006wc, Bartolo:2013exa, Firouzjahi:2014fda, Battefeld:2010rf}, particle production during inflation \cite{Barnaby:2010ke, Battefeld:2010vr, Battefeld:2013bfl} or turns in the inflaton trajectory in the landscape of heavy fields \cite{Cespedes:2012hu, Achucarro:2012fd, Cespedes:2013rda, Saito:2013aqa}.
\newline
\\
The clear differences between the scalar and tensor power spectra in Figure~\ref{violentpowstensca} can be explained by examining the behaviour of the slow-roll parameter $ \varepsilon $.
For the single-field case only, the scalar power spectrum follows as $ \mathcal{P}_{\zeta} \sim H^2/\varepsilon $, whereas the tensor power spectrum follows as $ \mathcal{P}_{\mathrm{T}} \sim H^2 $.
As a result, a sharp dip in the scalar power spectrum is observed at the same time when a sharp peak occurs in the slow-roll parameter, however, the tensor power spectrum is largely unaffected.
\newline
\\
Using both power spectra, the tensor-to-scalar ratio can be calculated and its behaviour as a function of the wavenumber is presented in the right-hand side panel of Figure~\ref{violentpowstensca}.
Although a very large unrealistic tensor-to-scalar ratio is generated in the region of the Planck pivot scale, this can be useful in constraining the parameters in the nonminimal coupling $ F(\varphi) $.

\subsubsection{Benchmark 2}
\begin{figure}[h]
\begin{center}
\scalebox{0.75}{\includegraphics{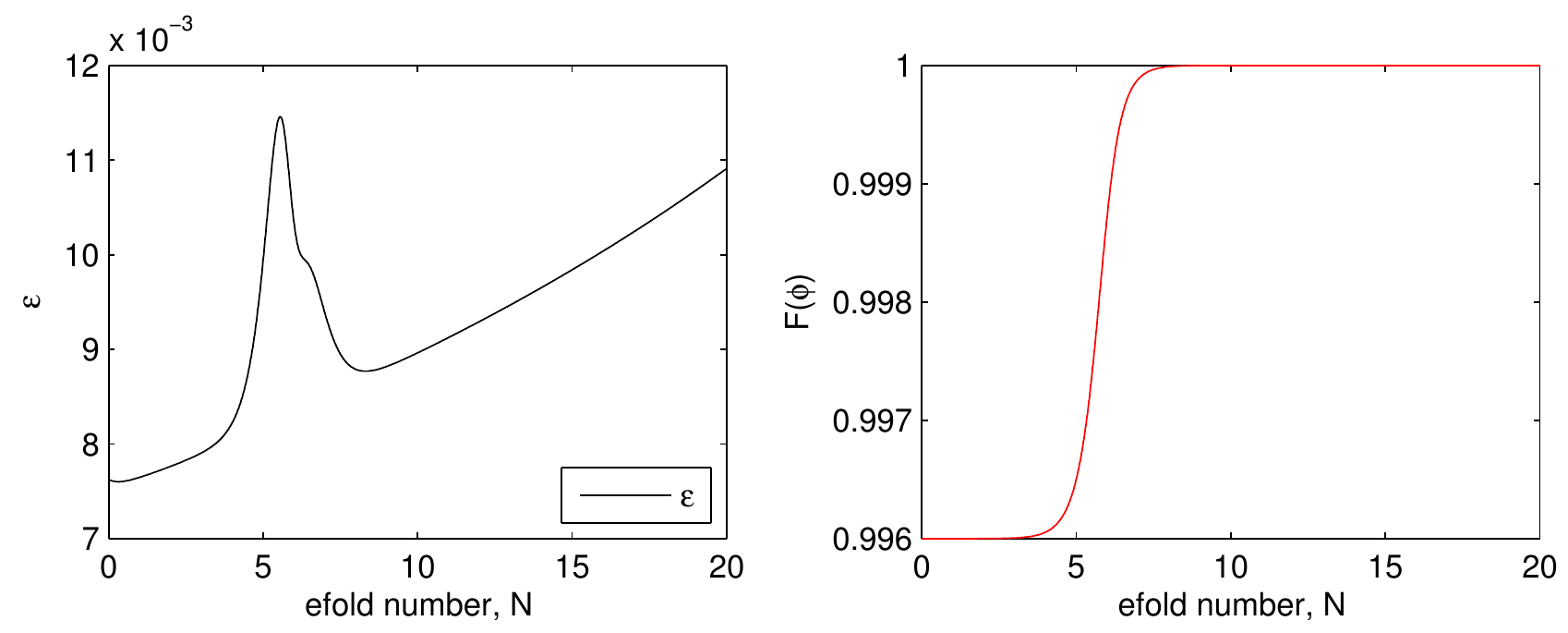}}
\caption{
Background dynamics for benchmark 2 model: the evolution of the slow-roll parameter $ \varepsilon $ (\emph{left panel}) and the corresponding $ F(\varphi) $ (\emph{right panel}) for the first 20 $ e $-folds of inflation.
The initial field value is $ \varphi_{\mathrm{ini}} = 16.5783 $ with a mass of $ m_\varphi = 2.1\times 10^{-5} $ and the parameters required in the nonminimal coupling $ F(\varphi) $ are $ \beta = 0.046 $, $ \gamma = 0.145 $ and $ \varphi_\ast = 15.8 $.
}
\label{bg4bicep2}
\end{center}
\end{figure}

In this model we show that by selecting the appropriate model parameters, it is possible to find an agreement between the results from the Planck and BICEP2 experiments.
The Planck experiment has placed an upper bound on the tensor-to-scalar ratio of $ r \leq 0.11 $ at the pivot scale $ k_0 = 0.002 \, \mathrm{Mpc}^{-1} $, whereas BICEP2 has stated a value of $ r = 0.2 $ at the scale $ k_{\mathrm{BICEP2}} = 0.0048 \, \mathrm{Mpc}^{-1} $.
The two observations would not agree with the standard power law power spectrum.
However, through the transition in the ef\mbox{}fective Planck mass, the disagreement between the two experiments can be weakened.
The parameters for such a model are $ m_\varphi = 6.9 \times 10^{-6} $, $ \beta = 0.002 $, $ \gamma = 0.111 $ and $ \varphi_\ast = 15.49 $, with the initial field value $ \varphi_\mathrm{ini} = 16.2271 $; the gravitational coupling $ F(\varphi) $ and the evolution of the slow-roll parameter $ \varepsilon $ is displayed in Figure~\ref{bg4bicep2}.
\newline
\\
In Figure~\ref{bg4bicep2}, we see that the slow-roll parameter creates a peak due to the increase in the coupling at approximately $ N = 6 $.
As previously seen in benchmark 1, a small ef\mbox{}fect on the slow-roll parameter $ \varepsilon $ greatly af\mbox{}fects the tensor-to-scalar ratio; notice that, with this combination of parameters, the initial value of F($\varphi$) = 0.996, deviating by less that 0.5\% from minimal coupling, compared with $ \sim 10\% $ deviation in benchmark 1.
\begin{figure}[h]
\begin{center}
\scalebox{0.75}{\includegraphics{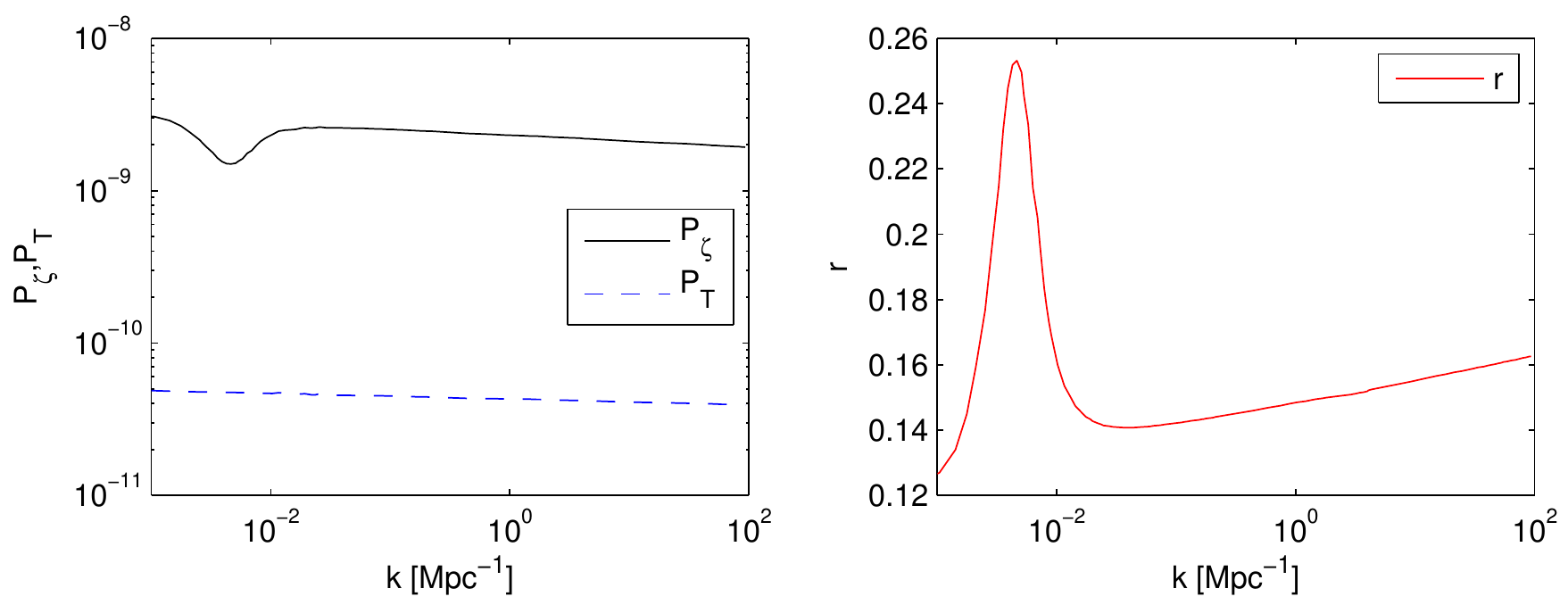}}
\caption{
The power spectra (\emph{left panel}) for the scalar ($ \mathcal{P}_{\zeta} $, solid black) and tensor ($ \mathcal{P}_{\mathrm{T}} $, blue dashed) perturbations and the associated tensor-to-scalar ratio $ r $ (\emph{right panel}) against wavenumber for benchmark 2.
Parameters used in the nonminimal coupling are $ \beta = 0.002 $, $ \gamma = 0.111 $ and $ \varphi_\ast = 15.49 $. 
The starting field value is $ \varphi_{\mathrm{ini}} = 16.2271 $ and field mass is $ m_\varphi = 6.9 \times 10^{-6} $.
}
\label{powstensca4bicep2}
\end{center}
\end{figure}
The resulting power spectra for the scalar and tensor perturbations are displayed in the left panel of Figure~\ref{powstensca4bicep2} and using the two power spectra, the tensor-to-scalar ratio can be calculated; 
the tensor-to-scalar ratio versus wavenumber is shown in the right panel of Figure~\ref{powstensca4bicep2}.
As expected, a reduction in the power of the scalar power spectrum is observed whereas the tensor power spectrum is unaf\mbox{}fected.
In Figure~\ref{powstensca4bicep2}, we have demonstrated that the tensor-to-scalar ratio can be restricted to the allowed values constrained by BICEP2.
The maximum value of the spectral index in the vicinity of the feature is obtained by means of \cite{Stewart:1993bc}
\be
n_\mathrm{s}(k) \simeq 1+ \frac{\mathrm{d} \ln \mathcal{P}_{\zeta}(k)}{\mathrm{d} \ln{k}}~,
\ee
and for $ k > 10^{-2} \, \mathrm{Mpc}^{-1} $, the spectral index is $ n_\mathrm{s} \simeq 0.98 $.
One can obtain a rough estimate on the maximum magnitude for the nonlinearity parameter $f_{\mathrm{NL}} $ in the squeezed limit from that \cite{Maldacena:2002vr}
\be
| f_{\mathrm{NL}} |_{\mathrm{max}} \simeq \frac{5}{12} | 1 - n_\mathrm{s} | \simeq 0.45~.
\ee
The dimensionless parameter $ f_{\mathrm{NL}} $ measures the amplitude of non-Gaussianity --- see the following references: \cite{Gangui:1993tt, Wang:1999vf, Komatsu:2001rj, Babich:2004gb}.
Thus, even in the vicinity of the transition, this model remains consistent with the Planck limit for local non-Gaussianity of $ f^{\mathrm{loc}}_{\mathrm{NL}} = 2.7 \pm 5.8 $ \cite{Ade:2013ydc}.

\subsubsection{Benchmark 3}
\begin{figure}[h]
\begin{center}
\scalebox{0.75}{\includegraphics{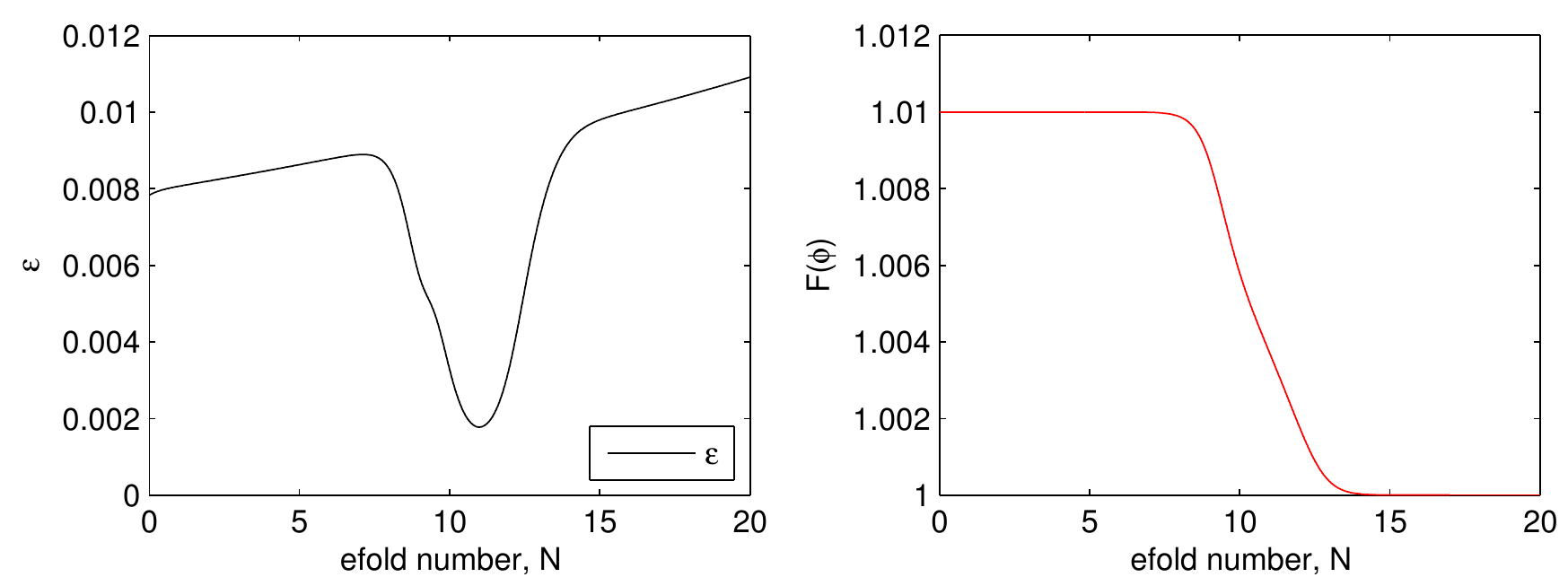}}
\caption{
The evolution of the slow-roll parameter $ \varepsilon $ (\emph{left panel}) and nonminimal coupling $ F(\varphi) $ (\emph{right panel}) for the first 20 $ e $-folds of inflation for benchmark 3.
In this case, the initial field value is $ \varphi_{\mathrm{ini}} = 15.9055 $ and mass is $ m_\varphi = 6.5 \times 10^{-6} $.
The parameters required in $ F(\varphi) $ are $ \beta = -\,0.005 $, $ \gamma = 0.100 $ and $ \varphi_\ast = 14.64 $.
}
\label{dipbg}
\end{center}
\end{figure}

\begin{figure}[h]
\begin{center}
\scalebox{0.75}{\includegraphics{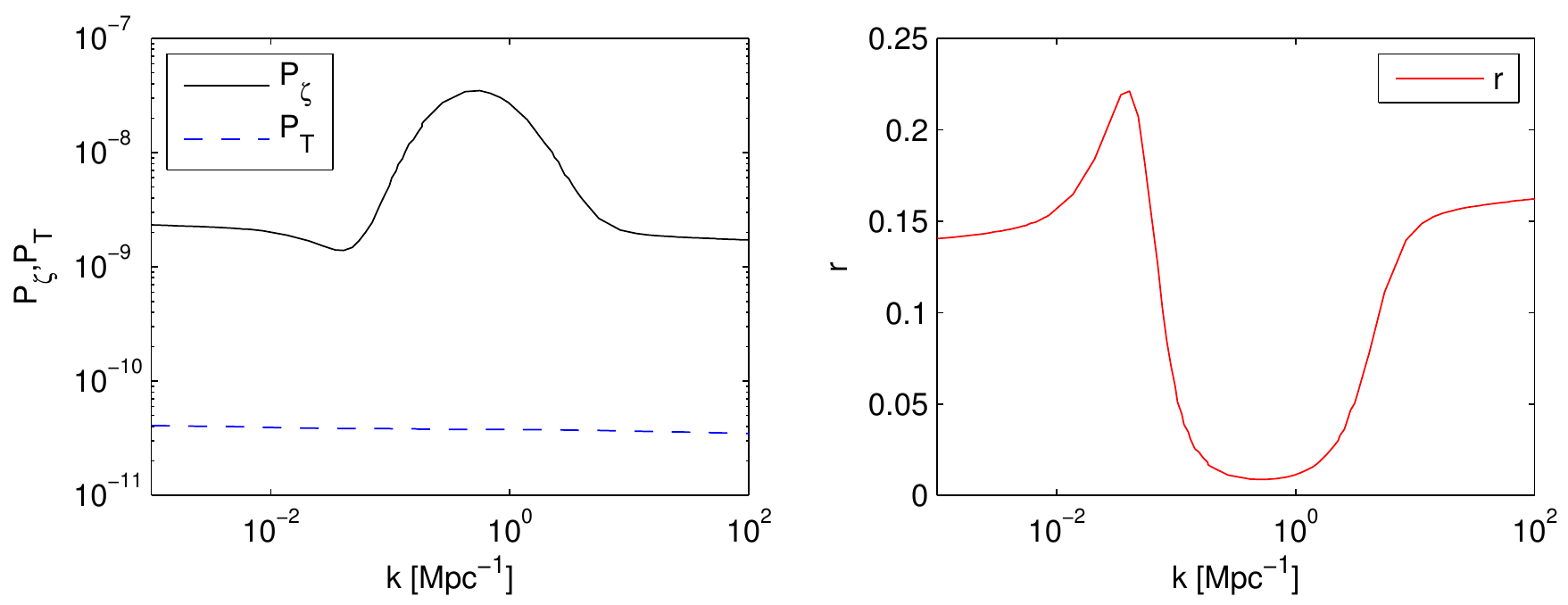}}
\caption{
The power spectra (\emph{left panel}) for the scalar ($ \mathcal{P}_{\zeta} $, solid black) and tensor $ \mathcal{P}_{\mathrm{T}} $, blue dashed) perturbations and the associated tensor-to-scalar ratio $ r $ against wavenumber, for benchmark 3.
Model parameters used are $ \beta = -\,0.005 $, $ \gamma = 0.100 $ and $ \varphi_\ast = 14.64 $. 
The starting field value is $ \varphi_{\mathrm{ini}} = 15.9055 $ and field mass is $ m_\varphi = 6.5 \times 10^{-6} $.
}
\label{dippowstensca}
\end{center}
\end{figure}

Finally, we have also studied the case when the factor $ \beta $ within the gravitational coupling $ F(\varphi) $ is of the opposite sign i.e. $ \beta $ is negative.
For this benchmark model, the parameters used in the nonminimal coupling are $ \beta = -\,0.005 $, $ \gamma = 0.100 $ and $ \varphi_\ast = 14.64 $.
In addition, the starting field value is $ \varphi_{\mathrm{ini}} = 15.9055 $ and mass used is $ m_{\varphi} = 6.5 \times 10^{-6} $.
Both the evolution of the slow-roll parameter $ \varepsilon $ and coupling $ F(\varphi) $ for this choice of parameters are displayed in Figure~\ref{dipbg}.
\newline
\\
From Figure~\ref{dipbg}, we see the change in sign in $ \beta $ causes a dip in the evolution of the slow-roll parameter, and from this, we expect a bump in the scalar power spectrum; 
the power spectra for both the scalar and tensor perturbations, and the tensor-to-scalar ratio with respect to wavenumber are presented in Figure~\ref{dippowstensca}.
We see the sign change in $ \beta $ accompanied by this choice in model parameters causes the dip to last for approximately 7 $ e $-folds.
\newline
\\
The reduction in the slow-roll parameter has had no considerable ef\mbox{}fect on the tensor power spectrum.
Furthermore, this reduction is the source of enhancement in the scalar power spectrum in the range: $ 10^{-2} \, \mathrm{Mpc}^{-1} < k < 10^1 \, \mathrm{Mpc}^{-1} $, which in turn suppresses the tensor-to-scalar ratio in the wavenumber range: $ 10^{-1} \, \mathrm{Mpc}^{-1} < k < 10^1 \, \mathrm{Mpc}^{-1} $.

\newpage
\subsection{Two-field model}
\label{sec:chap3twofield}
\begin{figure}[t]
  \begin{center}
    \scalebox{0.85}{\includegraphics{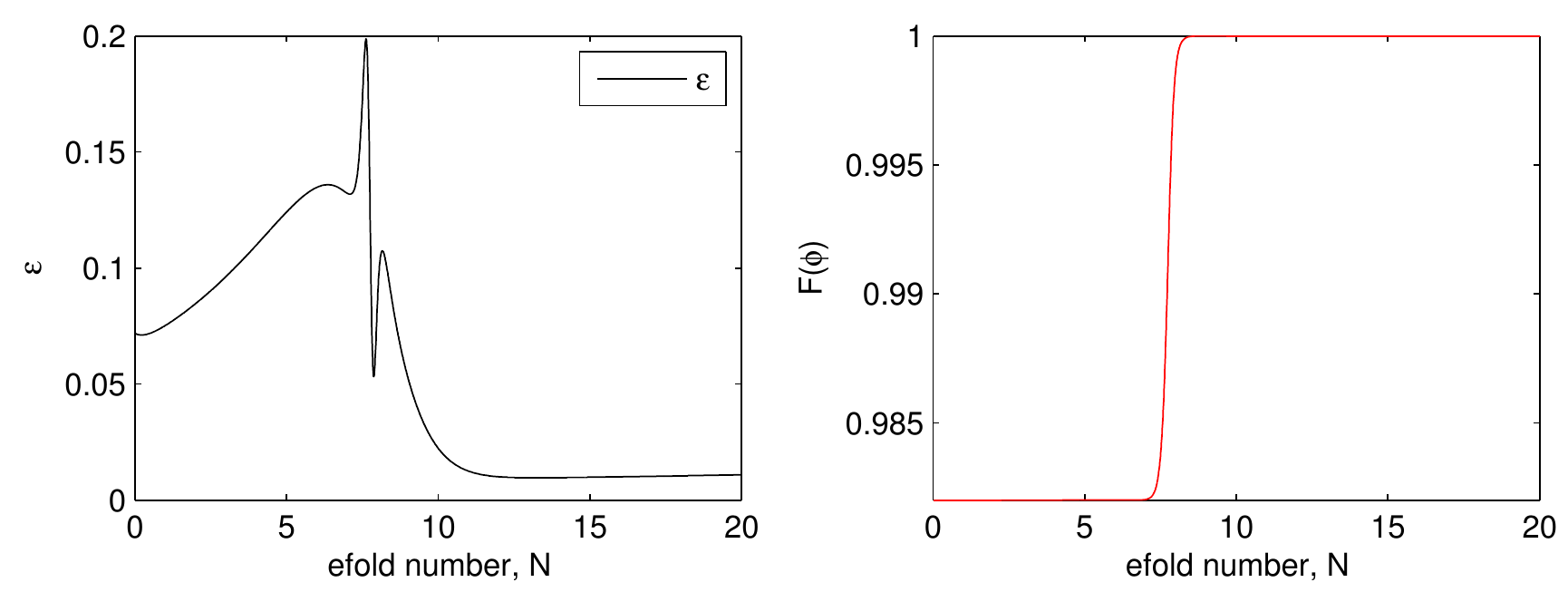}}
    \caption{Evolution  of  the slow-roll  parameter  $ \varepsilon  $
    (\emph{left}) and  the ef\mbox{}fective  Planck mass $  F$ (\emph{right})
    for  the first  $ 20  $ $e$-folds  of inflation  in  the two-field
    model.  The  model parameters are  $m_{\chi}=5.6\times 10^{-6}$, $
    m_\varphi =  4.48 \times 10^{-5}  $, $\beta=0.009$, $\gamma=0.111$
    and $ \varphi_\ast = 1.00  $, with $ \chi_{\mathrm{ini}} = 15.489$
    and $\varphi_{\mathrm{ini}}=4.5$. \label{2fbkgr}}
  \end{center}
\end{figure}

\begin{figure}[t]
  \begin{center}
    \scalebox{0.85}{\includegraphics{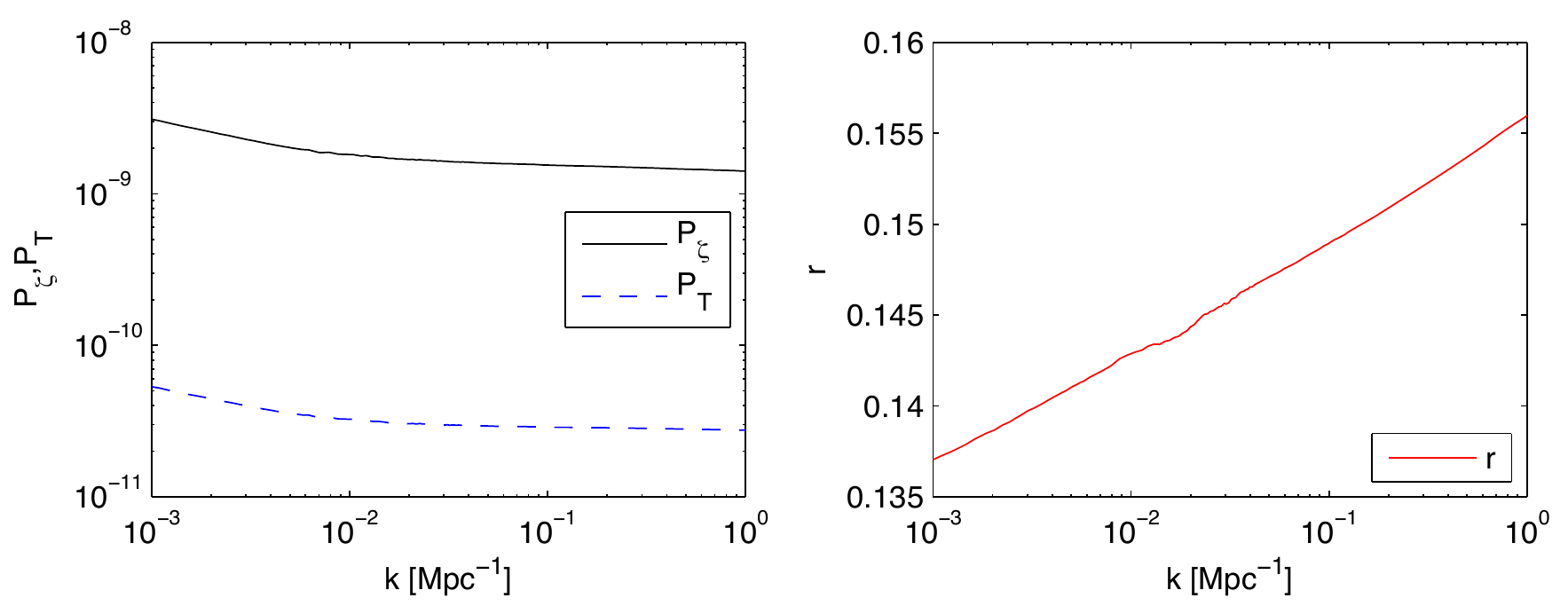}}
    \caption{Power      spectra      (\emph{left})      of      scalar
    ($\mathcal{P}_{\zeta}$,  solid  black)  and  tensor  perturbations
    ($\mathcal{P}_{\mathrm{T}}$,  blue   dashed)  and  the  associated
    tensor-to-scalar ratio  $r$ (\emph{right}) against  wavenumber $k$
    (${\rm Mpc^{-1}}$) for the  two-field model.  The model parameters
    are  $m_{\chi}=5.6\times  10^{-6}$,  $  m_\varphi  =  4.48  \times
    10^{-5} $, $\beta=0.009$, $\gamma=0.111$ and $ \varphi_\ast = 1.00
    $, with  $\chi_{\mathrm{ini}}=4.5$ and $  \varphi_{\mathrm{ini}} =
    15.489 $.\label{2fpowr}}
  \end{center}
\end{figure}

To extend the single-field model given by the action in Eq.~\eqref{action}, a second minimally-coupled field as stated in Eq.~\eqref{actionchi} is added.
This addition was made in order to test the results generated by the single-field model, i.e. the distinctive feature appearing in the primordial scalar power spectrum as seen in Figure.~\ref{powstensca4bicep2}, for example.
In this two-field model, the step transition again occurs in the Brans-Dicke field $ \varphi $, however the inflaton is played by the auxiliary field $ \chi $.
The Einstein field equations takes the form
\be
  G_{\mu \nu} = \frac{1}{F} \bigg[ T^{(\chi)}_{\mu \nu}
  					+ \partial_{\mu} \varphi \, \partial_{\nu} \varphi
  					- \frac{1}{2} g_{\mu \nu} g^{\alpha \beta} \partial_{\alpha} \varphi \, \partial_{\beta} \varphi
					+ \nabla_{\mu} \nabla_{\nu} F
  					- g_{\mu \nu} \Box F
  					- g_{\mu \nu} U \bigg]~,
\ee
where
\be
  T^{(\chi)}_{\mu \nu} = - \frac{2}{ \sqrt{-g} }
  \frac{ \delta S^{(\chi)} }{ \delta g^{\mu \nu} }
\ee
is the energy-momentum tensor of  the field $\chi$. 
\newline
\\
Varying the full action (a combination of Eqs.~\eqref{action} and \eqref{actionchi}) with respect to the two scalar fields yields their equations of motion as shown below
\begin{subequations}
\begin{align}
  2 \varpi \Box \varphi\ & =\ F_{,\varphi} g^{\mu\nu}T^{(\chi)}_{\mu\nu}
  - \varpi_{,\varphi} g^{\mu \nu} \partial_{\mu} \varphi \, \partial_{\nu} \varphi
  - 4 F_{,\varphi} U
  + 2F U_{,\varphi}~,
  \\
  \Box \chi \ & =\ V_{,\chi}~.
\end{align}
\end{subequations}
Applying the FRW metric to these equations produces the background equations of motion
\begin{subequations}
\begin{align}
-\,2 \varpi ( \ddot{\varphi} + 3H \dot{\varphi} ) & = F_{,\varphi} ( \dot{\chi}^2 - 4 V )
  										- \varpi_{,\varphi} \dot{\varphi}^2
  										- 4 F_{,\varphi} U
  										+ 2F U_{,\varphi}~,			\\
\ddot{\chi} + 3H \dot{\chi} + V_{,\chi} & =0~.
\end{align}
\end{subequations}
The Friedmann equations take the following forms:
\begin{subequations}
\begin{align}
  H^2 &\ = \ \frac{1}{3F} \bigg[ \frac{1}{2} \dot{\varphi}^2
  + \frac{1}{2} \dot{\chi}^2 + U + V - 3H \dot{F} \bigg]~,
  \\
  -2 \dot{H} &\ =\  \frac{1}{F} \bigg[  \dot{\varphi}^2
  +  \dot{\chi}^2 +  \ddot{F} - H \dot{F}  \bigg]~.	
\end{align}
\end{subequations}
The    relevant    perturbation   equations    may    be   found    in
\cite{Kaiser:2010yu, White:2012ya} and are given by
\begin{subequations}
\begin{align}
  \delta \ddot{\varphi} & + \bigg[ 3H
  + \frac{\varpi_{,\varphi}}{\varpi} \dot{\varphi} \bigg]
  \delta \dot{\varphi}
  \nonumber
  \\
  & +  \bigg[ \frac{1}{2}
  \bigg( \frac{F_{,\varphi}}{\varpi} \bigg)_{,\varphi} T^{(\chi)}
  + \frac{1}{2} \bigg( \frac{\varpi_{,\varphi}}{\varpi} \bigg)_{,\varphi}
  \dot{\varphi}^2
  - \frac{1}{2} \bigg( \frac{1}{\varpi} ( 4F_{,\varphi}U
  - 2FU_{,\varphi}  ) \bigg)_{,\varphi} + \frac{k^2}{a^2} \bigg]
  \delta \varphi
  \nonumber \\
  & - ( \dot{\Psi} + 3 \dot{\Phi} ) \dot{\varphi}
  + \frac{1}{\varpi} \bigg[ F_{,\varphi} T^{(\chi)} - 4F_{,\varphi} U
  + 2F U_{,\varphi}  \bigg] \Psi
  \nonumber \\
  & + \frac{1}{2 \varpi} F_{, \varphi} \delta T^{(\chi)} = 0~,	
  \\				
  \delta \ddot{\chi} & +  3H \delta \dot{\chi}
  + \frac{k^2}{a^2} \delta \chi -
  ( \dot{\Psi} + 3 \dot{\Phi}) \dot{\chi}
  + \, 2V_{,\chi} \Psi + V_{,\chi \chi}  \delta \chi = 0~,
\end{align}
\end{subequations}
where $ T^{(\chi)} $ is the trace of the energy-momentum tensor of the field $ \chi $ and $ \delta T^{(\chi)} $ is its perturbation.
The right-hand side of the perturbed Einstein equations in Eq.~\eqref{Einseq} are given by
\begin{subequations}
\begin{align}
  \delta \rho & =  \frac{1}{F} \bigg[ \dot{\varphi} \delta \dot{\varphi}
  - \dot{\varphi}^2 \Psi + \dot{\chi} \delta \dot{\chi}
  - \dot{\chi}^2 \Psi
  + ( U_{,\varphi} \delta \varphi + V_{,\chi} \delta \chi ) 	
  + \, 3 \dot{F} ( \dot{\Phi} + 2H \Psi )
  \nonumber	\\
 & \qquad - 3H ( \delta \dot{F} + H \delta F )
 - \frac{k^2}{a^2} \delta F \bigg]~,	
  \\
  \delta q & =  - \frac{1}{F} \bigg[ \dot{\varphi} \delta \varphi
  + \dot{\chi} \delta \chi + \delta \dot{F}
  - \dot{F} \Psi - H \delta F \bigg]~,
  \\
  \delta p & =  \frac{1}{F} \bigg[ \dot{\varphi} \delta \dot{\varphi}
  - \dot{\varphi}^2 \Psi + \dot{\chi} \delta \dot{\chi}
  - \dot{\chi}^2 \Psi
  - ( U_{,\varphi} \delta \varphi + V_{,\chi} \delta \chi ) 		
  - \, p_{\rm tot} \delta F + \delta \ddot{F} + 2H \delta \dot{F}
  \nonumber	\\
  & \qquad - \dot{F} \dot{\Psi} - 2\dot{F} \dot{\Phi}
  - \, 2( \ddot{F} + 2H \dot{F} ) \Psi
  + \frac{k^2}{a^2} \delta F \bigg]~,
\end{align}
\end{subequations}
where $ p_{\mathrm{tot}} $ is the total ef\mbox{}fective pressure from the two fields.
It is defined as
\be
p_{\mathrm{tot}} = \frac{1}{F} \bigg[ \frac{1}{2} \dot{\varphi}^2 
  + \frac{1}{2} \dot{\chi}^2 - U - V + \ddot{F} + 2 H \dot{F} \bigg]~.
\ee
\begin{figure}[t]
  \begin{center}
    \scalebox{0.85}{\includegraphics{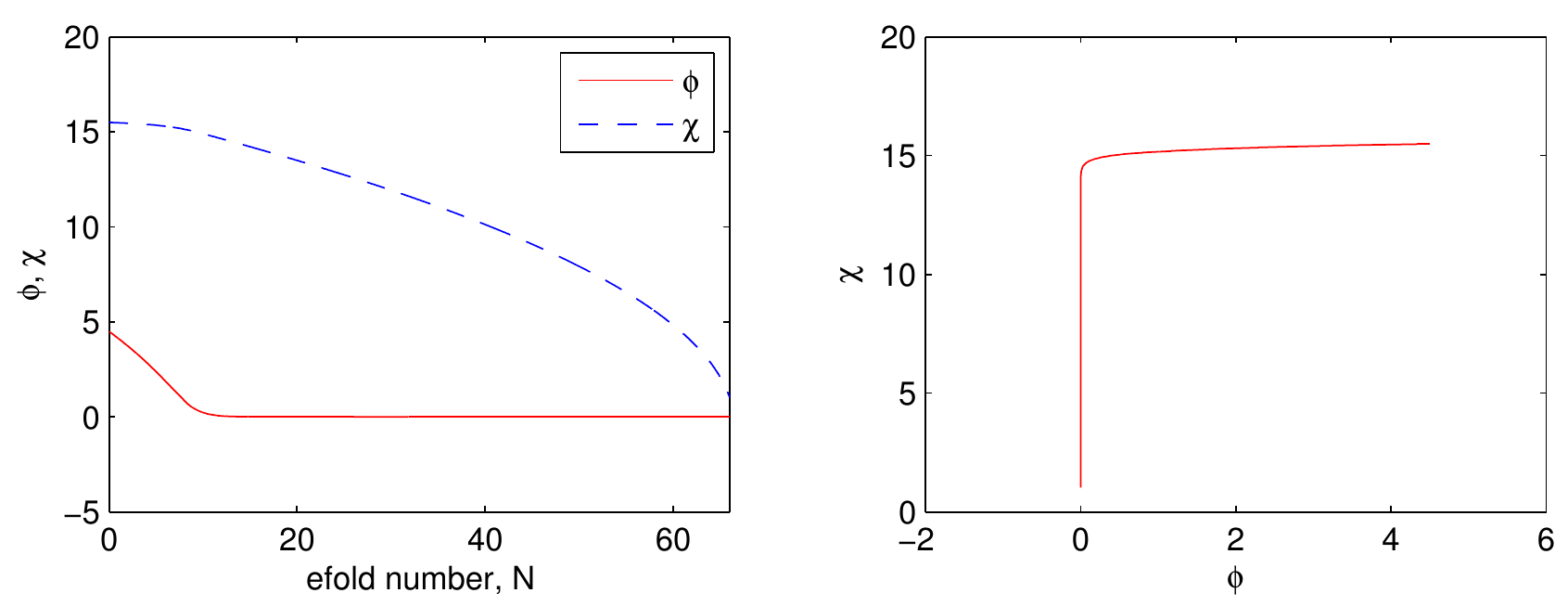}}
    \caption{Evolution of $\varphi$ and $\chi$ versus $e$-fold number $N$ (\emph{left panel}), and the field trajectory in the $ \varphi - \chi $ plane for the same $e$-foldings, illustrating the sharp turn at $ N \sim 10 $ (\emph{right panel}).
\label{twofevo}}
\end{center}
\end{figure}
For this model the parameters in the coupling $ F(\varphi) $ are chosen as $ \beta = 0.009 $, $ \gamma = 0.111 $  and $ \varphi_{\ast} = 1.0 $.
The initial field values are $ \varphi_{\mathrm{ini}} = 4.5 $  and $ \chi_{\mathrm{ini}} = 15.489 $, and their respectively masses are $ m_{\varphi} = 4.48 \times  10^{-5}$ and $ m_{\chi} = 5.6 \times 10^{-6} $.
As a result, the background dynamics of the two fields $ \varphi $ and $ \chi $ are displayed in Figure~\ref{twofevo} and it is clear that the final 50 $ e $-folds of inflation are driven by the field $ \chi $. 
The evolution  of  the  slow-roll parameter $\varepsilon$ and the ef\mbox{}fective Planck mass $F$ are shown in Figure~\ref{2fbkgr}. 
\newline
\\
In Figure~\ref{2fbkgr} we see the feature from the sharp transition in $ F(\varphi) $ appears upon the rolling of the Brans-Dicke field towards the origin.
This occurs at approximately 7 $e$-folds from the start of inflation.
We see that the fluctuation in the slow-roll parameter $ \varepsilon $ is of comparable nature to that displayed in benchmark 1. 
\newline
\\
We observe from Figure~\ref{2fpowr} a smooth enhancement in the tensor power spectrum for longer wavelengths i.e. $ k \lesssim 10^{-2} \, \mathrm{Mpc}^{-1} $.
This can be understood by the overall reduction in the slow-roll parameter after the departure of the second field.
The $k$-dependent tilt of the tensor power spectrum is given by \cite{Mukhanov:1990me, Hwang:1996xh, Hwang:2001pu}
\be
n_\mathrm{t} = -2 \varepsilon - \frac{\dot{\varphi}F_{,\varphi}}{H F}~.
\ee
The gradient in the nonminimal coupling $ F(\varphi) $ is zero before and after the transition, therefore, the slope of the tensor power spectrum is provided solely by the slow-roll parameter $ \varepsilon $.
This parameter is considerably larger pre-transition.
In contrast to the single-field model, we do not see the effect of the sharp feature in the effective Planck mass appearing in the scalar power spectrum for the two-field model considered here.
The sharp turn in the field trajectory occurring at $ N \sim 10 $ leads to the conversion of isocurvature modes into curvature modes, and essentially washes out any features in the scalar power spectrum.
In turn, this results in the enhancement of the scalar power spectrum at $ k \lesssim 10^{-2} \, \mathrm{Mpc}^{-1} $, leaving the horizon before the turn.
It is important to note that the conversion of isocurvature modes in two-field models causes the curvature perturbations to be frame dependent \cite{White:2012ya}.
We also see from Figure~\ref{2fpowr} that the tensor-to-scalar ratio over wavenumber $ k $ has the same form as that in a minimally-coupled single field case.

\section{Conclusion}
\label{sec:chap3sec4}
In this work, we have studied the ef\mbox{}fects of sharp transitions within the ef\mbox{}fective Planck mass during the inflationary epoch, specifically examining its ef\mbox{}fects on the primordial power spectra.
We have focused on single-field models before extending into a two-field model, which contains a secondary scalar field behaving as the inflaton field.
In the single-field models, we find that such a transition gives rise to strong features in the scalar power spectrum, with these features occurring at the scales corresponding to those leaving the horizon during the transition, which are within our observational window.
We see these features are reflected in the tensor-to-scalar ratio at the same scales.
In contrast to \cite{Adams:2001vc}, we do not find oscillatory behaviour in the scalar power spectrum for these type of transitions.
However this is to be expected as we do not study the same theory.
\newline
\\
We have presented a single-field case in which both the Planck and BICEP2 measurements agree, hence reducing tension between the two.
Moreover, we have also studied the single-field case with the inclusion of a minimally-coupled field.
In the two-field case, due to the conversion of isocurvature modes into curvature modes, we see dampened features in the scalar power spectra and in the tensor-to-scalar ratio.
To further this study the generation of the CMB angular power spectrum and comparison to data will be required.

\chapter{Disformal couplings}
In the last chapter we studied scalar-tensor theories in the context of inflation; these theories can also be studied in the realm of dark energy.
We will now make the transition to later times in cosmic history and consider the era of dark energy.
Additionally in this chapter we will consider the background only and therefore, no perturbations will be discussed.
\newline
\\
As mentioned in Chapter 3, in scalar-tensor theories, there are two representations of the action, namely the Jordan and Einstein frames.
We now remind the reader of the frame definitions: the Jordan frame is a frame in which the scalar field and the Ricci scalar are coupled, whereas the Einstein frame is one where the Einstein-Hilbert action is as standard.
Due to dif\mbox{}ferences in representation, the question of which frame, Einstein or Jordan, is ``most physical'' arises; this question is addressed in the following references: \cite{Faraoni:1999hp, Flanagan:2004bz, Bettoni:2013diz, Postma:2014vaa}.
When presented initially with the Jordan frame, in order to obtain the Einstein frame the standard practice is to perform a conformal transformation of the metric; conformal transformations have been studied extensively in \cite{Faraoni:1998qx, Kaiser:2010ps, Faraoni:2006fx, Dabrowski:2008kx}.
However, one can study a more general metric transformation between the two frames. 
\newline
\\
It was f\mbox{}irst demonstrated in 1993 by Bekenstein \cite{Bekenstein:1992pj} that the most general metric transformation between two frames which obeys both causality and the weak equivalence principle is
\be \label{bekensteinrel}
\tilde{g}_{\mu \nu} = C(\phi) g_{\mu \nu} + D(\phi) \partial_\mu \phi \, \partial_\nu \phi~.
\ee
This is known as the ``disformal'' transformation, which contains derivatives of the scalar f\mbox{}ield. 
The f\mbox{}irst term in Eq.~\eqref{bekensteinrel} is the conformal coupling between the scalar and matter f\mbox{}ields, and the disformal coupling is represented by the second term.
Disformal couplings have attracted a signif\mbox{}icant amount of attention since Bekenstein's paper in 1993, \cite{Kaloper:2003yf, Koivisto:2008ak, Zumalacarregui:2010wj, Brax:2012hm, vandeBruck:2012vq, Koivisto:2012za, Deruelle:2014zza}.
These couplings have been studied as an alternative to dark matter, for example the Tensor-Vector-Scalar gravitational theory (TeVeS) \cite{Bekenstein:2004ne}. 
It has led to the proposal of a new relativistic formulation of the modif\mbox{}ied Newtonian Dynamics (MOND) labeled bimetric MOND (BIMOND) \cite{Milgrom:2009gv}.
Other topics studied with disformal couplings are in bimetric theories of gravity with the inclusion of the Eddington-Born-Infeld (EBI) action \cite{Banados:2008rm, Banados:2008fi, Skordis:2009bf}.
The inf\mbox{}lationary scenario with the disformal relation has also been investigated \cite{Kaloper:2003yf}.
The disformal relation has also appeared in variable speed of light theories \cite{Clayton:1998hv, Clayton:2001rt}.
\newline
\\
In this chapter, we consider a scalar-tensor theory with disformally coupled f\mbox{}ields to study the ef\mbox{}fects of this type of coupling on the properties of the Cosmic Microwave Background (CMB).
In order to check consistency throughout this theory, both a f\mbox{}luid description and a kinetic theory was f\mbox{}irst established before discussing observables.
Observables have been chosen to be expressed in terms of the metric in which an observer experiences, in this case, the baryonic metric.
The baryonic metric is a metric in which baryonic matter is uncoupled to the scalar field. 
This choice was decided upon to avoid the frame-independent formulation for disformally coupled f\mbox{}ields.
The subject of frame-independence in conformally coupled theories is discussed in \cite{Catena:2006bd, Chiba:2013mha}.
\newline
\\
The chapter is laid out as follows: in Section~\ref{sec:chap4sec1} the action and def\mbox{}inition of a disformal coupling are presented, to be followed by the background equations of motion.
Following this, the f\mbox{}luid description and kinetic theory are presented in Sections~\ref{sec:chap4sec2} and \ref{sec:chap4sec3}.
A discussion on observable quantities is contained in Section~\ref{sec:chap4sec4}, leading into the presentation of results in Section~\ref{sec:chap4sec5}.
Finally, our f\mbox{}indings are summarised in Section~\ref{sec:chap4sec6}.

\section{The model}
\label{sec:chap4sec1}
The action that we consider is
\be 
\label{eframeaction}
S = \int \mathrm{d}^{4} x \, \sqrt{-g} \left[ \frac{M_{\rm Pl}^2}{2} R 
								-\frac{1}{2}g^{\mu\nu} \partial_\mu\phi \, \partial_\nu\phi 
								- V(\phi) \right] 
								 + \sum S_{i} \,[\,{\tilde g}_{\mu\nu}^{i},\chi_i \,]~,
\ee
where $ R $ is the Ricci scalar calculated with respect to the metric $ g_{\mu \nu} $, $ S_{i} $ denotes the action of the various matter f\mbox{}ields $ \chi_i $ propagating on dif\mbox{}ferent geodesics via the relation
\be
\label{gmunu}
{\tilde g}_{\mu\nu}^{i} = C_{i}(\phi) g_{\mu\nu}+D_{i}(\phi)\partial_\mu \phi \, \partial_\nu \phi~.
\ee
It is important to note that this action is expressed in the Einstein frame.
From now on, the reduced Planck mass are set to one, $ M_{\mathrm{Pl}} = 1 $.
\newline
\\
A cosmological setting is considered and the metric $ g_{\mu \nu} $ is of FRW form; for the FRW line element, we refer the reader to Eq.~\eqref{frw} in Section~\ref{sec:chap1prelim}.
By the variation of the action in Eq.~\eqref{eframeaction} with respect to the scalar f\mbox{}ield and using Eq.~\eqref{frw} yields the Klein-Gordon equation for the field $ \phi $
\be
\label{kg}
\ddot{\phi} + 3H\dot{\phi} + V' = \sum Q_{i}~,
\ee
and with respect to the metric $ g_{\mu \nu} $, the Einstein f\mbox{}ield equations
\begin{align}
\label{einsteinfield}
G^{\mu \nu} &= T^{\mu \nu}~,		\nonumber	\\
			&= T_{\phi}^{\mu \nu} + T_i^{\mu \nu}~,
\end{align}
with $ G^{\mu \nu} $ as the Einstein tensor and $ T^{\mu \nu} $ representing the total energy-momentum tensor of the system.
It is composed of two energy-momentum tensors: the scalar f\mbox{}ield and of the matter species
\begin{align}
T^{\mu \nu}_{\phi} &= \frac{2}{\sqrt{-g}} \frac{ \delta S_{\phi}}{ \delta g_{\mu \nu} }~,		\\
\label{einsteintensor}
T^{\mu \nu}_i &= \frac{2}{\sqrt{-g}} \frac{ \delta S_i}{ \delta g_{\mu \nu} }~,
\end{align}
where $ S_{\phi} $ is the action of the scalar field.
We shall assume that the energy-momentum tensor for each individual species is of perfect f\mbox{}luid form and hence $ T \indices{^{\mu}_{\nu}} = \mathrm{diag}( -\rho, p, p, p) $, 
where $ \rho $ and $ p $ are the energy density and pressure of the f\mbox{}luid respectively.
Using this and the Einstein f\mbox{}ield equations in Eq.~\eqref{einsteinfield} in a FRW spacetime yields the conservation equation for the matter species
\begin{align}
\label{rho_i} 
\dot{\rho}_{i} + 3H(\rho_{i} + p_{i}) &= - Q_{i}\dot{\phi}~,
\end{align}
where the couplings, $ Q_i $ are found by varying the action Eq.~\eqref{eframeaction} are given by
\be
\label{Q}
Q_{i} = \frac{C_{i}'}{2C_{i}} T_{i} + \frac{D_{i}'}{2C_{i}} \phi,_{\mu}\phi,_{\nu}T^{\mu \nu}_{i} 
	 		 - \nabla_{\mu} \bigg( \frac{D_{i}}{C_{i}} \phi_{\nu}T^{\mu \nu}_{i} \bigg)~,
\ee
with $ T_{i} $ representing the trace of the energy-momentum tensor of the species denoted by $ i $.
The following notation is used: dots represent derivatives with respect to cosmic time $ t $, primes are those to the scalar f\mbox{}ield $ \phi $ and commas denote partial derivatives.
The Hubble parameter is $ H = \dot {a}/a $.
In the following we will consider two fluids labelled $ i = 1, 2 $ and each is disformally coupled with a different magnitude given by $ D_i $.
From Eqs.~\eqref{rho_i} and \eqref{Q}, the couplings $ Q_1 $ and $ Q_2 $ are
\begin{subequations}
\begin{align}
Q_{1}  = {} & \frac{\mathcal{A}_{2}}{\mathcal{A}_{1} \mathcal{A}_{2} - D_{1} D_{2} \rho_{1} \rho_{2}} \bigg( \mathcal{B}_{1} - \frac{ \mathcal{B}_{2} D_{1} \rho_{1} }{\mathcal{A}_{2}} \bigg)~,	\\
Q_{2}  = {} & \frac{\mathcal{A}_{1}}{\mathcal{A}_{1} \mathcal{A}_{2} - D_{1} D_{2} \rho_{1} \rho_{2}} \bigg( \mathcal{B}_{2} - \frac{ \mathcal{B}_{1} D_{2} \rho_{2} }{\mathcal{A}_{1}} \bigg)~,
\end{align}
\end{subequations}
where
\begin{subequations}
\begin{align}
\mathcal{A}_{i} & = C_{i} + D_{i} ( \rho_{i} - \dot{\phi}^2 )~,	 \\
\mathcal{B}_{i} & = \bigg[ \frac{C_{i}'}{2} \bigg(-1 + 3 \frac{p_i}{\rho_i} \bigg) - \frac{D_{i}'}{2} \dot{\phi}^2		\nonumber	\\
							& \qquad \qquad + D_i \left\{ 3H\bigg( 1+ \frac{p_i}{\rho_i} \bigg) \dot{\phi} 
							+ V' + \frac{C_{i}'}{C_{i}} \dot{\phi}^2 \right\} \bigg] \rho_{i}~.
\end{align}
\end{subequations}
In order to achieve the dynamics needed for slow-roll, a slowly varying potential is required. 
For this purpose, the exponential potential was chosen
\be
V(\phi) = V_0 e^{-\lambda \phi}~,
\ee
where $ V_0 $ is a constant and $ \lambda = 1 $. 
We now choose to study the ef\mbox{}fects of the disformal coupling without the presence of the conformal term; $ C(\phi) = 1 $, 
and $ D(\phi) $ is treated as a constant energy scale: $ D(\phi) = M^{-4} $ where $ M $ is a constant.
\newline
\\
Before proceeding into the study of the CMB, two descriptions must be made consistent; these are the f\mbox{}luid description (used so far) and the kinetic theory.
By doing this, we can recover the equations found through the variation of the action in Eq.~\eqref{eframeaction} and have a more complete description of this disformal theory.

\section{Fluid description}
\label{sec:chap4sec2}
For simplicity, one f\mbox{}luid will be considered in this section; this can be easily generalised to multiple f\mbox{}luids. 
We will revert back to the usage of multiple f\mbox{}luids after this discussion. 
\newline
\\
The f\mbox{}luid considered will have an arbitrary equation of state $ w $ and the disformal coupling between the f\mbox{}luid and the scalar f\mbox{}ield is given by $ M $, the constant energy scale.
In the Jordan frame, the scalar f\mbox{}ield is decoupled from the matter f\mbox{}ields and so the f\mbox{}luid considered is of perfect form
\be
\label{jordanfluid}
\tilde{T}^{\mu \nu} = ( \tilde{\rho} + \tilde{p} ) \tilde{u}^{\mu} \tilde{u}^{\nu} + \tilde{p} \tilde{g}^{\mu \nu}~,
\ee
resulting in $ \tilde{T} \indices{^\mu_\nu} = \mathrm{diag}( -\tilde{\rho}, \tilde{p}, \tilde{p}, \tilde{p} ) $.
The 4-velocity vectors of a comoving observer in the Jordan frame $ \tilde{u}^{\mu} $ are def\mbox{}ined as
\be
\tilde{u}^{\mu} = \frac{ \mathrm{d} x^{\mu} }{\mathrm{d} \tilde{\tau} } = \bigg( \bigg(1- \frac{\dot{\phi}^2}{M^4} \bigg)^{\! -\frac{1}{2}}, 0, 0, 0 \bigg)~,
\ee
where the relation between the proper time of the observer $ \tilde{\tau} $ and the time variable of the Einstein frame $ t $ is
\be
\mathrm{d} \tilde{\tau} = \bigg( 1 - \frac{\dot{\phi}^2}{M^4} \bigg)^{\! \frac{1}{2}} \mathrm{d} t~.
\ee
Likewise, the f\mbox{}luid considered in the Einstein frame is also of perfect fluid form
\be
\label{einsteinfluid}
T^{\mu \nu} = ( \rho + p ) u^{\mu} u^{\nu} + p g^{\mu \nu}~,
\ee
with $ T\indices{^\mu_\nu} = \mathrm{diag}( -\rho, p, p, p) $ and the Einstein frame 4-velocity vectors of a comoving observer $ u^{\mu} $ are
\be
u^{\mu} = \frac{ \mathrm{d} x^{\mu} }{\mathrm{d} \tau } = (1, 0, 0, 0)~.
\ee
Tildes will be employed to distinguish the Jordan frame from the Einstein frame. In addition to this, the contraction of metrics with various ranked tensors must be clarif\mbox{}ied.
In the Einstein frame, all tensorial quantities are contracted with the Einstein frame metric
\be
A_{\mu} = g_{\mu \alpha} A^{\alpha}~, \qquad T \indices{^{\mu}_{\nu}} = g_{\nu \alpha} T^{\alpha \mu}~.
\ee
Similarly, all tensorial quantities in the Jordan frame are contracted with the Jordan frame metric
\be
\tilde{A}_{\mu} = \tilde{g}_{\mu \alpha} \tilde{A}^{\alpha}~, \qquad \tilde{T} \indices{^{\mu}_{\nu}} = \tilde{g}_{\nu \alpha} \tilde{T}^{\alpha \mu}~.
\ee
The f\mbox{}luid in the Jordan frame satisf\mbox{}ies the standard energy conservation equation 
\be 
\label{jordanconservation}
\tilde{\nabla}_{\mu} \tilde{T}^{\mu \nu} = 0~,
\ee
for which the covariant derivative is compatible with the metric $ \tilde{g}_{\mu \nu}  $, i.e. $ \tilde{\nabla}^\mu \tilde{g}_{\mu \nu}=0 $.
The energy-momentum tensor can be expressed in terms of the Jordan frame action:
\be
\label{jordantensor}
\tilde{T}^{\mu \nu} = \frac{2}{\sqrt{-\tilde{g}}} \frac{ \delta S }{ \delta \tilde{g}_{\mu \nu} }~.
\ee
At the level of the background, the scalar f\mbox{}ield is is only time dependent.
\newline
\\
A relationship between the Jordan and Einstein frame energy-momentum tensors can be obtained by using Eqs.~\eqref{einsteintensor} and \eqref{jordantensor}
\begin{subequations}
\begin{align}
\label{genmota}
T^{\mu \nu} &=	 \sqrt{\frac{-\tilde{g}}{-g}}	\, \frac{\delta \tilde{g}_{\alpha \beta} }{\delta g_{\mu \nu} } \, \tilde{T}^{\alpha \beta}~,		 \\ 
\label{mota}
T^{\mu \nu} & = C^3 \sqrt{ 1 + \frac{D}{C} g^{\mu \nu} \phi_{,\mu} \phi_{,\nu} } \, \tilde{T}^{\mu \nu}~,
\end{align}
\end{subequations}
and in the FRW spacetime, this relation is simplif\mbox{}ied to
\begin{align}
T^{\mu \nu} &= \sqrt{ 1 - \frac{\dot{\phi}^2}{M^4} } \, \tilde{T}^{\mu \nu}~.
\end{align}
Note here that the dot represents the derivative with respect to $ t $ where $ g_{00} = -1 $, i.e. the time derivative in the Einstein frame.
\newline
\\
By simply taking the trace of the energy-momentum tensors for both frames and using Eq.~\eqref{mota}, the relationship between the two equations of state is
\be\label{eos}
\frac{p}{\rho} = \frac{\tilde{p}}{\tilde{\rho}} \bigg(  1 - \frac{\dot{\phi}^2}{M^4}  \bigg)~.
\ee
The conservation equation in Eq.~\eqref{jordanconservation} can be rewritten using the relation between the two equations of state in Eq.~\eqref{eos}, this yields
\be
\label{joreinconservation}
\dot{\rho} + 3H \bigg( \rho + \frac{M^4}{M^4 - \dot{\phi}^2} \,p \bigg) = \frac{\dot{\phi} \ddot{\phi}}{M^4 - \dot{\phi}^2} \, \rho~.
\ee
Using the Klein-Gordon equation in Eq.~\eqref{kg} and Eq.~\eqref{joreinconservation}, the conservation equation found from using the variational principle stated in Eq.~\eqref{rho_i} is retrieved and hence, a consistent macroscopic description is achieved.

\section{Kinetic theory}
\label{sec:chap4sec3}
To make this disformal theory fully coherent, a microscopic description must be attained. 
In order to do this, a statistical approach is needed --- see \cite{ehlers, stewart, 1982ApJ...257..578R, Uzan:1998mc}.
\newline
\\
The energy-momentum tensor can be expressed in terms of an integral of the distribution function $ f $ over phase space. 
The distribution functions for the two frames, $ \tilde{f} $ and $ f $, are given in terms of their position and momenta vectors \cite{Catena:2006bd, Durrer2008}:
\be
\label{neinstein}
\mathrm{d}{\tilde N} = \mathrm{d}x^1 \mathrm{d}x^2 \mathrm{d}x^3 \mathrm{d}{\tilde P}_1 \mathrm{d}{\tilde P}_2 \mathrm{d}{\tilde P}_3 {\tilde f}
\ee
and 
\be
\label{njordan}
\mathrm{d}N = \mathrm{d}x^1 \mathrm{d}x^2 \mathrm{d}x^3 \mathrm{d}P_1 \mathrm{d}P_2 \mathrm{d}P_3 f~,
\ee
where $ \mathrm{d} \tilde{N} $ and $ \mathrm{d} N $ are the number of particles in a differential phase space in the Jordan and Einstein frames respectively, $ \mathrm{d}x^1 \mathrm{d}x^2 \mathrm{d}x^3 $ is the position phase space and $ \mathrm{d}{\tilde P}_1 \mathrm{d}{\tilde P}_2 \mathrm{d}{\tilde P}_3 $ is the momenta phase space for the Jordan frame (and similarly for the Einstein frame but with tildes). 
This formalism was also used for the conformal transformation case in \cite{Misner:1974qy}.
\newline
\\
Likewise with the f\mbox{}luid description given in the previous section, each frame has its own def\mbox{}inition in kinetic theory \cite{ehlers, stewart}  :
\begin{align} 
\label{em1}
T^{\mu\nu} = \int \frac{\mathrm{d}^3 P}{\sqrt{-g}} \frac{P^\mu P^\nu}{P^0} f~, \\
\label{em2}
{\tilde T}^{\mu\nu} = \int \frac{\mathrm{d}^3 {\tilde P}}{\sqrt{-{\tilde g}}} \frac{{\tilde P}^\mu {\tilde P}^\nu}{\tilde P^0} {\tilde f}~.
\end{align}
Notice that the integrals are given in terms of the 4-momenta of that frame (tildes for Jordan and none for Einstein) and are as follows
\be
P^\mu = \frac{\mathrm{d}x^\mu}{\mathrm{d}\lambda}~, \qquad {\tilde P}^\mu = \frac{\mathrm{d}x^\mu}{\mathrm{d}\tilde\lambda}~,
\ee
resulting in
\be
\label{momtransformed}
\tilde{P}^{\mu} = \frac{\mathrm{d} x^{\mu}}{\mathrm{d} \lambda} \frac{\mathrm{d} \lambda}{\mathrm{d} \tilde{\lambda}} = P^{\mu} \frac{\mathrm{d} \lambda}{\mathrm{d} \tilde{\lambda}}~,
\ee
where $ \lambda $ and $ \tilde{\lambda} $ are the affine parameters in the Einstein and Jordan frames, respectively.
Within the Jordan frame, after decoupling the photons will obey the collisionless Boltzmann equation \cite{Dodelson}
\be
\frac{\mathrm{d} \tilde{f} }{\mathrm{d} t} = 0~.
\ee
By using the relation between the two energy-momentum tensors Eq.~\eqref{genmota}, their def\mbox{}initions: Eqs.~\eqref{em1} and \eqref{em2} yields
\be
\frac{ \mathrm{d} \tilde{\lambda} }{ \mathrm{d} \lambda }\int \frac{ \mathrm{d}^{3} P}{ \sqrt{-g} } \, f = \int \frac{ \mathrm{d}^{3} \tilde{P}}{ \sqrt{-\tilde{g}} } \, \sqrt{\frac{-\tilde{g}}{-g}} \, \tilde{f}~.
\ee
This relation can be applied to the distribution functions of the two frames Eqs.~\eqref{neinstein} and \eqref{njordan}, and results in the following:
\be
\int \mathrm{d} \tilde{N} = \frac{\mathrm{d} \tilde{\lambda}}{\mathrm{d} \lambda} \int \mathrm{d}N~.
\ee
From this relation, it is clear that the two quantities $ \mathrm{d} N $ and $ \mathrm{d} \tilde{N} $ are not frame-invariant and ultimately, in the Einstein frame the Boltzmann equation takes the form
\be
\label{df/dt}
\frac{\mathrm{d}}{\mathrm{d}t}\bigg( \frac{\mathrm{d} \tilde{\lambda}}{ \mathrm{d} \lambda } f \bigg) = 0~.
\ee
The frame variance is due to the presence of the disformal coupling and is unlike to conformal transformations \cite{Catena:2006bd}.
\newline
\\
We now revert back to understanding how disformal couplings af\mbox{}fects the properties of the CMB; 
we know that photons will travel on geodesics in the Jordan frame and hence the following equations are true
\be
\label{geodesic}
\tilde{P}^{\mu} \tilde{\nabla}_{\mu} \tilde{P}^{\nu} = 0~, \qquad \tilde{g}_{\mu \nu} \tilde{P}^{\mu} \tilde{P}^{\nu} = 0~.
\ee
Note that the second of these equations can be rewritten as $ \tilde{g}_{\mu \nu} P^{\mu} P^{\nu} = 0 $ by using Eq.~\eqref{momtransformed}.
Adding to this, we require the relation between the af\mbox{}f\mbox{}ine parameters def\mbox{}ined in the Jordan and Einstein frame.
In order to calculate this, we can use the fact that the energy-momentum tensors in Eqs.~\eqref{em1} and \eqref{em2} can be rewritten for a system of point particles \cite{Carroll}
\begin{align}
\label{em3}
T^{\mu \nu} &= \sum \frac{P^{\mu} P^{\nu}}{P^0} \delta^{\,(3)} ( \bold{x} - \bold{x}' )~,		\\
\label{em4}
\tilde{T}^{\mu \nu} &= \sum \frac{\tilde{P}^{\mu} \tilde{P}^{\nu}}{\tilde{P}^0} \delta^{\,(3)} ( \bold{x} - \bold{x}' )~.
\end{align}
Similarly to the expressions of the distribution functions stated in Eqs.~\eqref{neinstein} and \eqref{njordan}, the position vectors of the points $ \bold{x} $, 
are the same in both frames,
and hence are the same Dirac delta functions.
Using the energy-momentum tensor relationship between the two frames, their respective 4-momenta and the discrete point particle representation given in Eqs.~\eqref{mota}, \eqref{momtransformed}, and \eqref{em4}, 
we f\mbox{}ind the relation between the Jordan and Einstein frame af\mbox{}f\mbox{}ine parameters to be
\begin{align}
T^{\mu \nu} &= \sqrt{ 1 - \frac{\dot{\phi}^2}{M^4} } \tilde{T}^{\mu \nu}~,		\nonumber \\
			&= \sqrt{ 1 - \frac{\dot{\phi}^2}{M^4} } \frac{\tilde{P}^{\mu} \tilde{P}^{\nu}}{\tilde{P}^0} \delta^{\,(3)} ( \bold{x} - \bold{x}' )~,	\nonumber \\
			&= \sqrt{ 1 - \frac{\dot{\phi}^2}{M^4} } \bigg( \frac{ \mathrm{d} \lambda }{\mathrm{d} \tilde{\lambda} } \bigg) \frac{P^{\mu} P^{\nu}}{P^0} \delta^{\,(3)} ( \bold{x} - \bold{x}' )~,	\nonumber \\
T^{\mu \nu}	&= \sqrt{ 1 - \frac{\dot{\phi}^2}{M^4} } \bigg( \frac{ \mathrm{d} \lambda }{\mathrm{d} \tilde{\lambda} } \bigg) T^{\mu \nu}~,		\nonumber \\
\label{lambdas}
\therefore	\qquad	\frac{\mathrm{d} \tilde{\lambda} }{\mathrm{d} \lambda} &= \sqrt{ 1 - \frac{\dot{\phi}^2}{M^4} }~.
\end{align}
To obtain various conservations through kinetic theory, the Liouville equation is needed
\be
\label{liouvilleeq}
\hat{\mathcal{L}}f = \mathcal{C}[f]~,
\ee 
where $ \hat{\mathcal{L}} $ is the Liouville operator and $ \mathcal{C}[f] $ contains the collision terms.
The Liouville operator acting upon a distribution function $ f $ is def\mbox{}ined as
\be
\label{liouville}
\hat{\mathcal{L}} f = \frac{\mathrm{d} f}{\mathrm{d} \lambda} = \frac{\mathrm{d} x^{\mu}}{\mathrm{d} \lambda} \frac{ \partial f }{ \partial x^{\mu} } 
						+ \frac{\mathrm{d} P^{\mu}}{\mathrm{d} \lambda} \frac{ \partial f }{ \partial P^{\mu} }~.
\ee
The geodesic equation in Eq.~\eqref{geodesic} can be rewritten as
\be
\label{rewritegeo}
\frac{\mathrm{d} \tilde{P}^{\mu} }{\mathrm{d} \tilde{\lambda} } + \tilde{\Gamma}^{\mu}_{\alpha \beta} \tilde{P}^{\alpha} \tilde{P}^{\beta} = 0~,
\ee
where the connection is given by the Christof\mbox{}fel symbols. 
In the equation given above, $ \tilde{\Gamma}^{\mu}_{\alpha \beta} $ are the Christof\mbox{}fel symbols in the Jordan frame.
For completeness, when $ C(\phi) = 1 $ and $ D(\phi) = M^{-4} $ the nonzero Jordan frame Christof\mbox{}fel symbols in terms of the Einstein frame are
\begin{subequations}
\begin{align}
{\tilde \Gamma}^0_{00} &= \frac{\dot{\phi} \ddot{\phi}}{M^4 - \dot{\phi}^2}~,	\\
{\tilde \Gamma}^0_{ij} &= \frac{M^4}{M^4 - \dot{\phi}^2} a^2 H \delta_{ij}~,		\\
{\tilde \Gamma}^i_{j0} &= H \delta^{i}_{j}~.
\end{align}
\end{subequations}
Using Eqs.~\eqref{momtransformed}, \eqref{liouville} and \eqref{rewritegeo} --- expressed in the Jordan frame, we can write down the Einstein frame Liouville operator:
\begin{align}
\hat{\mathcal{L}} f &= \frac{\mathrm{d} f}{\mathrm{d} \lambda} 		\nonumber	\\
				&= P^0 \frac{ \partial f }{\partial t } 
					- \frac{ \partial f }{ \partial P^0 } \left\{ P^0 \, \frac{\mathrm{d}^{2} \lambda}{\mathrm{d} \tilde{\lambda}^{\,2}} \bigg( \frac{ \mathrm{d} 												\tilde{\lambda} }{\mathrm{d} \lambda} \bigg)^2 + \tilde{\Gamma}^0_{\alpha \beta} P^{\alpha} P^{\beta}  \right\}~.
\end{align}
Now referring back to the Einstein frame Boltzmann equation in Eq.~\eqref{df/dt}, we can use this to calculate ($ \mathrm{d} f / \mathrm{d} \lambda $)
\begin{align}
0 &= \frac{\mathrm{d}}{\mathrm{d}t} \bigg( \frac{\mathrm{d} \tilde{\lambda} }{\mathrm{d} \lambda} f \bigg)~,		\nonumber	\\
			&= \frac{\mathrm{d}}{\mathrm{d}t} \bigg( \frac{\mathrm{d} \tilde{\lambda} }{\mathrm{d} \lambda} \bigg) f + \frac{\mathrm{d}f}{\mathrm{d}t} \bigg( \frac{\mathrm{d} \tilde{\lambda} }{\mathrm{d} \lambda} \bigg)~,	\nonumber	\\
\therefore \qquad \frac{\mathrm{d}f}{\mathrm{d}t} &= -\, \frac{\mathrm{d} \ln (\frac{\mathrm{d} \tilde{\lambda}}{\mathrm{d} \lambda}) }{\mathrm{d}t} f~.	\\
\frac{\mathrm{d}f}{\mathrm{d} \lambda} &= -\, \frac{\mathrm{d} \ln (\frac{\mathrm{d} \tilde{\lambda}}{\mathrm{d} \lambda}) }{\mathrm{d}t} P^0 f~.
\end{align}
Applying the relation between the two af\mbox{}f\mbox{}ine parameters in Eq.~\eqref{lambdas} and the Christof\mbox{}fel symbols stated earlier to the Liouville equation in Eq.~\eqref{liouvilleeq}, yields the Einstein frame Boltzmann equation
\begin{align}
\label{simplebolt}
P^0\frac{\partial f}{\partial t} - H P P^0 \frac{\partial f}{\partial P} &= \frac{\dot{\phi}\ddot{\phi}}{M^4 - \dot{\phi}^2}P^0f~,
\end{align}
where we have also used the geodesic equation $ \tilde{g}_{\mu \nu} P^{\mu} P^{\nu} = 0 $ \footnote{We remind the reader that we rewrite the geodesic equation in Eq.~\eqref{geodesic} by using Eq.~\eqref{momtransformed}} which leads to
\be
\frac{\mathrm{d} P^0}{\mathrm{d} P} = \frac{M^4}{M^4 - \dot{\phi}^2} \frac{a^2 P}{P^0}~,
\ee
with $ P^2 = \delta_{ij} P^i P^j $.
We will require def\mbox{}initions of the energy density, pressure and number density derived from the statistical representation of the energy-momentum tensor; they are given as follows
\begin{subequations}
\label{stattherm}
\begin{align}
\rho &= \frac{1}{(2 \pi)^3} \int \frac{\mathrm{d}^{3} P}{\sqrt{-g}} \,P^0 f~,		\\
p &= \frac{1}{(2 \pi)^3} \int \frac{\mathrm{d}^{3} P}{\sqrt{-g}} \, \frac{a^{2} \delta_{ij} P^i P^j}{3P^0} f ~,			\\
n &= \frac{1}{(2 \pi)^3} \int \frac{\mathrm{d}^{3} P}{\sqrt{-g}} \,f~,		
\end{align}
\end{subequations}
where $ n $ is the particle number density.
By integrating Eq.~\eqref{simplebolt} in momentum space, we obtain again the same equation as Eq.~\eqref{joreinconservation} and hence, this kinetic theory
is consistent with both the f\mbox{}luid and action approaches.
\newline
\\
When Eq.~\eqref{simplebolt} is compared to the the standard collisionless Boltzmann equation, we see an extra term present on the right-hand side. 
The presence of this additional term is a result of photons not travelling on geodesics in the Einstein frame.
We can interpret this modif\mbox{}ication as an ``ef\mbox{}fective'' collision term as it contains no derivatives of the distribution function $ f $.
\newline
\\
In addition to its ability to produce the conservation equation, the Liouville equation in Eq.~\eqref{liouvilleeq} can also be used to construct an equation describing the evolution of the particle number density
\be
\label{numdensint}
\int \frac{\mathrm{d}^{3} P}{\sqrt{-g}} \frac{\hat{\mathcal{L}} f}{P^0} = \int \frac{\mathrm{d}^{3} P}{\sqrt{-g}} \frac{\mathcal{C}[f]}{P^0}~,
\ee
with attention placed upon the factor of $ 1/P^0 $ on both sides of the equation.
Applying this factor to the Einstein frame Boltzmann equation in Eq.~\eqref{simplebolt} yields
\be
\label{numint}
\frac{\partial f}{\partial t} - H P \frac{\partial f}{\partial P} = \frac{\dot{\phi}\ddot{\phi}}{M^4 - \dot{\phi}^2} f~,
\ee
and by integrating Eq.~\eqref{numint} over momentum space whilst utilising the def\mbox{}initions stated earlier in Eqs.~\eqref{stattherm} yields the equation for the particle number density in the Einstein frame, which is as follows
\be
\label{particlenum}
\dot n + 3 H n = - \frac{\dot F}{2F} n~, 
\ee
with $F = (d\tilde \lambda/d\lambda)^2$. 
This equation implies that the particle number in the Einstein frame is not conserved and hence the CMB is not a blackbody (and does not obey the adiabaticity condition in Eq.~(19) in \cite{Lima:1995kd} --- this relation is also used in \cite{Lima:2000ay}).
As a further check, the equation of state for the Einstein frame obtained from kinetic theory is 
\begin{align}
g_{\mu \alpha} T^{\alpha \nu} &= g_{\mu \alpha} \int \frac{ \mathrm{d}^{3} P}{\sqrt{-g}} \frac{P^{\alpha}P^{\nu}}{P^0} f~,	\nonumber	\\
						&= \bigg( \tilde{g}_{\mu \alpha} - \frac{\phi,_{\mu} \phi,_{\alpha}}{M^4} \bigg) \int \frac{ \mathrm{d}^{3} P}{\sqrt{-g}} \frac{P^{\alpha}P^{\nu}}{P^0} f~,	\nonumber	\\
-\rho + 3p					&= -\frac{\dot{\phi}}{M^4} \int \frac{ \mathrm{d}^{3} P}{\sqrt{-g}} P^0 f~,		\nonumber	\\
-\rho + 3p					&= -\frac{\dot{\phi}}{M^4} \rho~,		\nonumber	\\
\therefore \qquad 	3p		&= \bigg( 1 - \frac{\dot{\phi}^2}{M^4} \bigg) \rho~,
\end{align}
where $ \tilde{g}_{\mu \nu} P^{\mu} P^{\nu} = 0 $.
We f\mbox{}ind it is the same as that from the f\mbox{}luid description Eq.~\eqref{eos}.
\newline
\\
From this, we can state that the equation of state is not frame-independent in the presence of disformal couplings. 
This is in exception to when the pressure in the Jordan frame is zero, which in turn makes the Einstein frame pressure zero.
In addition, the distribution function in the context of disformal couplings, is also revealed not to be a frame-independent quantity.

\section{Observables}
\label{sec:chap4sec4}
We shall return to the original problem with two species: radiation and matter, with matter containing both baryonic and dark, that are separately disformally coupled to the scalar f\mbox{}ield.
\newline
\\
In this scenario, it is important to note that there are three distinctive frames:
\begin{itemize}
\item The Einstein frame, which takes the action of the form stated in Eq.~\eqref{eframeaction}. 
\item The Òradiation frameÓ. This is a frame in which the CMB radiation is uncoupled from the scalar f\mbox{}ield and due to this, 
matter is in general coupled to the f\mbox{}ield.
\item The Jordan frame. In this frame, radiation is in general coupled to the scalar, with matter (encompassing both baryonic and dark) uncoupled. 
\end{itemize}
For the f\mbox{}inal two frames described, the gravity-scalar part of the action will have a nonstandard form. 
Note that if the strength of the disformal couplings in both matter and radiation are the same, both the radiation and Jordan frames coincide.
From the previous section, we know that the distribution function and the equation of state is dif\mbox{}ferent for each frame, and this af\mbox{}fects observables. 
\newline
We will perform all calculations in the Einstein frame for simplicity. 
Once this is done, observables will be expressed in terms of the Jordan frame in order to compare with data. 
The Jordan frame is the most natural choice as this is the frame where matter (and hence experimental equipment) is uncoupled.
\newline
\\
To search for disformal couplings we employ two tests.
Two such tests chosen are the redshift-temperature and the $ \mu $-distortion;
both the temperature of the CMB radiation and spectral distortions have been previously used to place constrains on extensions to the standard theory \cite{Lima:2000ay, Mirizzi:2005ng}. 
\newline
\\
The CMB temperature evolution has been studied in the redshift range $ 0 \leq z \leq 3 $ by the following authors \cite{Mather:1998gm, Luzzi:2009ae, Noterdaeme:2010tm, Avgoustidis:2011aa}.
In order to study the temperature-redshift relation, an expression of the redshift is required. 
The def\mbox{}inition of the redshift is given in terms of the f\mbox{}luid's velocity vectors and 4-momenta; it is def\mbox{}ined as \cite{Uzan}
\be
\label{redshift}
1 + z = \frac{ (u_{\mu} P^{\mu})^{(\mathrm{m})}_{\mathrm{em}} }{ (u_{\mu} P^{\mu})^{(\mathrm{m})}_{\mathrm{obs}} }~,
\ee
where $ z $ is the redshift, $ u_{\mu} $ are the 4-velocity vectors and $ P^{\mu} $ are the 4-momenta.
The subscripts ``em'' and ``obs'' stand for emission and observation respectively, with superscript ``m'' representing the matter frame.
In this expression, all quantities are those which belong to the Jordan frame; the 4-velocity is $ u_{\mu} = (-c_{\mathrm{obs}}, 0, 0, 0) $ and $ P^{\mu} $ is the measured 4-momenta of the photon.
The measured speed of light $ c_{\mathrm{obs}} $ is given in Eq.~\eqref{measspeed}.
\newline
\\
Using the requirements of radiation in the Jordan frame (Eq.~\eqref{geodesic}), we know that photons will follow disformal geodesics. 
Equation~\eqref{geodesic} in combination with Eq.~\eqref{redshift} reveals that the redshift relation in terms of the baryonic metric is the same as that found for General Relativity:
\be
1 + z = \frac{a_0}{a}~,
\ee
where $ a_0 $ is the present day scale factor.
An ef\mbox{}fective temperature is assigned to the energy density of the CMB radiation in the Jordan frame 
$ \rho_\gamma^{(\mathrm{m})} $, through the relation $ \rho \propto T^4 $ \cite{Chluba:2011hw}.
\newline
\\
The second test involves the study of $ \mu $-distortions.
Previously, we stated that the distribution function $ f $ alone is not conserved but instead obeys Eq.~\eqref{df/dt}.
We make the assumption that the CMB when f\mbox{}irst produced is to a very high accuracy a blackbody. 
The distribution function in the radiation-dominated epoch is of Planckian form
\be
f(\vec{\nu},\,T) = \frac{1}{e^{\, \nu/T } - 1 }~,
\ee
and in addition to this, $ \dot{\phi} = 0 $.
We further ignore the, if any, spectral distortions that were generated during the early Universe. 
Over time the contribution made by the derivative of $ \phi $ will become more signif\mbox{}icant and will af\mbox{}fect the form of the distribution function:
\be
f_{\mathrm{obs}} = f_{\mathrm{ini}} \bigg( \frac{\mathrm{d} \lambda}{\mathrm{d} \tilde{\lambda}} \bigg)_0~,
\ee
where $ f_{\mathrm{obs}} $ and $ f_{\mathrm{ini}} $ are the observed and initial-time distribution functions and subscript ``0'' indicates at observation i.e. at present time.
The presence of the factor $ \mathrm{d} \lambda/ \mathrm{d} \tilde{\lambda} $ can be quantif\mbox{}ied as an ef\mbox{}fective chemical potential $ \mu $ in a distribution function. 
Hence, the observed distribution function can be written in the form
\be
f(\vec{\nu},\,T) = \frac{1}{e^{\, \nu/T + \mu} - 1 }~.
\ee
The chemical potential should be viewed as a way to quantify the deviation from the Planckian spectrum caused by the disformal coupling.
\newline
\\
Given that observationally $ \mu $-distortions are very small i.e. $ \mu \ll 1 $ --- observations include those made by COBE FIRAS \cite{Fixsen:1996nj, Mather:1993ij} and PIXIE \cite{Kogut:2011xw}, see Dent et al.~for a discussion on $ \mu $-distortions \cite{2012PhRvD..86b3514D} --- we can use this in combination with Eq.~\eqref{df/dt} to f\mbox{}ind that
\be
\label{mudist}
\mu \approx ( c_{\mathrm{obs}} - 1 ) ( 1 - e^{-\nu/T})~,
\ee
where $ c_{\mathrm{obs}} $ is the measured (dimensionless) speed of light.
By using the following line elements for the observer and photons
\begin{align}
\label{linematter}
\mathrm{d} s_{\mathrm{obs}}^2 & = g_{\mu \nu}^{\mathrm{m}} \mathrm{d} x^{\mu} \mathrm{d} x^{\nu} 
							= -\mathrm{d} t^2 + \mathrm{d} \vec{x}_{\mathrm{obs}}^{\,2}~,		\\
\label{linephot}
\mathrm{d} s_{\gamma}^2 & = g_{\mu \nu}^{\gamma} \mathrm{d} x^{\mu} \mathrm{d} x^{\nu} 
							= 0~,
\end{align}
and the metrics as stated in Eq.~\eqref{gmunu} with $ C = 1 $ and $ D_i = M_i^{-4} $ where $ i $ indicates the matter species i.e. matter and radiation, we can rewrite the photon metric in terms of the matter metric
\be
g_{\mu \nu}^{\gamma} = g_{\mu \nu}^{\mathrm{m}} + \bigg( \frac{1}{M_{\gamma}^4} - \frac{1}{M_{\mathrm{m}}^4} \bigg) \phi_{,\mu} \phi_{,\nu}~.
\ee
From the line elements stated in Eqs.~\eqref{linematter} and \eqref{linephot}, this leads to
\begin{align}
\mathrm{d} s_{\gamma}^2 & = \bigg[ g_{\mu \nu}^{\mathrm{m}} + \bigg( \frac{1}{M_{\gamma}^4} - \frac{1}{M_{\mathrm{m}}^4} \bigg) \phi_{,\mu} \phi_{,\nu} \bigg] \mathrm{d} x^{\mu} \mathrm{d} x^{\nu}~,		\nonumber	\\
0	& = -\mathrm{d} \tau_{\mathrm{m}}^2 + \mathrm{d} \vec{x}_{\mathrm{obs}}^{\,2} + \bigg( \frac{1}{M_{\gamma}^4} - \frac{1}{M_{\mathrm{m}}^4} \bigg) \mathrm{d} \phi^2~,		\nonumber	\\
\Rightarrow \quad 	\frac{ \mathrm{d} \vec{x}_{\mathrm{obs}}^{\,2} }{ \mathrm{d} \tau_{\mathrm{m}}^2 } & = 1 - \bigg( \frac{1}{M_{\gamma}^4} - \frac{1}{M_{\mathrm{m}}^4} \bigg) \frac{ \mathrm{d} \phi^2 }{ \mathrm{d} \tau_{\mathrm{m}}^2 }~,		\nonumber	\\
\therefore \qquad 	c_{\mathrm{obs}}^2 & = 1 - \bigg( \frac{1}{M^4_{\gamma}} - \frac{1}{M^4_{\mathrm{m}}} \bigg) \bigg( \frac{\mathrm{d} \phi}{\mathrm{d} \tau_{\mathrm{m}}} \bigg)^2~,
\end{align}
where we have used the fact that when an observer is at rest $ \mathrm{d} s_{\mathrm{obs}}^2 = - \mathrm{d} t^2 = - \mathrm{d} \tau_{\mathrm{m}}^2 $, and $ \tau_{\mathrm{m}} $ is the proper time of the observer.
Alternatively, the observed speed of light can be expressed in terms of Einstein time variables
\be
\label{measspeed}
c_{\mathrm{obs}}^2 = \frac{1 - \dot{\phi}^2/M^4_{\gamma}}{1 - \dot{\phi}^2/M^4_{\mathrm{m}}}~.
\ee
There are limits which have been experimentally found by COBE and FIRAS \cite{Fixsen:1996nj}: $ | \mu | < 9 \times 10^{-5} $. 
This will be used to constrain the $ M_{\mathrm{m}} \times M_{\gamma} $ parameter space.

\section{Results}
\label{sec:chap4sec5}
\begin{figure}
\begin{center}
\scalebox{0.7}{\includegraphics{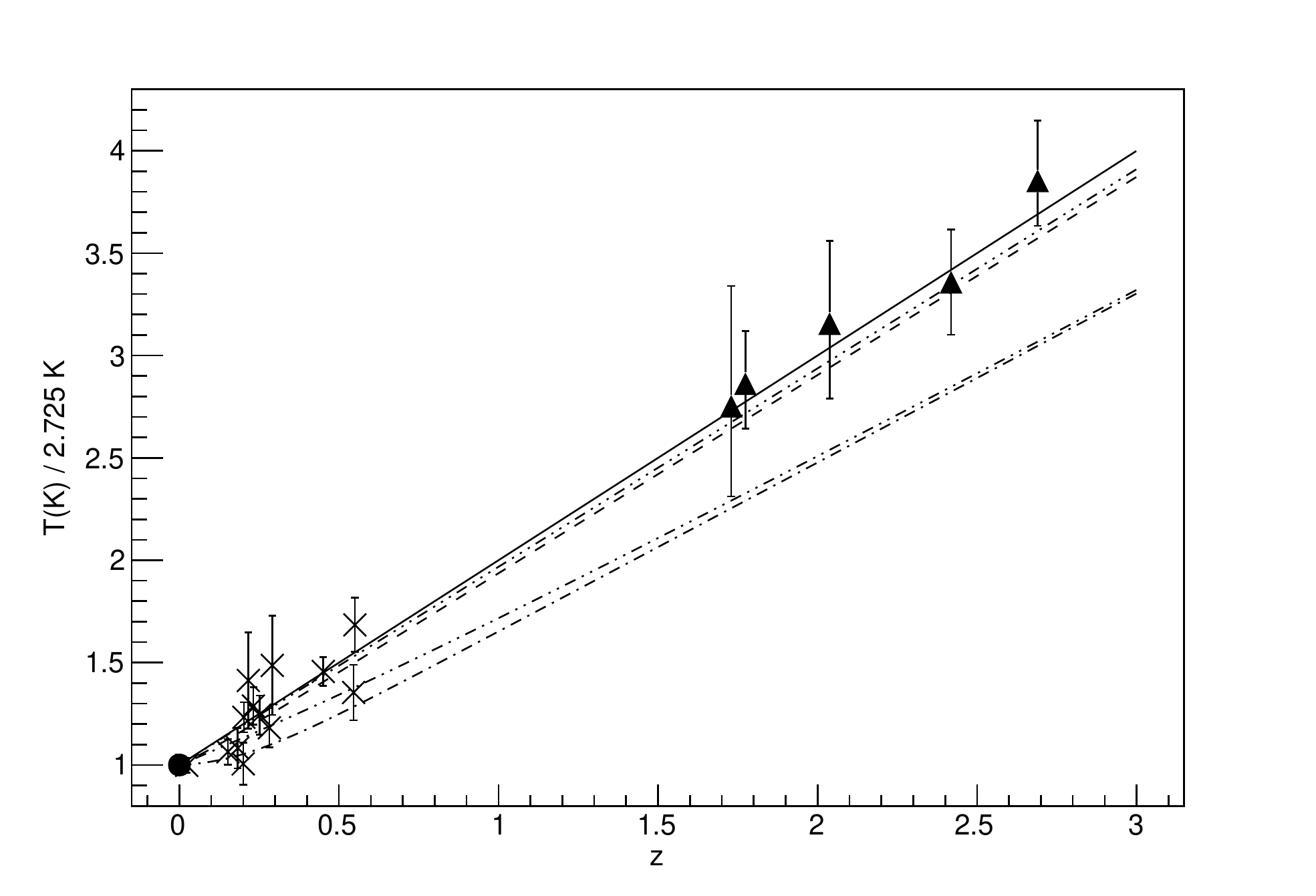}}
\caption{Plot of the ratio $T(z)/T_0$ where $T_0 = 2.725 \mathrm{K} $, against redshift, $ z $, for the exponential potential. The black circle marked at $ z = 0 $ is measured by COBE \cite{Mather:1998gm}. Dif\mbox{}ferent sets of measurements and their corresponding errors: Those marked with crosses are given by \cite{Luzzi:2009ae} and triangles are by \cite{Noterdaeme:2010tm}. 
There are f\mbox{}ive lines marked on this plot and each has an associated $ M_{\gamma} $: Solid line is $ M_{\gamma} \rightarrow \infty $, dashed line $M_{\gamma}=2.203\times10^{-5}$~eV, dashed line with one dot $M_{\gamma}=3\times10^{-5}$~eV, dashed line with two dots $M_{\gamma}=1.5\times10^{-3}$~eV and dashed line with three dots $M_{\gamma}=2.2188\times10^{-3}$~eV. For this plot, we f\mbox{}ix $M_{\rm m} = 0.05$~eV.}
\label{temperature}
\end{center}
\end{figure}

The Klein-Gordon equation and the conservation equations for the various species in Eqs.~\eqref{kg} and \eqref{rho_i} are integrated such that the present day density parameter values are $\Omega_{\rm m,0} = 0.3$ (cold dark matter and baryons), $\Omega_{\phi,0}=0.7$ (dark energy) \cite{Komatsu:2010fb} and current temperature of the CMB is $ T_0 = 2.725 \mathrm{K} $ \cite{Mather:1998gm}. 
The initial f\mbox{}ield value is f\mbox{}ixed at $ \phi_\mathrm{ini} = 1.5 $ and $ V_0 $ is allowed to vary in order to obtain the our present day Universe.
The two methods stated in the previous section: temperature-redshift and $\mu$-distortions, are then calculated. 
\newline
\\
The expected temperature evolution as a function of redshift is numerically calculated for a range of $ M_\gamma $; our results are shown in Figure~\ref{temperature}. 
Figure~\ref{temperature} displays the values of $ M_\gamma $ which are in agreement and those not with current measurements from \cite{Mather:1998gm, Luzzi:2009ae, Noterdaeme:2010tm, Avgoustidis:2011aa}.
Also included is the prediction of General Relativity, for $ M \rightarrow \infty $.
A large $ M_\gamma $ results in a weak coupling between the scalar f\mbox{}ield and radiation. 
We f\mbox{}ind that this weak coupling distorts the linear temperature-redshift relation found in General Relativity. 
However, the coupling has a damping ef\mbox{}fect on the f\mbox{}ield which increases as the coupling increases (equivalent to $ M_\gamma $ decreasing). 
When the coupling is very large, the dampening is so severe that it results in $ \dot{\phi} \rightarrow 0 $. 
Due to disformal couplings being composed of f\mbox{}ield derivatives, it is clear that when this limit is reached the coupling will vanish (shown by the tripledot-dashed line in Figure~\ref{temperature}).
\begin{figure}
\begin{center}
\scalebox{0.27}{\includegraphics{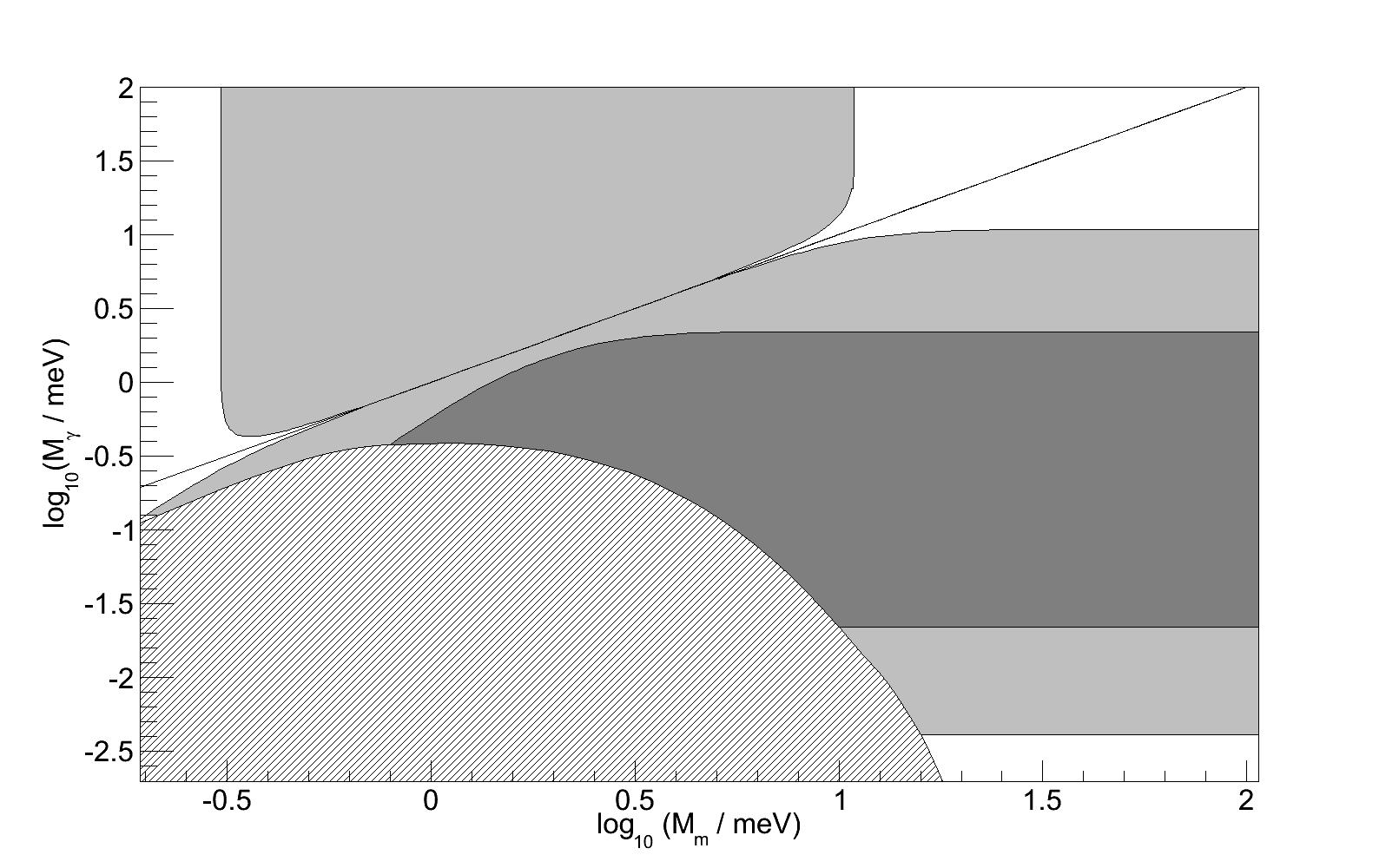}}
\caption{
Bounds on the $M_{\mathrm{m}} \times M_{\gamma}$ parameter space. 
The solid diagonal line represents $M_{\mathrm{m}} = M_{\gamma}$ .
The darkest shaded region is excluded by the CMB temperature evolution alone, we show the exclusion region above 68\% C.L. .
The light-gray shaded regions are excluded by the measured constraints on the $\mu$-distortions \cite{Fixsen:1996nj}. 
The line-shaded (hatched) area is excluded from our search of the parameter space as the numerics are unreliable in that region. 
Note that the regions excluded by the $\mu$-distortion never touch the line $ M_{\gamma} = M_{\mathrm{m}} $, since in this limit the $\mu$-distortion vanishes.
}
\label{exclus}
\end{center}
\end{figure}
\newline
\\
A chi-squared minimisation method with 68\% (1$\sigma$) conf\mbox{}idence level is used to obtain an excluded region in $ M_{\gamma} \times M_{\mathrm{m}} $ parameter space. 
By comparing the expression for the $ \mu $-distortion in Eq.~\eqref{mudist} against measured constraints in \cite{Fixsen:1996nj}, 
we f\mbox{}ind them to be similar when under the assumption that the frequencies $ \nu $ are of order $ T_0 $.
Bounds on the $ M_{\mathrm{m}} \times M_\gamma $ parameter space acquired from both methods are presented in Figure~\ref{exclus}.
\newline
\\
We f\mbox{}ind that the constraints found from $ \mu $-distortions are much stronger than those from the temperature-redshift relation. 
Adding to this, the overall shape from both methods are alike, which draws the conclusion that there must be a dif\mbox{}ference in the two couplings for any disformal ef\mbox{}fect to be observable.

\section{Conclusion}
\label{sec:chap4sec6}
From this study of disformal transformations, there are two points worth emphasising: 
f\mbox{}irstly, we f\mbox{}ind that for a given f\mbox{}luid, the equation of state under disformal transformations is in general, a frame dependent quantity and secondly, 
this coupling 
has an effect upon the distribution function, and consequently making the distribution frame dependent.
By considering CMB radiation coupled disformally to a scalar f\mbox{}ield, we found a modif\mbox{}ication to the  distribution function of the CMB photons. 
This modif\mbox{}ication disappears when both baryon and radiation coupling strengths to the scalar f\mbox{}ield, are equal.
\newline
\\
Throughout this work, all observable quantities were expressed in the Jordan frame; in this frame, the baryonic matter and scalar f\mbox{}ield are uncoupled. 
This work can be further developed. For instance, the couplings strengths for both $ M_{\mathrm{m}} $ and $ M_\gamma $ should be in general, allowed to be dependent on the scalar f\mbox{}ield.
At the start of this work the conformal factor was set as $ C(\phi) = 1 $, with the purpose to study a purely disformal scenario.
The most natural follow-up to this study is to examine the scalar f\mbox{}ield dynamics and ultimately, the constraints posed by the temperature-redshift relation and $ \mu $-distortion, with the inclusion of the conformal coupling.
In addition, conformal couplings require screening in higher density regions of the Universe due to a f\mbox{}ifth force arising in such models, so naturally disformal couplings with a screening mechanism should also be studied \cite{Koivisto:2012za}.
In these models, the quantity $ \dot{\phi} $ is expected to be much smaller and this will in turn, diminish contributions from the disformal coupling.
A study into the evolution of the f\mbox{}irst order perturbations is essential as disformal couplings will af\mbox{}fect the CMB temperature anisotropies and structure formation.

\chapter{Conclusion}
There are two periods of accelerated expansion within the lifetime of our Universe: the first is inflation which occurs at early times, and second is a result of dark energy at later times.
Currently standing, there is an abundance of concrete models that are able to describe the dynamics during both of these periods.
Many theoretical cosmologists have developed theories including those from string theory. 
One type of theory considered to be a more natural extension of General Relativity are scalar-tensor theories.
In this Thesis we have three cosmological studies with the first two in the context of inflation, and the final one in a dark energy scenario.
\newline
\\
In Chapter 2, three models of inflation were considered with all containing at least one nonminimally coupled scalar field.
The models taken into account included double inflation, followed by one with a kinetic coupling and lastly, a DBI field; these were chosen with the purpose to study the effects of their kinetic terms upon the power spectra evolution for the curvature and isocurvature perturbations.
We found that the inclusion of these noncanonical kinetic terms causes the isocurvature power spectrum to vary.
To be specific, if inflation is driven by the scalar field containing such noncanonical kinetic term, this would cause deviations of the isocurvature power spectrum  amplitude from the case of double inflation.
From our study, we found that a DBI field driving the inflationary period gives rise to the largest isocurvature perturbations, which will in turn affect predictions of the B-mode polarization in the CMB, which are currently being under investigation by experiments such as BICEP2 and Planck.
\newline
\\
In the following chapter, we studied the scalar and tensor power spectra generated during inflation with the effects of a sharp transition in the effective Planck mass.
We performed all calculations within the Jordan frame, i.e. physical frame.
In the single-field model we considered, the effects of the sudden transition can clearly be seen on the scalar power spectrum, however, not so evident on the tensor spectrum.
We have also shown that with the inclusion of this transition occurring at scales close to those leaving our horizon, and hence within our observable window of scales, it is possible to reduce tension between the Planck and BICEP2 observations of the tensor-to-scalar ratio.
To further this work, the addition of a minimally-coupled scalar field was made. 
However, the presence of the second field caused the suppression of all features resulting from the sharp transition, within the primordial power spectra.
\newline
\\
In Chapter 4, we move forward to later times in order to study dark energy with the addition of a nonconformal coupling, namely the disformal coupling.
We considered the model in which all matter species, including radiation and matter, are disformally coupled to the scalar field.
To ensure the theory is consistent throughout before pursuing observable quantities, the fluid description and kinetic theory were established.
From this, we found that the equation of state for the species in question, is a frame dependent quantity.
In addition to this, the distribution function of the CMB changes between frames under disformal transformations (but is invariant under conformal transformations).
This modification ends when both the radiation and matter species are coupled with equal magnitude to the scalar field.
In order to relate this coupling to observations, we choose to perform calculations in the physical frame.
For the case where only disformal couplings are present, we have also placed constraints on the strengths of both the radiation and matter couplings through both the temperature-redshift relation and $ \mu $-distortions.
\newline
\\
To extend the study in Chapter 2, the full primordial power spectra for the curvature and isocurvature modes for all models considered will need to be produced, and this to be followed by research into the pre- and reheating mechanism.
Post-inflation dynamics will also be required for the theory studied in Chapter 3.
One logical extension to the project on disformal couplings, in particular to radiation, is to consider the perturbations generated.
All the models considered in this Thesis will require the generation of the CMB angular power spectrum, and then for this to be compared to experimental data.
\newline
\\
Within the last three years, there have been many experiments releasing their data including those from Planck and BICEP2, all of which will place further constraints on theoretical models, and possibly rule some out.
The results of the BICEP2 experiment state that B-mode polarization has been detected.
However, there are questions about their analysis and as such, has led to suspicions about their findings on gravitational waves.
At this current time, Planck has not released their results on polarization, but are in slight disagreement with those from BICEP2.
With the first two inflationary studies presented in this Thesis, we see that it is possible to find models producing isocurvature modes relating to these B-mode polarizations, and it is possible to resolve disagreements between the two major experiments.
For the case of dark energy, we still have no clue what it is, and with the proposal of unusual couplings between matter species, it will lead to more research within this area.
With all this said, we will always require experiments with ever increasing precision to probe the CMB, and hopefully reveal that our Universe is not
governed by the dynamics of minimally-coupled scalar fields.


\appendix
\chapter{Numerical method}
This appendix describes the numerical methods that were required throughout this Thesis.
We have used the method outlined by \cite{Tsujikawa:2002qx}, but there are many other approaches available in the literature, for example see references \cite{Salopek:1990jq, Bardeen:1983qw}.

\section{Initial conditions}
The initial conditions for the background fields are chosen such that they produce an inflationary period which lasts for a minimum of 60 $ e $-foldings; the end of inflation is recorded when the slow-roll parameter $ \varepsilon $ reaches 1.
In addition, the derivatives of the background fields are given their slow-roll values.
The Bunch-Davies vacuum fluctuation of the scalar field in terms of conformal time is given by \cite{LiddLyth} 
\be
\label{bd1}
\delta \phi_{\, \mathrm{ini}} = \frac{1}{a\sqrt{2k}} e^{-i k \eta }~,
\ee
and is the oscillatory initial condition used in the numerical runs, the derivative of Eq.~\eqref{bd1} with respect to conformal time is
\be
\delta \phi'_{\, \mathrm{ini}} = - \frac{1}{a\sqrt{2k}} ( \mathcal{H} + i k ) e^{-i k \eta }~,
\ee
where $ \mathcal{H} = a'/a $ is the conformal time Hubble parameter.
For the case of a DBI field, this initial conditions changes slightly to include the Lorentz-like factor $ \gamma $ \cite{vandeBruck:2010yw}
\begin{subequations}
\begin{align}
\delta \chi_{\, \mathrm{ini}} & = \frac{1}{a \gamma \sqrt{2k}} e^{-i \frac{k}{\gamma} \eta }~,		\\
\delta \chi'_{\, \mathrm{ini}} & = - \frac{1}{\sqrt{2k}} \bigg( \mathcal{H} + \frac{\gamma'}{\gamma} + i \frac{k}{\gamma} \bigg) e^{-i \frac{k}{\gamma} \eta }~.
\end{align}
\end{subequations}

\section{Set up}
Most, if not all differential equations mentioned in this Thesis e.g. background Klein-Gordon and field perturbation equations, are of second-order.
In order to solve these equations numerically, each second-order differential equation will need to be rewritten to form two first-order differential equations.
The commonly used integration variable is the $e$-fold number as stated in Chapter 1 where $ N = \ln{a} $.
For example, to solve the background Klein-Gordon equation for a minimally coupled scalar field with respect to the number of $e$-foldings,
the variables are
\be
\bold{y} =
\begin{pmatrix}
\phi   \\
\phi_{,N}   
\end{pmatrix}~,
\ee
and the set of differential equations which need to be solved are
\be
\bold{y}_{,N} =
\begin{pmatrix}
\phi_{,N}   \\
- \bigg( 3 + \frac{H_{,N}}{H} \bigg) \phi_{,N} - \frac{V'(\phi)}{H^2}
\end{pmatrix}~,
\ee
where $ \,_{,N} $ indicates derivative with respect to $ e $-fold number.
The differential equations in this Thesis were solved using Runge-Kutta 5th order and 8th order methods.
It must be noted that for the case of the DBI field, it is best to perform the integral in logarithmic time, $ x = \ln{t} $.
\newline
\\
The code begins with the condition
\be
\frac{k}{aH} \gg 1~,
\ee
which ensures that the field perturbations are deep within the horizon.
Note that the wavenumber $ k $ is the same as that arising in the perturbation equations.
\newline
\\
We treat the perturbations as stochastic variables deep within the horizon.
In addition to this, the modes must be uncorrelated deep within the horizon and to simulate this we must perform two numerical runs \cite{Tsujikawa:2002qx}: the first run is set where the gauge-invariant perturbation of the field $ \phi $ is set in the Bunch-Davies vacuum and the gauge-invariant perturbations of the field $ \chi $ is set to zero, and for the second run the conditions are swapped.
\newline
\\
Each of these runs are used to evaluate the curvature perturbation, either $ \zeta $ or $ \mathcal{R} $, and then combined to form the curvature power spectrum
\be
\mathcal{P}_{\zeta} = \frac{k^3}{2 \pi^2} ( |\zeta_1|^2 + |\zeta_2|^2 )~,
\ee
where the subscripts $ 1, 2 $ denote the run number.
This method is the same for the entropy power spectrum.

\backmatter

\newpage
\addcontentsline{toc}{chapter}{Bibliography}

\bibliographystyle{hieeetr}

\bibliography{refs.bib}

\end{document}